\documentclass[AMA,Times1COL]{WileyNJDv5} 
\RequirePackage{amsthm,amsmath,amsfonts,amssymb}
\RequirePackage{graphicx}

\usepackage{rotating}
\usepackage{multirow}
\usepackage{bm}
\usepackage{makecell}
\usepackage{mathtools}
\usepackage{url}
\usepackage{dsfont}
\usepackage{enumerate}
\mathtoolsset{showonlyrefs}
\allowdisplaybreaks

\makeatletter
\newcommand*{\Rom}[1]{\expandafter\@slowromancap\romannumeral #1@}
\makeatother

\newcommand*{\rom}[1]{\romannumeral #1}
\DeclareMathOperator*{\argmax}{arg\,max}
\DeclareMathOperator*{\argmin}{arg\,min}

\newcommand{\tp}{\mathbb{P}}

\newcommand{\te}{\mathbb{E}}
\newcommand{\cov}{\textup{Cov}}

\newcommand{\btheta}{\bm{\theta}}
\newcommand{\bepsilon}{\bm{\epsilon}}
\newcommand{\mC}{\mathcal{C}}
\newcommand{\mE}{\mathcal{E}}

\newcommand{\mL}{\mathcal{L}}

\newcommand{\bbeta}{\bm{\beta}}

\newcommand{\bbetak}[1]{\bm{\beta}^{(#1)}}

\newcommand{\bbetaks}[1]{\bm{\beta}^{(#1)*}}
\newcommand{\hbetak}[1]{\widehat{\beta}^{(#1)}}
\newcommand{\hbeta}{\widehat{\beta}}
\newcommand{\hbbeta}{\widehat{\bm{\beta}}}
\newcommand{\hbbetak}[1]{\widehat{\bm{\beta}}^{(#1)}}
\newcommand{\hDelta}{\widehat{\bm{\Delta}}}
\newcommand{\bDeltak}[1]{\bm{\Delta}^{(#1)}}
\newcommand{\hDeltak}[1]{\widehat{\bm{\Delta}}^{(#1)}}
\newcommand{\tDelta}{\widetilde{\bm{\Delta}}}
\newcommand{\bmu}{\bm{\mu}}
\newcommand{\bDelta}{\bm{\Delta}}
\newcommand{\bTheta}{\bm{\Theta}}
\newcommand{\hTheta}{\widehat{\bm{\Theta}}}
\newcommand{\bgamma}{\bm{\gamma}}
\newcommand{\hgamma}{\widehat{\bm{\gamma}}}
\newcommand{\htau}{\widehat{\tau}}

\newcommand{\infnorma}[1]{\left\|#1\right\|_{\infty}}

\newcommand{\infnorm}[1]{\|#1\|_{\infty}}
\newcommand{\twonorm}[1]{\|#1\|_{2}}
\newcommand{\onenorm}[1]{\|#1\|_{1}}
\newcommand{\zeronorm}[1]{\|#1\|_{0}}
\newcommand{\diag}{\text{diag}}

\newcommand{\maxnorm}[1]{\|#1\|_{\max}}

\newcommand{\norma}[1]{\left|#1\right|}
\newcommand{\norm}[1]{|#1|}
\newcommand{\bx}{\bm{x}}

\newcommand{\yk}[1]{y^{(#1)}}
\newcommand{\byk}[1]{\bm{y}^{(#1)}}

\newcommand{\bY}{\bm{Y}}
\newcommand{\bX}{\bm{X}}
\newcommand{\bSigma}{\bm{\Sigma}}
\newcommand{\hSigma}{\widehat{\bm{\Sigma}}}
\newcommand{\huSigma}{\widehat{\Sigma}}

\newcommand{\bmuks}[1]{\bm{\mu}^{(#1)*}}

\newcommand{\bz}{\bm{z}}

\newcommand{\<}{\langle}
\renewcommand{\>}{\rangle}
\newcommand{\bu}{\bm{u}}

\newcommand\smallo{
  \mathchoice
    {{\scriptstyle\mathcal{O}}}
    {{\scriptstyle\mathcal{O}}}
    {{\scriptscriptstyle\mathcal{O}}}
    {\scalebox{.7}{$\scriptscriptstyle\mathcal{O}$}}
}

\articletype{Article Type}%

\received{Date Month Year}
\revised{Date Month Year}
\accepted{Date Month Year}
\journal{Journal}
\volume{00}
\copyyear{2023}
\startpage{1}

\begin{document}

\title{$\ell_1$-penalized Multinomial Regression: Estimation, inference, and prediction, with an application to risk factor identification for different dementia subtypes}

\author[1]{Ye Tian}
\author[2]{Henry Rusinek}
\author[2]{Arjun V. Masurkar}
\author[3]{Yang Feng}

\authormark{TIAN \textsc{et al.}}
\titlemark{$\ell_1$-penalized Multinomial Regression: Estimation, inference, and prediction, with an application to risk factor identification for different dementia subtypes}

\address[1]{\orgdiv{Department of Statistics}, \orgname{Columbia University}, \orgaddress{\state{New York}, \country{USA}}}

\address[2]{\orgdiv{Grossman School of Medicine}, \orgname{New York University}, \orgaddress{\state{New York}, \country{USA}}}

\address[3]{\orgdiv{Department of Biostatistics, School of Global Public Health}, \orgname{New York University}, \orgaddress{\state{New York}, \country{USA}}}

\corres{Yang Feng, Department of Biostatistics, School of Global Public Health, New York University, USA \email{yang.feng@nyu.edu}}



\abstract[Abstract]{High-dimensional multinomial regression models are very useful in practice but have received less research attention than logistic regression models, especially from the perspective of statistical inference. In this work, we analyze the estimation and prediction error of the contrast-based $\ell_1$-penalized multinomial regression model and extend the debiasing method to the multinomial case, providing a valid confidence interval for each coefficient and $p$-value of the individual hypothesis test. We also examine cases of model misspecification and non-identically distributed data to demonstrate the robustness of our method when some assumptions are violated. We apply the debiasing method to identify important predictors in the progression into dementia of different subtypes. Results from extensive simulations show the superiority of the debiasing method compared to other inference methods.}



\maketitle



\section{Introduction}\label{sec: introduction}

High-dimensional data, where the number of variables is comparable to or larger than the sample size, are ubiquitous in many applications, including health care and genomics. As a result, the topic has generated intensive recent research, such as the Lasso \citep{tibshirani1996regression}, the non-concave penalty including SCAD \citep{fan2001variable} and MCP \citep{zhang2010nearly}, the general $\ell_p$ penalty ($0 < p \leq 1$) \citep{raskutti2011minimax}, elastic net \citep{zou2005regularization}, adaptive Lasso \citep{zou2006adaptive}, group Lasso \citep{yuan2006model}, variable selection consistency \citep{zhao2006model, loh2017support}, and the general penalized M-estimators \citep{negahban2012unified, loh2015regularized}. 

Most previous works focus on estimating the sparse parameter and predicting new observations in high-dimensional classification or regression models. In practice, a valid statistical inference procedure is also essential. However, for penalized high-dimensional models like the Lasso,  inference can be very challenging because the limiting distribution of the estimator is untraceable \citep{fu2000asymptotics}. The most popular inference method is based on random sampling or bootstrapping \citep{chatterjee2011bootstrapping, chatterjee2013rates, sartori2011penalized}, such as the residual bootstrapping (usually for fixed design) and the vector bootstrapping (usually for random design). Another approach is the multiple sample splitting \citep{wasserman2009high, meinshausen2009p}, which can be used to control both the false discovery rate (FDR) and the family-wise error rate (FWER). Post-selection inference, conducted after model selection, is another popular approach \citep{zhang2022post}. As pointed out in \cite{taylor2015statistical}, the inference procedure needs to be adjusted to accommodate the model selection process. \cite{lee2016exact} proposed an inference method by exactly characterizing the distribution of the statistic conditioned on the model selection result. Another approach is the debiasing or desparsifying method \citep{javanmard2014confidence, van2014asymptotically, zhang2014confidence}, where confidence intervals and $p$-values are obtained by adding a correction term (derived from KKT conditions) to the Lasso estimator to eliminate the bias caused by penalization.

Many of the aforementioned approaches can be extended to generalized linear models or general M-estimators. However, the theoretical analysis of the logistic regression model (a \textit{single-response} model) may not carry over to the high-dimensional multinomial regression model, which is essentially a \textit{multi-response} model. In particular, for inference, more work needs to be done to extend previous methods and theoretical analysis to the multinomial regression model.

In this paper, we focus on the following high-dimensional multinomial regression model. Suppose we have a sample $\{(\bx_i, y_i)\}_{i=1}^n \overset{\textup{i.i.d.}}{\sim} (X, Y)$ generated from the distribution
\begin{equation}\label{eq: multinomial model}
	\tp(Y=k|X=\bx) = \frac{\exp((\bbetaks{k})^\top \bx)}{\sum_{k=1}^K \exp((\bbetaks{k})^\top \bx)}, \, k = 1, \ldots, K, \quad X \sim \tp_X,
\end{equation}
where $\bx, \bbetaks{k} \in \mathbb{R}^p$, $k = 1:K$, $K \geq 2$ and $\bbetaks{K} = \bm{0}_p$. Here, the class $K$ is set to be the ``reference class'', and each $\bbetaks{k}$ represents the contrast coefficient between class $k$ and the reference class. When $K = 2$, the model reduces to the logistic regression model. We assume that $\zeronorm{\bbetaks{k}} \leq s \ll p$ for all $k = 1:(K-1)$. This means that although there may be more features than observations, only a few contribute to the model. Under this high-dimensional setting, we impose the Lasso penalty on the contrast coefficients and solve the following optimization problem:
\begin{equation}
 	\{\hbbetak{k}\}_{k=1}^{K-1} =\argmin_{\{\bbetak{k}\}_{k=1}^{K-1}}\Bigg\{\underbrace{-\frac{1}{n}\sum_{k=1}^{K-1}\sum_{i=1}^n \yk{k}_i(\bbetak{k})^\top \bx_i + \frac{1}{n}\sum_{i=1}^n\log\Bigg[\sum_{k=1}^K \exp((\bbetak{k})^\top \bx_i)\Bigg]}_{\mL_n(\{\bbetak{k}\}_{k=1}^{K-1})} + \lambda\sum_{k=1}^{K-1}\onenorm{\bbetak{k}}\Bigg\}, \label{eq: lasso estimation} 
\end{equation}
where $\yk{k}_i = \mathds{1}(y_i = k)$ is an indicator of whether the $i$-th observation belongs to class $k$. Our goal is to estimate the sparse contrast coefficients $\{\bbetaks{k}\}_{k=1}^{K-1}$ and derive a valid confidence interval for each coordinate (or equivalently, deriving a $p$-value for the hypothesis test on each coordinate) by extending the debiased Lasso to the multinomial regression setting. 

In many implementations of the penalized multinomial regression (for example, R package \texttt{glmnet} and Python package \texttt{scikit-learn}), an over-parameterized version of \eqref{eq: lasso estimation} is considered, where each class is assumed to be associated with a sparse coefficient vector. This is not equivalent to the contrast-based penalization problem \eqref{eq: lasso estimation}. Although the over-parameterization with the penalty does not present the identifiability issue, it can be difficult to interpret these coefficient vectors in practice. As illustrated in the empirical studies in Section \ref{sec: numerical}, the contrast-based penalization method outperforms the over-parameterized version by yielding lower estimation error.

\subsection{Related works}\label{subsec: intro hd}
Besides the literature mentioned above, there are a few other works specifically focused on high-dimensional multinomial regression. \cite{pencer2016feature} proposed an intuitive extension of the debiased Lasso to the multinomial case without theoretical justification. We build on the same idea to extend the debiasing method by ``vectorizing the coefficient matrix in multinomial regression models'' and provide a theoretical guarantee for the technique. More recently, \cite{abramovich2021multiclass} studied the misclassification error rate on sparse high-dimensional multinomial regression and derived some minimax results. Our work differs from theirs in at least two significant ways. First, they assumed the coefficient matrix with columns $\{\bbetaks{k}\}_{k=1}^{K-1}$ to be ``row-wise sparse'', meaning that many of the non-zero components of contrast coefficients associated with different classes overlap. We assume each column of the coefficient matrix, i.e., each of $\{\bbetaks{k}\}_{k=1}^{K-1}$, to be sparse, and their non-zero components can be very different. Second, while their focus is on prediction performance, we address estimation, prediction, and inference, with a primary focus on inference.

In this paper, we focus on the high-dimensional regime $n \gg s$, $s\ll p$, and $p$ can grow exponentially as a function of $n$, commonly referred to as the \textit{sparse high-dimensional regime}. Another popular high-dimensional regime in the literature is the \textit{fixed-ratio high-dimensional regime}, where $n/p \rightarrow c_1$, $s/p \rightarrow c_2$ with some constants $c_1, c_2\in (0, 1)$. We briefly compare these two regimes and discuss their differences in three perspectives below\footnote{Thanks to one reviewer for the suggestion to add these discussions.}.
\begin{enumerate}[(i)]
    \item Studies of the fixed-ratio regime can provide a precise characterization of the mean-square error (MSE) and conditions for variable selection or signal reconstruction (e.g., \citealp{donoho2009message, donoho2010message, donoho2010message2}). Within this regime, the bias caused by penalties such as the Lasso can be explicitly quantified and managed to describe the asymptotic behavior of the penalized estimator and its MSE (e.g., \citealp{javanmard2014hypothesis, huang2017controlling, huang2020asymptotic}). However, this often requires stronger distribution assumptions or prior knowledge about the nuisance parameters in the model, such as the covariance matrices or noise variances. Additionally, the results are typically asymptotic, and finite-sample bounds are usually challenging to obtain. In contrast, in the sparse high-dimensional regime, after removing the non-vanishing part of the bias from the penalized estimator, the remaining terms become negligible as $n \rightarrow \infty$, which allows us to obtain the asymptotic normality, making the inference procedure cleaner and more tractable. Moreover, it is easier to provide non-asymptotic results of estimation error and misclassification error in this regime. 
    \item Different applications might motivate the analysis of different regimes. For instance, in compressed sensing, signals can be dense, making the fixed-ratio regime a natural choice for study \citep{kabashima2009typical}. In genomic, clinical, and biomedical research, where there can be many biomarkers and diagnostic variables but limited observations from certain populations (e.g., \citealp{ghosh2005classification, ryali2010sparse}), the number of variables typically exceeds the sample size, and signals are usually very sparse, fitting the sparse high-dimensional regime well. In our dementia study application (to be introduced in the next subsection), the sample size of specific dementia subtypes is very small, and the number of important predictors is generally much smaller than the total number of predictors. Therefore, we focus on the sparse high-dimensional regime in this paper.
\end{enumerate}
Overall, the two regimes complement each other, and the results obtained in each regime are not generally comparable. In addition, \cite{cai2017confidence} proved that an optimal adaptive inference procedure (i.e., one adaptive to unknown nuisance parameters such as the sparsity level $s$ and the covariance matrix of $X$) does not exist for high-dimensional linear regression models when $n \lesssim (s\log p)^2$. Therefore, most studies on inference in the sparse high-dimensional regime (e.g., \citealp{zhang2014confidence, van2014asymptotically, javanmard2014confidence, ning2017general, ma2024statistical}) focus on the case where $n \gg (s\log p)^2$. We conjecture that the same conclusion holds for the high-dimensional multinomial regression. In this paper, we focus on the same regime $n \gg (s\log p)^2$ and develop an adaptive inference procedure ready for practical use. If our conjecture is true, then $n \gg (s\log p)^2$ is the necessary condition for an adaptive inference procedure to work effectively, further justifying our study of this regime.

\subsection{A real example: Identifying important variables associated with different subtypes of MCI/dementia in NACC database}\label{subsec: intro example}
Before introducing the details of the inference method, we first look at a real example that motivates the development of our methodology. This example extends the study conducted by  \cite{tian2022risk}. 

As new approaches to dementia treatment and prevention are developed, it is increasingly important to identify which group of older adults are at a high risk of developing dementia. In the past decade, research on dementia risk prediction has rapidly intensified. A recent literature review focuses on 39 studies that model late-life dementia risk with a follow-up duration of 1-39 years \citep{hou2019models}. Age, sex, education, MMSE, neuropsychological test score, body mass index, and alcohol intake are the most common predictors included in these recently developed models. Most models have about 10 predictor variables, 4 years prediction interval, and moderate predictive ability (Figure 2, \cite{hou2019models}).

Using an $\ell_1$-penalized multinomial regression model improves upon prior work. Due to the high dimensionality of the dataset, many previous works pre-select a few potentially important variables, then fit the model only on these variables \citep{battista2017optimizing, james2021performance, thabtah2022detection}. The $p$-values calculated on the sub-model may be unreliable since the variable selection process can be subjective. With penalization, there is no need to manually pre-select potentially important variables before modeling.

The Uniform dataset (UDS, versions 1-3) collected by the National Alzheimer Coordinating Center (NACC) from June 2005 to November 2021 is used for modeling. We are interested in identifying significant predictors for the progression from a cognitively normal (CN) status to four common subtypes of mild cognitive impairment (MCI) and dementia, which include Alzheimer's disease (AD), frontotemporal lobar degeneration (FTLD), Lewy body/Parkinson's disease (LBD/PD), and vascular brain injury (VBI). The primary etiologic diagnosis at each patient's last visit is the response variable. We include only participants who visited the center at least twice and had a normal diagnosis on their initial visit. Without further data cleaning and pre-processing, there were 12797 participants, 1936 predictors, and 1 response variable. At their follow-up visits, among these participants, 1682 (13.14\%) were diagnosed with AD, 42 (0.33\%) with FTLD, 133 (1.04\%) were diagnosed with LBD/PD, 261 (2.04\%) were diagnosed with VBI, and 10679 (83.45\%) remained normal (CN). After exclusionary pre-processing steps, including removing highly correlated predictors (over 95\% correlation) and predictors with more than 10\% missing values, there remained 466 predictors for model fitting. We used the same strategy to impute the missing values as in \cite{tian2022risk}. Since multiple imputations were used, we applied the Licht-Rubin procedure \citep{rubin2004multiple, licht2010new} to combine the inference results obtained in each replication through our inference method. Due to the imbalance of different subtypes, the sample sizes of dementia/MCI subtypes like FTLD, LBD/PD, and VBI are smaller than the number of predictors, which can make the ordinary multinomial regression model without penalization very unstable. 

The contrasts, odds ratios, and $p$-values of significant predictors between AD and CN obtained by the debiasing method (to be introduced), are reported in Table \ref{table: ad}. The results for other subtypes compared to CN will be presented in Section \ref{subsec: real example numerical}. Notably, the number of days from the initial visit to the most recent visit is a strong predictor for AD against CN. With other factors fixed, a long time from the initial visit indicates a higher likelihood of progressing into AD instead of remaining normal. APOE4 allele carriage is a significant risk factor, which is well-documented \citep{corder1993gene}. Psychometric factors at baseline include vegetable and animal naming (measures of semantic fluency), Trails B (a measure of executive function and attention), and immediate recall of the Craft story (a measure of attention). Subjective cognitive decline has long been associated with an increased risk of MCI/dementia \citep{reisberg1982global, reisberg2010outcome}, especially when reinforced by informant report of memory decline \citep{jessen2014conceptual}, and this is also detected as an important predictor in this study.

\begin{table}[!h]
\caption{Contrasts, odds ratios, and $p$-values of top 10 predictors between AD and CN obtained by the debiased $\ell_1$-penalized multinomial regression.  Brackets indicate dummy variables for categorical variables.}
\label{table: ad}
\centering
\begin{tabular}{l|rrrl}
\hline
Predictor        & \multicolumn{1}{l}{Coefficient} & \multicolumn{1}{l}{Odds ratio} & \multicolumn{1}{l}{$p$-value} & Description   \\ \hline
NACCDAYS       & 0.6245                          & 1.8674                         & $< 10^{-4}$                & Days from initial visit to most recent visit \\\hline  
NACCAPOE4      & 1.4692                          & 4.3456                         & $< 10^{-4}$                & APOE genotype [4 = (e4, e4)] \\\hline  
BIRTHYR        & -0.9941                         & 0.37                           & $< 10^{-4}$                & Subject's year of birth       \\\hline  
VEG            & -0.2555                         & 0.7745                         & $< 10^{-4}$                & Total number of vegetables named in 60 seconds   \\\hline  
DECIN1         & 0.6145                          & 1.8487                         & $< 10^{-4}$                & \makecell[l]{Does the co-participant report a decline in subject's memory? [1= Yes]}   \\\hline  
ANIMALS        & -0.2141                         & 0.8073                         & $< 10^{-4}$                & Total number of animals named in 60 seconds \\\hline  
COGSTAT1       & -0.6955                         & 0.4988                         & $< 10^{-4}$                & \makecell[l]{Per clinician, based on the neuropsychological examination, the subject's cognitive status is \\deemed [1 = better than normal for age]} \\\hline  
NACCNE4S1      & 0.5057                          & 1.6581                         & $< 10^{-4}$                & Number of APOE e4 alleles   \\\hline  
TRAILB         & 0.2074                          & 1.2305                         & $< 10^{-4}$                & {Trail Making Test Part b - Total number of seconds to complete}   \\\hline  
CRAFTVRS\_abn1 & 4.0484                          & 57.3031                        & 0.0002             & \makecell[l]{Craft Story 21 Recall (Immediate) - Total story \\units recalled, verbatim scoring [abn1 = Abnormal]}  \\ \hline
\end{tabular}
\end{table}

\subsection{Organization of this paper}
The remaining parts of this paper are organized as follows. In Section \ref{sec: method}, we introduce the $\ell_1$-penalized multinomial regression model and discuss the details of estimation, prediction, statistical inference, and the analysis under model misspecification and non-identically distributed data. Detailed algorithms for computation and the associated theories are provided. In Section \ref{sec: numerical}, we demonstrate the effectiveness of the inference method compared to other popular approaches, such as the bootstrap method and multiple splitting method. We also compare the estimation mean square errors of our contrast-based method and the over-parameterized method, as well as the misclassification error rate of our method against other prediction benchmark methods. Moreover, we provide additional results on the NACC real dataset. In Section \ref{sec: discussion}, we conclude the paper and discuss the limitations of the method, along with future research directions. Additional details omitted in the main text, extra numerical results, and all technical proofs can be found in the appendix.

\subsection{Notations}
We summarize the mathematical notations here and make some remarks on certain terminologies. 

For a set $S$, we use $|S|$ to denote its cardinality. For a vector $\bm{x} = (x_1, \ldots, x_p) \in \mathbb{R}^p$, we define its $\ell_0$-pseudo-norm, $\ell_1$-norm, $\ell_2$-norm and $\ell_{\infty}$-norm as $\zeronorm{\bm{x}} = |\{j \in [p]: x_j \neq 0\}|$, $\onenorm{\bm{x}} = \sum_{j=1}^p|x_j|$, $\twonorm{\bm{x}} = \sqrt{\sum_{j=1}^px_j^2}$ and $\infnorm{\bm{x}} = \max_{j=1:p}|x_j|$, respectively. For a symmetric real matrix $\bm{A}$, we use $\lambda_{\min}(\bm{A})$ and $\lambda_{\max}(\bm{A})$ to denote the smallest and largest eigenvalues of $\bm{A}$, respectively. $\tp$ and $\te$ represent the probability and expectation, respectively. For two real series $\{a_n\}_{n=1}^{\infty}$ and $\{b_n\}_{n=1}^{\infty}$, we write $a_n \lesssim b_n$ when there exists a sufficiently large absolute constant $C > 0$ such that $a_n \leq Cb_n$ for all $n$, and we write $a_n \gtrsim b_n$ when there exists a sufficiently large absolute constant $C > 0$ such that $a_n \geq Cb_n$ for all $n$. $a_n \ll b_n$ or $a_n = \smallo(b_n)$ means $a_n/b_n \rightarrow 0$, and $a_n \asymp b_n$ means $a_n/b_n, b_n/a_n \leq C< \infty$. For a series of random variables $\{x_n\}_{n=1}^{\infty}$ and a distribution law $\mu$, we write $x_n \overset{\textup{d}}{\to} \mu$ if the distribution of $x_n$ weakly converges to  $\mu$ as $n \rightarrow \infty$. For a series of random variables $\{x_n\}_{n=1}^{\infty}$ and a random variable $x$, we write $x_n \overset{\tp}{\rightarrow} x$ if for any $\epsilon > 0$, $\tp(|x_n - x| > \epsilon) \rightarrow 0$ as $n \rightarrow \infty$. For two random variable series $\{x_n\}_{n=1}^{\infty}$ and $\{y_n\}_{n=1}^{\infty}$ with $y_n \neq 0$, we write $x_n = \mathcal{O}_{\tp}(y_n)$ if for any $\epsilon > 0$, $\exists C > 0$, such that $\tp(|x_n/y_n| > C) < \epsilon$ holds for all $n \geq 1$. We use $C$, $C'$, $C''$, $c$, $\ldots$, to denote absolute constants which could differ from line to line.

Throughout this paper, we may not distinguish between the terms ``debiased Lasso'' \citep{javanmard2014confidence, zhang2014confidence} and ``desparsified Lasso'' \citep{van2014asymptotically}. All the confidence intervals and $p$-values mentioned in this paper are in the asymptotic sense. In other words, they are approximate confidence intervals and $p$-values, which only work well when the sample size $n$ is large.

\section{Methods and theories}\label{sec: method}


\subsection{Estimation and prediction}\label{subsec: method estimation prediction}

The optimization problem \eqref{eq: lasso estimation} can be solved with a cyclic coordinate descent algorithm similar to the one described in \cite{friedman2010regularization}. Cyclic coordinate descent algorithms for $\ell_1$-penalized problems like \eqref{eq: lasso estimation} have been shown to converge to the global minimizer in literature \citep{tseng2009coordinate, saha2013nonasymptotic}. The iterative procedure with a fixed penalty parameter $\lambda$ is detailed in Algorithm \ref{algo: lasso estimation}. In practical implementation, we start with a large $\lambda$ and apply the ``warm-start'' strategy by calculating subsequent solutions for a decreasing sequence of $\lambda$ to speed up computation. The penalty parameter $\lambda$ can be chosen through a cross-validation procedure \citep{hastie2009elements}.

We concatenate the coefficients $\{\bbetaks{k}\}_{k=1}^{K-1}$ into a single long vector $\bbeta^*$ of dimension $(K-1)p$. Similarly, we concatenate the estimates $\{\hbbetak{k}\}_{k=1}^{K-1}$ into a $(K-1)p$-dimensional vector $\hbbeta$. Given any estimate $\bbeta = \{\bbetak{k}\}_{k=1}^{K-1}$, we define the conditional probability $\tp(Y=k|X=\bx) = \frac{\exp(\bx^\top \bbetak{k})}{1+\sum_{k=1}^{K-1}\exp(\bx^\top \bbetak{k})}$ as $p_k(\bx;\bbeta)$ for $k = 1:(K-1)$.

\begin{algorithm}[!h]
\begin{algorithmic}
\caption{Cyclic coordinate descent algorithm to solve \eqref{eq: lasso estimation}}
\label{algo: lasso estimation}
\State \textbf{Input:} Data $\{\bx_i, y_i\}_{i=1}^n$, penalty parameter $\lambda$, initial estimates $\{\hbbeta^{(k)}\}_{k=1}^{K-1}$, maximum iteration number $T$
\For{$t_1 = 1$ to $T$}
	\For{$k = 1$ to $K-1$}
		\For{$t_2 = 1$ to $T$}
			\State $w_i^{(k)} \leftarrow p_k(\bx_i;\hbbeta)$ for $i=1:n$
			\For{$j = 1$ to $p$}
				\State $v_j^{(k)} \leftarrow \frac{1}{n}\sum_{i=1}^n w_i^{(k)}x_{ij}^2$
				\State $r_{ji}^{(k)} \leftarrow \frac{\yk{k}_i - p_k(\bx_i;\hbbeta)}{w_i^{(k)}}$
				\State $\hbetak{k}_j \leftarrow \frac{\mathcal{S}\big(n^{-1}\sum_{i=1}^n w_i^{(k)}x_{ij}r_{ji}^{(k)} + v_j^{(k)}\hbetak{k}_j, \lambda \big)}{v_j^{(k)}}$, where $\mathcal{S}(a,b) = \text{sign}(a)\max\{|a|-b,0\}$
			\EndFor
			\If{converged}
				\State break
			\EndIf
		\EndFor
	\EndFor
	\If{converged}
		\State break
    \EndIf
\EndFor
\State \textbf{Output:} Final estimates $\{\hbbeta^{(k)}\}_{k=1}^{K-1}$
\end{algorithmic}
\end{algorithm}

The point estimate obtained from \eqref{eq: lasso estimation} enjoys certain non-asymptotic properties under specific assumptions.

\begin{assumption}\label{asmp: x}
	$\bx_i \overset{\textup{i.i.d.}}{\sim} X$ with $\te X = \bm{0}_p$ and $\cov(X) = \bSigma_X$ which satisfies $\lambda_{\min}(\bSigma_X) \geq c > 0$ for some constant $c$. And for any $\bu \in \mathbb{R}^p$, $\te e^{\bu^\top X} \leq \exp(\twonorm{\bu}^2C/2)$ for some constant $C > 0$, i.e., $\bu^\top X$ is sub-Gaussian with a bounded variance proxy.
\end{assumption}

\begin{assumption}\label{asmp: sparsity beta}
	$\max_k\zeronorm{\bbetaks{k}} \leq s \leq p$, $\max_{k}\twonorm{\bbetaks{k}} \leq C < \infty$, with some constant $C > 0$.
\end{assumption}

\begin{theorem}[Upper bound of the estimation error]\label{thm: estimation error beta}
	Let $\lambda = C\sqrt{\frac{\log p}{n}}$ with a large constant $C>0$. Under Assumptions \ref{asmp: x} and \ref{asmp: sparsity beta}, when $n \geq C'K^5s\log (Kp)$ with a large constant $C' > 0$, we have
	\begin{align}
		\twonorm{\hbbeta - \bbeta^*} \leq C'''K^{5/2}\sqrt{\frac{s\log (Kp)}{n}}, \quad \onenorm{\hbbeta - \bbeta^*} \leq C'''K^3s\sqrt{\frac{\log (Kp)}{n}},
	\end{align}
	with probability at least $1-C''''(Kp)^{-1}$.
\end{theorem}

To see the complete picture, we have the following lower bound for the estimation error.

\begin{theorem}[Lower bound of the estimation error]\label{thm: lower bdd estimation error beta}
	Under Assumption \ref{asmp: x}, when $n \geq CKs(\log (ep/s))(\log K)$, $\log p \geq C\log K$, and $\log p \geq C\log s$, with a large constant $C>0$, we have
	\begin{equation}
		\inf_{\hbbeta}\sup_{\bbeta \in \mathcal{B}}\tp\left(\twonorm{\hbbeta-\bbeta} \geq C'K\sqrt{\frac{s\log (Kp)}{n}}\right) \geq \frac{1}{4}, \quad \inf_{\hbbeta}\sup_{\bbeta \in \mathcal{B}}\tp\left(\onenorm{\hbbeta-\bbeta} \geq C'K^{3/2}s\sqrt{\frac{\log (Kp)}{n}}\right) \geq \frac{1}{4},
	\end{equation}
	where $\mathcal{B} = \{\bbeta=\{\bbetak{k}\}_{k=1}^{K-1}: \zeronorm{\bbetak{k}} \leq s, \max_{k}\twonorm{\bbetak{k}} \leq C''\}$, and $C', C'' > 0$ are some constants.
\end{theorem}

\begin{remark}
The failure probability $1/4$ indicates the possibility that the estimation error exceeds the lower bound, and it can be further refined to any constant smaller than $1/2$. Theorem \ref{thm: estimation error beta} establishes upper bounds that hold with probability $1-\smallo(1)$, while the lower bounds aim to identify the rate at which the $1-\smallo(1)$ probability does not hold (i.e., there is a constant probability of failure). Thus, the smaller the gap between the upper and lower bounds, the closer the method is to the optimal one. Comparing Theorems \ref{thm: estimation error beta} and \ref{thm: lower bdd estimation error beta}, we observe that the dependence on $n$, $s$, and $p$ matches in upper and lower bounds. However, the dependence on the number of classes $K$ in Theorem \ref{thm: estimation error beta} may not be optimal. 
\end{remark}

\begin{remark}
	It might be surprising that the order of $K$ in the $\ell_2$-estimation error is greater than $1/2$. Intuitively, there are $n$ samples, and we are estimating a $(K-1)p$-dimensional parameter with at most $(K-1)s$ non-zero components. According to the classical Lasso theory, the optimal $\ell_2$-estimation error of $\bbeta$ should be of order $\sqrt{\frac{Ks\log (Kp)}{n}}$. An intuitive interpretation is that the samples of different classes do not contribute equally to the estimation of all $\bbetaks{k}$'s. This can be seen from the expression of the conditional distribution \eqref{eq: multinomial model}. The denominator on the right-hand side of \eqref{eq: multinomial model} is the same for different $k$'s, but the numerator only contains one specific $\bbetaks{k}$. In the simplest setting, suppose the sample sizes of different classes are the same. If the contribution of samples from class $k$ to the estimation of $\bbetaks{k}$ dominates the contribution of samples from other classes, then the $\ell_2$-estimation error of $\bbetaks{k}$ could be $\mathcal{O}_{\tp}\Big(\sqrt{\frac{s\log p}{n/K}}\Big) = \mathcal{O}_{\tp}\Big(\sqrt{\frac{Ks\log (Kp)}{n}}\Big)$ because the effective sample size is $n/K$ instead of $n$ and $\log p \gtrsim \log K$. Summing up the errors of all $(K-1)$ $\bbetaks{k}$'s, we obtain the rate $\mathcal{O}_{\tp}\Big(K\sqrt{\frac{s\log (Kp)}{n}}\Big)$ for the $\ell_2$-estimation error bound.
\end{remark}

With the estimate $\hbbeta = \{\hbbetak{k}\}_{k=1}^{K-1}$, we can construct a classifier to predict the label for a new observation $\bx$:
\begin{equation}
	\mC_{\hbbeta}(\bx) = \argmax_{k=1:K}p_k(\bx;\hbbeta).
\end{equation}
It is well-known that the Bayes classifier $\mC_{\bbeta^*}(\bx) = \argmax_{k=1:K}p_k(\bx;\bbeta^*)$ achieves the smallest misclassification error rate. For any classifier $\mC$, denote the misclassification error rate $\tp(\mC(X) \neq Y)$ as $R(\mC)$. The difference between the error rates of a classifier $\mC$ and the Bayes classifier $\mC_{\bbeta^*}$ is called the excess misclassification error rate. We have the following upper bound for the \textit{excess misclassification error rate} of $\mC_{\hbbeta}$.

\begin{theorem}[Upper bound of the excess misclassification error]\label{thm: error}
	Under the same conditions imposed in Theorem \ref{thm: estimation error beta}, we have
	\begin{equation}
		R(\mathcal{C}_{\hbbeta}) - R(\mathcal{C}_{\bbeta^*}) \leq CK^{5/2}\sqrt{\frac{s\log (Kp)}{n}},
	\end{equation}
	with probability at least $1-C'(Kp)^{-1}$, where $C, C' > 0$ are some constants.
\end{theorem}

\subsection{Statistical inference}\label{subsec: method inf}
As we mentioned in the introduction, it is well-known that the Lasso estimator is biased, with its bias being non-negligible compared to its variance and difficult to quantify. Therefore, the Lasso estimator \eqref{eq: lasso estimation} cannot be directly used to construct a confidence interval or conduct a hypothesis test. In this paper, we modify the desparsified Lasso proposed in \cite{van2014asymptotically} to adapt to the multinomial regression setting, allowing for valid statistical inference.

For presentation convenience, we denote $\bSigma = \te [\nabla^2 \mL_n(\bbeta^*)]$ and $\bTheta = \bSigma^{-1}$, both of which play an essential role in the statistical inference. Next, we derive the debiased Lasso estimator in a heuristic way.

Since \eqref{eq: lasso estimation} is a convex optimization problem and the objective function is coercive, the optimal solution $\hbbeta$ must satisfy the first-order condition
\begin{equation}\label{eq: debiased 1}
	\nabla \mL_n(\hbbeta) + \lambda \bm{\kappa} = \bm{0}_{(K-1)p}, \quad \bm{\kappa} = (\kappa_1, \ldots, \kappa_{(K-1)p})^\top,
\end{equation}
where $\kappa_j = \text{sign}(\hbeta_j)$ when $\hbeta_j \neq 0$ and $\kappa_j \in \partial \norm{\beta}\big|_{\beta=0} = [-1, 1]$. $\partial \norm{\beta}\big|_{\beta=0}$ is the subgradient of $\norm{\cdot}$ at $0$. By plugging the expression of $\nabla \mL_n(\hbbeta)$ based on \eqref{eq: lasso estimation}, we have
\begin{equation}
	\frac{1}{n}\sum_{i=1}^n \begin{pmatrix}
		[p_1(\bx_i;\hbbeta)-p_1(\bx_i;\bbeta^*)]\bx_i \\
		\vdots \\
		[p_{K-1}(\bx_i;\hbbeta)-p_{K-1}(\bx_i;\bbeta^*)]\bx_i 
	\end{pmatrix} - 
	\frac{1}{n}\sum_{i=1}^n \begin{pmatrix}
		\epsilon_i^{(1)}\bx_i \\
		\vdots \\
		\epsilon_i^{(K-1)}\bx_i 
	\end{pmatrix} + \lambda \bm{\kappa} = \bm{0}_{(K-1)p},
\end{equation}
where we define the ``residual" of the $i$-th observation corresponding to class $k$ as $\epsilon_i^{(k)} = \yk{k}_i - p_k(\bx_i;\bbeta^*)$, with $\yk{k}_i = \mathds{1}(y_i = k)$, $i = 1:n$, $k = 1:(K-1)$. By \eqref{eq: debiased 1}, we replace $\lambda \bm{\kappa}$ with $-\nabla \mL_n(\hbbeta)$ in the equation above, apply the mean-value theorem, and use vectorized notations to obtain
\begin{equation}\label{eq: decomp B}
	\bm{B}_n(\hbbeta-\bbeta^*) + \frac{1}{n}\begin{pmatrix}
		\bX^\top(\bm{y}^{(1)}-\bm{p}_1(\bX;\hbbeta)) \\
		\vdots \\
		\bX^\top(\bm{y}^{(K-1)}-\bm{p}_{K-1}(\bX;\hbbeta))
	\end{pmatrix} = \frac{1}{n}\begin{pmatrix}
		\bX \bm{\epsilon}^{(1)} \\
		\vdots \\
		\bX \bm{\epsilon}^{(K-1)}
	\end{pmatrix},
\end{equation}
where $\bX = (\bx_1^\top, \ldots, \bx_n^\top) \in \mathbb{R}^{n \times p}$ is the concatenated predictor matrix, $\bm{y}^{(k)}=(\yk{k}_1, \ldots, \yk{k}_n)^\top \in \mathbb{R}^n$ is the concatenated response vector corresponding to class $k$, $\bm{p}_k(\bX, \hbbeta) = (p_k(\bx_1, \hbbeta), \ldots, p_k(\bx_n, \hbbeta))^\top$ is the concatenated conditional probability for class $k$, and $\bm{\epsilon}^{(k)} \coloneqq (\epsilon_1^{(k)}, \ldots, \epsilon_n^{(k)})^\top$ is the concatenated residual vector corresponding to class $k$, with $k = 1:(K-1)$. Here, $\bm{B}_n = \frac{1}{n}\sum_{i=1}^n \int_0^1 \bm{B}(\bx_i; \bbeta^* + t(\hbbeta-\bbeta^*)) dt$, where $\forall \bx \in \mathbb{R}^p$, $\bbeta = \{\bbetak{k}\}_{k=1}^{K-1} \in \mathbb{R}^{(K-1)p}$,
\begin{equation}
    \bm{B}(\bx; \bbeta) \coloneqq \begin{pmatrix}
	p_1(\bx;\bbeta)[1-p_1(\bx;\bbeta)]\bx\bx^\top & -p_1(\bx;\bbeta)p_2(\bx;\bbeta)\bx\bx^\top &\ldots &-p_1(\bx;\bbeta)p_{K-1}(\bx;\bbeta)\bx\bx^\top \\
	-p_1(\bx;\bbeta)p_2(\bx;\bbeta)\bx\bx^\top &p_2(\bx;\bbeta)[1-p_2(\bx;\bbeta)]\bx\bx^\top &\ldots &-p_2(\bx;\bbeta)p_{K-1}(\bx;\bbeta)\bx\bx^\top \\
	\vdots &\vdots  &\vdots &\vdots \\
	  -p_1(\bx;\bbeta)p_{K-1}(\bx;\bbeta)\bx\bx^\top &-p_2(\bx;\bbeta)p_{K-1}(\bx;\bbeta)\bx\bx^\top &\ldots &p_{K-1}(\bx;\bbeta)[1-p_{K-1}(\bx;\bbeta)]\bx\bx^\top
\end{pmatrix}  \in \mathbb{R}^{[(K-1)p] \times [(K-1)p]}. \label{eq: def B}
\end{equation}
With some algebraic calculations, it is not hard to show that $\te[\bm{B}(X; \bbeta)] = \bSigma$. Heuristically, since $\twonorm{\hbbeta - \bbeta^*}, \onenorm{\hbbeta - \bbeta^*} \overset{\tp}{\rightarrow} 0$ due to Theorem \ref{thm: estimation error beta}, by some concentration inequalities, we would expect $\bm{B}_n(\hbbeta - \bbeta^*) \approx \widetilde{\bm{B}}_n (\hbbeta - \bbeta^*) \approx \bSigma (\hbbeta - \bbeta^*)$ in the entry-wise convergence sense with high probability, where $\widetilde{\bm{B}}_n \coloneqq \frac{1}{n}\sum_{i=1}^n \bm{B}(\bx_i; \bbeta^*)$. Combing with \eqref{eq: decomp B}, it implies
\begin{equation}\label{eq: debias motive}
	\hbbeta  \approx \bbeta^* - \frac{1}{n}\bTheta\begin{pmatrix}
		\bX^\top(\bm{y}^{(1)}-\bm{p}_1(\bX;\hbbeta)) \\
		\vdots \\
		\bX^\top(\bm{y}^{(K-1)}-\bm{p}_{K-1}(\bX;\hbbeta))
	\end{pmatrix} + \frac{1}{n}\bTheta \begin{pmatrix}
		\bX \bm{\epsilon}^{(1)} \\
		\vdots \\
		\bX \bm{\epsilon}^{(K-1)}
	\end{pmatrix}.
\end{equation}
Noticing that the last term on the RHS of \eqref{eq: debias motive} has mean zero, it motivates us to define the ``debiased" Lasso estimator by removing the second bias term as
\begin{equation}\label{eq: debiased estimator}
	\widehat{\bm{b}} \coloneqq \hbbeta + \frac{1}{n}\bTheta\begin{pmatrix}
		\bX^\top(\bm{y}^{(1)}-\bm{p}_1(\bX;\hbbeta)) \\
		\vdots \\
		\bX^\top(\bm{y}^{(K-1)}-\bm{p}_{K-1}(\bX;\hbbeta))
	\end{pmatrix}.
\end{equation}
In fact, we can make the previous heuristics rigorous and show that
\begin{equation}\label{eq: clt heuristic}
	\sqrt{n}(\widehat{b}_j - \beta^*_j) = \textup{remaining bias terms} + \frac{1}{\sqrt{n}}\bTheta_j^\top \begin{pmatrix}
		\bX \bm{\epsilon}^{(1)} \\
		\vdots \\
		\bX \bm{\epsilon}^{(K-1)}
	\end{pmatrix},
\end{equation}
where remaining bias terms refer to the small offset terms caused by the previous ``$\approx$" argument. Note that by Lindeberg-Feller central limit theorem (e.g., Theorem 3.4.10 in \cite{durrett2019probability}), we have
\begin{equation}\label{eq: clt heuristic 2}
    \frac{1}{\sqrt{n}\sqrt{\bTheta_j^\top\bSigma\bTheta_j}}\bTheta_j^\top \begin{pmatrix}
		\bX \bm{\epsilon}^{(1)} \\
		\vdots \\
		\bX \bm{\epsilon}^{(K-1)}
	\end{pmatrix} \overset{\textup{d}}{\to} N(0, 1), \quad n \rightarrow \infty.
\end{equation}
As we will show later, in the regime that we are interested in, where $n \gg s^2(\log p)^2$, the remaining bias terms in \eqref{eq: clt heuristic} converge to zero in probability as $n \rightarrow \infty$. This leads to the asymptotic normality
\begin{equation}\label{eq: clt heuristic 4}
    \frac{\sqrt{n}(\widehat{b}_j - \beta^*_j)}{\sqrt{\bTheta_j^\top\bSigma\bTheta_j}} \overset{\textup{d}}{\to} N(0, 1),
\end{equation}
which can be used to construct confidence intervals for $\bbeta_j^*$'s and test statistics of the hypothesis test with the null $H_0: \bbeta_j^* = 0$. Note that the remaining bias terms are negligible due to the sparsity assumption $n \gg s^2(\log p)^2$. As discussed in Section \ref{subsec: intro hd}, this is not true in the fixed-ratio regime where $n/p$ converges to a fixed constant \citep{javanmard2014hypothesis}. The clean asymptotic normality \eqref{eq: clt heuristic 4} benefits from the sparse regime we study.

In practice, since $\bTheta$ and $\bSigma$ are unknown, we have to estimate them. For $\bSigma$, we can use the empirical estimator $\hSigma = \frac{1}{n}\sum_{i=1}^n \bm{B}(\bx_i;\hbbeta)$. For $\bTheta$, we estimate each row via the nodewise regression \citep{meinshausen2006high}. For convenience, we denote 
$$
\bTheta = \diag((\tau_1^*)^{-2}, \ldots, (\tau_{(K-1)p}^*)^{-2})\begin{pmatrix}
	1 &-\gamma^*_{1,1} &-\gamma^*_{1,2} &\ldots &-\gamma^*_{1,(K-1)p-1} \\
	-\gamma^*_{2,1} &1 &-\gamma^*_{2,2} &\ldots &-\gamma^*_{2,(K-1)p-1} \\
	\vdots &\vdots &\vdots &\vdots   &\vdots  \\
	-\gamma^*_{(K-1)p-1,1} &-\gamma^*_{(K-1)p-1,2} &-\gamma^*_{(K-1)p-1,3}  &\ldots &1
\end{pmatrix},
$$
where $\bgamma_j^* = (\gamma^*_{j, 1}, \ldots, \gamma^*_{j, (K-1)p-1})^\top = (\bSigma_{-j,-j})^{-1}\bSigma_{-j,j}$ by the formula of the inverse block matrix. Note that, unlike the case of GLMs \citep{van2014asymptotically}, here it is difficult to estimate $\bgamma_j^*$ through regression on the $j$-th column of a weighted $\bX$ matrix versus the other columns. Instead, we solve the following quadratic program (QP):
\begin{equation}\label{eq: gamma estimation}
	\hgamma_j = \argmin_{\bgamma \in \mathbb{R}^{(K-1)p-1}}\left\{\huSigma_{j,j} - \hSigma_{j,-j}\bgamma + \frac{1}{2}\bgamma^\top\hSigma_{-j,-j}\bgamma + \lambda_j\onenorm{\bgamma}\right\},
\end{equation}
and let
\begin{equation}
	\htau_j^2 = \huSigma_{j,j}-\hSigma_{j,-j}\hgamma_j,
\end{equation}
for $j = 1, \ldots, (K-1)p$. There are many ways to solve this QP problem. For example, we can rewrite it as a standard QP and solve it using any QP solvers. For other methods, see the introduction part of \cite{solntsev2015algorithm}. Here we solve it by the coordinate descent algorithm outlined in Algorithm \ref{algo: gamma estimation}.

\begin{algorithm}[!h]
\begin{algorithmic}
\caption{Coordinate descent algorithm to solve \eqref{eq: gamma estimation}}
\label{algo: gamma estimation}
\State \textbf{Input:} Empirical estimate $\hSigma$, penalty parameter $\lambda_j$, maximum iteration number $T$, initial estimate $\hgamma_j$, index $j$
\For{$t = 1$ to $T$}
	\For{$r = 1$ to $(K-1)p-1$}
		\State $\widehat{\gamma}_{j,r} \leftarrow \mathcal{S}\big((\huSigma_{-j,j})_r- (\hSigma_{-j,-j})_{r,-r}\hgamma_{j,-r}, \lambda \big)/\huSigma_{j,j}$, where $(\huSigma_{-j,j})_r$ is the $r$-th component of $\huSigma_{-j,j}$, $(\hSigma_{-j,-j})_{r, -r}$ is the $r$-th row of $\hSigma_{-j,-j}$ without the $r$-th column, and $\mathcal{S}(a,b) = \text{sign}(a)\max\{|a|-b,0\}$
	\EndFor
	\If{converged}
		\State break
	\EndIf
\EndFor
\State \textbf{Output:} Final estimates $\hgamma_j$
\end{algorithmic}
\end{algorithm}

Note that Algorithm \ref{algo: gamma estimation} only shows the procedure to calculate the solution to \eqref{eq: gamma estimation} for one $j$ and one specific $\lambda_j$. In practice, we run this algorithm for different $\lambda_j$ choices and iterate from $j=1$ to $j=(K-1)p$, and the iteration can be parallelized to speed up computation in practice. Each $\lambda_j$ can be selected through a cross-validation procedure.

With the estimate $\hSigma$ and $\hTheta$, if the estimated variance $\hTheta_j^\top \hSigma \hTheta_j$ converges to the true asymptotic variance $\bTheta_j^\top \bSigma \bTheta_j$ in \eqref{eq: clt heuristic 4} in probability, then by \eqref{eq: clt heuristic 4} and the Slutsky's lemma, we have
\begin{equation}\label{eq: clt heuristic 3}
	\frac{\sqrt{n}(\widehat{b}_j - \beta^*_j)}{\sqrt{\hTheta_j^\top\hSigma\hTheta_j}} \overset{\textup{d}}{\to} N(0, 1),
\end{equation}
which can be used to build the confidence interval for $\beta^*_j$ and compute the $p$-value for the hypothesis testing problem $H_0: \beta^*_j=0$ v.s. $H_1: \beta^*_j \neq 0$. We summarize the procedures of CI construction and $p$-value calculation in Algorithms \ref{algo: CI} and \ref{algo: hypothesis test}, respectively.

\begin{algorithm}[!h]
\begin{algorithmic}
\caption{Construct the $(1-\alpha)$-confidence interval of $\beta^*_j$}
\label{algo: CI}
\State \textbf{Input:} Debiased estimate $\widehat{\bm{b}}$, empirical estimate (of $\bSigma$) $\hSigma$, estimate (of $\Theta$) $\hTheta$ calculated from \eqref{eq: gamma estimation}, index $j$, CI level $(1-\alpha)$
\State \textbf{Output:} $(1-\alpha)$-confidence interval $\left[\widehat{b}_j - q_{\alpha/2}\sqrt{\frac{\hTheta_j^\top\hSigma\hTheta_j}{n}}, \widehat{b}_j + q_{\alpha/2}\sqrt{\frac{\hTheta_j^\top\hSigma\hTheta_j}{n}}\right]$, where $q_{\alpha/2}$ is the $\alpha/2$-upper tail quantile of $N(0,1)$
\end{algorithmic}
\end{algorithm}

\begin{algorithm}[!h]
\begin{algorithmic}
\caption{Calculate $p$-value for the hypothesis test $H_0: \beta^*_j=0$ v.s. $H_1: \beta^*_j \neq 0$}
\label{algo: hypothesis test}
\State \textbf{Input:} Debiased estimate $\widehat{\bm{b}}$, empirical estimate (of $\bSigma$) $\hSigma$, estimate (of $\Theta$) $\hTheta$ calculated from \eqref{eq: gamma estimation}, index $j$, significance level $\alpha$
\State \textbf{Output:} $p$-value $2\Phi \left(-\sqrt{n}|\widehat{b}_j|/\sqrt{\hTheta_j^\top\hSigma\hTheta_j}\right)$, where $\Phi$ is the CDF of $N(0,1)$
\end{algorithmic}
\end{algorithm}

We are now ready to rigorously state \eqref{eq: clt heuristic 3}. Before doing so, we impose the following sparsity assumption on $\bTheta$, which is necessary for the nodewise regression \eqref{eq: gamma estimation} to work. Similar conditions can be found in \cite{van2014asymptotically, ning2017general, guo2021inference}.

\begin{assumption}\label{asmp: sparsity gamma}
	Suppose the following conditions hold for $\bSigma = \te [\nabla^2 \mL_n(\bbeta^*)]$ and $\bTheta = \bSigma^{-1}$:
	\begin{enumerate}[(i)]
		\item $\max_{j=1:[(K-1)p]}\zeronorm{\bTheta_j} \leq Ks_0$, $\max_{j=1:[(K-1)p]}\twonorm{\bTheta_j} \leq C < \infty$ with some constant $C > 0$;
        \item $\max_{i=1:n}\max_{k=1:(K-1)}\big|\bx_i^\top \bbetaks{k}\big| \leq C' < \infty$ a.s. with some constant $C' > 0$.
	\end{enumerate}
\end{assumption}
\begin{remark}\label{rmk: asmp 3}
    The sparsity assumption of $\bTheta_j$ in Condition (\rom{1}) can be removed for a modified inference procedure, which, however, requires sample spitting. Condition (\rom{2}) can also be removed at the cost of a more stringent sample size requirement. Interested readers can find the details in Sections \ref{sec: relax sparsity appendix} and \ref{sec: relax a.s. boundedness appendix} of the appendix. 
\end{remark}

We are now ready to present the following theorem, which characterizes the asymptotic distribution of $\widehat{b}_j$.

\begin{theorem}\label{thm: clt}
	Let $\lambda = C\sqrt{\frac{\log (Kp)}{n}}$ in \eqref{eq: lasso estimation} and $\lambda_j = C\sqrt{\frac{\log (Kp)}{n}}$ in \eqref{eq: gamma estimation} with a large constant $C>0$. Under Assumptions \ref{asmp: x}-\ref{asmp: sparsity gamma}, when $n \gg K^{18}s^2(\log (Kp))^2(\log(np))^2 + K^{10}s_0^2(\log (Kp))^2$, we have
	\begin{equation}
		\frac{\sqrt{n}(\widehat{b}_{j}-\beta^*_{j})}{\sqrt{\hTheta_{j}^\top\hSigma\hTheta_{j}}}  \overset{\textup{d}}{\to}N(0, 1),
	\end{equation}
	for $j = 1, \ldots, (K-1)p$.
\end{theorem}

\begin{remark}
    As discussed in Section \ref{subsec: intro hd}, \cite{cai2017confidence} proved that the adaptive inference procedure does not exist unless $n \gg (s\log p)^2$ for high-dimensional linear regression models. We conjecture that a similar conclusion holds for multinomial regression models when $K$ is fixed. If this is true, when $s_0 \lesssim s$, the current sample size requirement in Theorem \ref{thm: clt} is optimal up to a $(\log(np))^2$ factor, because our inference procedure does not rely on the knowledge of $s$ or $\bSigma_X$ and is thus adaptive. However, the sample size requirement may not be optimal when $K$ grows with $n$. Interested readers are encouraged to relax the condition by improving our proof in the appendix.
\end{remark}

We also want to comment on the impact of estimation error of $\bSigma$ and $\bTheta$ on inference accuracy. As noted in our previous heuristic analysis, by \eqref{eq: clt heuristic 4} and the Slutsky's lemma, it suffices to find estimators $\hSigma$ and $\hTheta$ that can consistently estimate the asymptotic variance (even at an arbitrarily slow rate), i.e. $|\hTheta^\top_j \hSigma \hTheta_j - \bTheta^\top_j \bSigma \bTheta_j| \overset{\tp}{\rightarrow} 0$. In the proof of Theorem \ref{thm: clt}, we show that $|\hTheta^\top_j \hSigma \hTheta_j - \bTheta^\top_j \bSigma \bTheta_j|$ can be bounded by some term involving $\onenorm{\hTheta_j - \bTheta_j}$, $\twonorm{\hTheta_j - \bTheta_j}$, and $\maxnorm{\hSigma - \bSigma}$. It can be derived that $\onenorm{\hTheta_j - \bTheta_j}$, $\twonorm{\hTheta_j - \bTheta_j}$, and $\maxnorm{\hSigma - \bSigma}$ all converge to zero in probability, under the current assumptions. In this way, the estimation errors of $\bSigma$ and $\bTheta$ do not explicitly appear in $n^{-1/2}$-inference rate. However, the estimation errors of $\bSigma$ and $\bTheta$ still have some impact on the inference accuracy by contributing higher-order terms that shrink faster than $n^{-1/2}$, which are ignored in the use of Slutsky's lemma. Hence, as one reviewer pointed out, estimating the high-dimensional covariance matrix $\bSigma$ and its inverse $\bTheta$ is very difficult in general, but we only need consistency of the asymptotic variance estimator (where the estimation error rate can be arbitrarily slow) for the asymptotic normality to be true, which indeed holds under current assumptions.

\subsection{Model misspecification and non-identically distributed data}\label{subsec: model misspecification and non id}
In the previous sections, we assumed that the data $\{(\bx_i, y_i)\}_{i=1}^n \overset{\textup{i.i.d.}}{\sim} (X, Y)$ and $y_i$ given $\bx_i$ is generated from a multinomial regression model, which might be too restrictive in practice. In this subsection, we will investigate the performance of $\ell_1$-penalized multinomial regression \ref{eq: multinomial model} and the debiased Lasso estimator \ref{eq: debiased 1} in two situations:
\begin{enumerate}[(i)]
    \item The model is misspecified. More specifically, the data $\{(\bx_i, y_i)\}_{i=1}^n \overset{\textup{i.i.d.}}{\sim} (X, Y)$ is generated from an unknown distribution, where $Y$ given $X$ is not necessarily generated from a multinomial regression model.
    \item The data $\{(\bx_i, y_i)\}_{i=1}^n$ is independent but not identically distributed. More specifically, $y_i$ given $\bx_i$ is generated from a multinomial regression model, but $\bx_i$'s are not necessarily generated from the same distribution.
\end{enumerate}
Note that it is possible to consider a mix of two situations, but for simplicity, we will discuss each scenario individually.

\subsubsection{Model misspecification}\label{subsubsec: model misspecification}
First, we consider case (\rom{1}), where the distribution of $Y$ given $X$ is misspecified. Recall the conditional probability $p_k(\bx;\bbeta) = \frac{\exp(\bx^\top \bbetak{k})}{1+\sum_{k=1}^{K-1}\exp(\bx^\top \bbetak{k})}$ under the multinomial regression model, for $\bbeta = \{\bbetak{k}\}_{k=1}^{K-1}$ and $k = 1:(K-1)$. Consider the best approximater
\begin{equation}\label{eq: beta approx}
    \bbeta^* \coloneqq \argmin_{\bbeta = \{\bbetak{k}\}_{k=1}^{K-1} \in \mathbb{R}^{(K-1)p}} \te[-\log p_Y(X;\bbeta)], 
\end{equation}
where the expectation is with respect to the true underlying distribution of $(X, Y)$. Minimizing $\te[-\log p_Y(X;\bbeta)]$ can be shown to be equivalent to minimizing the cross-entropy or the Kullback-Leibler divergence between the true distribution and the multinomial logistic regression model with any marginal distribution of $X$. We will show that the same conclusions derived in the previous sections hold for the best approximator $\bbeta^*$, when the model is misspecified. 

Since $Y$ given $X$ is not necessarily generated from the multinomial regression model, we need to replace the empirical estimator $\hSigma = \frac{1}{n}\sum_{i=1}^n \bm{B}(\bx_i;\hbbeta)$ used in Section \ref{subsec: method inf} by $\hSigma = \frac{1}{n}\sum_{i=1}^n \widehat{\bm{B}}(y_i, \bx_i;\hbbeta)$ with $\hbbeta = \{\hbbetak{k}\}_{k=1}^{K-1}$ from \eqref{eq: lasso estimation}, where for any $\bx \in \mathbb{R}^p$,  $y \in [K]$, and $\bbeta = \{\bbetak{k}\}_{k=1}^{K-1} \in \mathbb{R}^{(K-1)p}$,
\begin{align}
    \widehat{\bm{B}}(y, \bx; \bbeta) &\coloneqq \begin{pmatrix}
	\bx\bx^\top\widehat{\epsilon}_1^2(y, \bx, \bbeta) & \bx\bx^\top\widehat{\epsilon}_1(y, \bx, \bbeta)\widehat{\epsilon}_2(y, \bx, \bbeta) &\ldots &\bx\bx^\top\widehat{\epsilon}_1(y, \bx, \bbeta)\widehat{\epsilon}_{K-1}(y, \bx, \bbeta) \\
	\bx\bx^\top\widehat{\epsilon}_1(y, \bx, \bbeta)\widehat{\epsilon}_2(y, \bx, \bbeta) &\bx\bx^\top\widehat{\epsilon}_2^2(y, \bx, \bbeta) &\ldots &\bx\bx^\top\widehat{\epsilon}_2(y, \bx, \bbeta)\widehat{\epsilon}_{K-1}(y, \bx, \bbeta) \\
	\vdots &\vdots  &\vdots &\vdots \\
	  \bx\bx^\top\widehat{\epsilon}_1(y, \bx, \bbeta)\widehat{\epsilon}_{K-1}(y, \bx, \bbeta) &\bx\bx^\top\widehat{\epsilon}_2(y, \bx, \bbeta)\widehat{\epsilon}_{K-1}(y, \bx, \bbeta) &\ldots &\bx\bx^\top\widehat{\epsilon}_{K-1}^2(y, \bx, \bbeta)
   \end{pmatrix},\\
   \widehat{\epsilon}_k(y, \bx, \bbeta) &\coloneqq \yk{k} - p_k(\bx; \bbeta), \quad \textup{ with }\,\, \yk{k} \coloneqq \mathds{1}(y = k). 
\end{align}

Then, we can show that the inference procedure leads to valid inference for the best approximator $\bbeta^*$ when the model is misspecified.

\begin{theorem}\label{thm: misspecification}
    Under Assumptions \ref{asmp: x}-\ref{asmp: sparsity gamma}, Theorems \ref{thm: estimation error beta}, \ref{thm: error}, and \ref{thm: clt} still hold for $\bbeta^*$ defined in \eqref{eq: beta approx}.
\end{theorem}

In the low-dimensional case, the convergence of MLE to the best approximator defined in \eqref{eq: beta approx} under model misspecification has been extensively studied in the classical inference literature (e.g., \citealp{cox1962further, huber1967behavior, kent1982robust, white1982maximum}). \cite{ning2017general} provides a similar result for a score test-based inference procedure for high-dimensional linear regression. Their result relies on the assumption that $Y - \te [Y|X]$ is sub-Gaussian with a bounded variance proxy. In the context of multinomial regression models, the sub-Gaussianity is implied automatically by the boundedness of $Y - \te [Y|X]$; thus, no additional assumption is required. Our result complements this line of literature by addressing the high-dimensional multinomial regression case.

\subsubsection{Non-identically distributed data}\label{subsubsec: non id}
Next, we consider case (\rom{2}), where the data $\{(\bx_i, y_i)\}_{i=1}^n$ is independent but not identically distributed. In this case, we redefine the covariance matrix of predictors by the average as $\bSigma_X \coloneqq \frac{1}{n}\sum_{i=1}^n\te(\bx_i \bx_i^\top)$, and the covariance matrix for the multinomial model by the average as $\bSigma \coloneqq \frac{1}{n}\sum_{i=1}^n\te [\bm{B}(\bx_i; \bbeta^*)]$, where $\bm{B}(\bx_i; \bbeta^*)$ is defined in \eqref{eq: def B}.  

In the case of non-identically distributed data, we need to replace Assumption \ref{asmp: x} with the following modified assumption.

\begin{assumption}\label{asmp: x non id}
	For all $i = 1:n$, $\te \bx_i = \bm{0}_p$ and $\cov(\bx_i)$ satisfies $\lambda_{\min}(\cov(\bx_i)) \geq c > 0$ for some constant $c$. And for all $i = 1:n$ and any $\bu \in \mathbb{R}^p$, $\te e^{\bu^\top \cov(\bx_i)} \leq \exp(\twonorm{\bu}^2C/2)$ for some constant $C > 0$.
\end{assumption}

The first observation is that the non-asymptotic analysis used to prove the estimation error and misclassification error bounds in Theorems \ref{thm: estimation error beta} and \ref{thm: error} can still apply with the redefined $\bSigma_X$ and $\bSigma$, because the main tools we used are concentration inequalities for sub-Gaussian and sub-exponential variables which can apply to the sum of non-identically distributed variables. Hence, we can obtain the same bounds as in Theorems \ref{thm: estimation error beta} and \ref{thm: error}. The second remark is that the asymptotic normality for the debiased Lasso estimator in Theorem \ref{thm: clt} was proved by Lindeberg-Feller central limit theorem, which applies to the sum of non-identically distributed variables satisfying certain conditions. It is not difficult to show that the required conditions (often called Lindeberg's conditions) indeed hold; therefore, the same results in Theorem \ref{thm: clt} can be obtained.

We summarize these intuitions into the following theorem.

\begin{theorem}\label{thm: non id}
    Under Assumption \ref{asmp: sparsity beta}, \ref{asmp: sparsity gamma}, and \ref{asmp: x non id}, Theorems \ref{thm: estimation error beta}, \ref{thm: error}, and \ref{thm: clt} still hold for the non-identically distributed data $\{(\bx_i, y_i)\}_{i=1}^n$.
\end{theorem}

Theorem \ref{thm: non id} takes the heterogeneity of the data into account without requiring the data to be i.i.d. generated from the same distribution. The heterogeneity in data is very common in many applications, such as healthcare and medicine \citep{celi2013big}, finance and economics \citep{stock2020introduction}, and social sciences \citep{smith2004developing}. A further interesting extension is to deal with dependent observations. There are a few recent studies on the high-dimensional regression for dependent data with certain structures \citep{han2020high, ing2020model, fan2022adaptive}, and most of them focus on estimation and prediction. It remains unknown how to generalize our framework for high-dimensional multinomial regression to the dependent case.

\section{Numerical studies}\label{sec: numerical}
We have presented partial results of the application of $\ell_1$-penalized multinomial regression on the NACC dataset in Section \ref{subsec: intro example}. In this section, we explore the approach under different simulation settings and discuss the remaining real-data results. Our focus will be on statistical inference, for example, confidence interval (CI) construction and hypothesis testing. 

Under each model, we will first compare the estimation error of our contrast-based parameterization and the over-parameterization approach \citep{friedman2010regularization} on the sparse contrast coefficients. Then, we will compare the misclassification error rates of the contrast-based $\ell_1$-penalized multinomial regression and other benchmark methods, including Support Vector Machines (SVMs), Linear Discriminant Analysis (LDA), Random Forests, and $k$-Nearest Neighbors ($k$NN). We fit SVMs with RBF kernel and run $k$NN with the number of neighbors $k = \lfloor\sqrt{n}\rfloor$, i.e., the largest integer that is smaller than or equal to $\sqrt{n}$. We will also compare the CIs produced by the debiased-Lasso method with the CIs produced by the vector bootstrap method \citep{sartori2011penalized} in terms of coverage probability and length. In addition, we will compare the debiased-Lasso method with the vector bootstrap method in individual testing by the type-I error rate and the power. Finally, we adjust the $p$-values produced by the debiased-Lasso with a Bonferroni correction and compare the method with the multiple splitting approach \citep{meinshausen2009p}, for multiple testing. We will describe the evaluation metrics used in detail, including average coverage probability, type-I error rate, and family-wise error rate (FWER), in Section \ref{subsubsec: numerical model 1} for Model $1$. For Models $2$-$8$, we only state the main findings, and the details of these metrics are omitted. Due to space constraints, we only display the results of Models 1-4 in the main text, and leave the results of Models 5-8 to Section \ref{subsec: additioal simulation appendix} of the appendix.

All experiments are conducted in R \citep{corer2020}. The contrast-based $\ell_1$-penalized multinomial regression with the debiasing method is implemented in our new R package \texttt{pemultinom}, which is available on CRAN. The over-parameterized $\ell_1$-penalized multinomial regression is implemented through package \texttt{glmnet} \citep{friedman2010regularization}. We use cross-validation to find the best penalty parameter in all penalized regressions, including the estimation of the contrast coefficients and the nodewise regression. For the vector bootstrap method, the same number of observations are resampled with replacement from the data in each of the $200$ replications. The empirical quantiles are used to calculate the CI \citep{sartori2011penalized}. We use vector bootstrap because it works better than the residual resampling bootstrap under random design \citep{chatterjee2011bootstrapping}. For the multiple-splitting method \citep{meinshausen2009p}, $200$ splits are used in each simulation. SVMs, Random Forests, and $k$NN are implemented using R packages \texttt{e1071}, \texttt{randomForests}, and \texttt{caret}, respectively. All the codes to reproduce the results in this paper can be found at \url{https://github.com/ytstat/multinom-inf}. 

\subsection{Simulations}\label{subsec: simulation}

\subsubsection{Model 1}\label{subsubsec: numerical model 1}
In this model, for independent observations $\{(\bx_i, y_i)\}_{i=1}^n$, we first generate the predictor $\bx_i$ from $N(\bm{0}_p, \bSigma_{p \times p})$ with $\bSigma = (0.75^{|i-j|})_{p \times p}$, then generate $y_i$ from \eqref{eq: multinomial model} with $K=3$ classes and contrast coefficients $\bbetaks{1} = (1, 1, 1, \bm{0}_{p-3})$ and $\bbetaks{2} = (0, 0, 0, 1, 1, 1, \bm{0}_{p-6})$, where $p = 200$. We consider different sample size settings $n = 100, 200, 400$. 

\begin{table}[!h]
\caption{The sum of squared $\ell_2$-estimation error of $\bbetaks{1}$ and $\bbetaks{2}$ in Model $1$, for the contrast-based penalization and the over-parameterization method.}
\label{table: model 1 sse}
\centering
\begin{tabular}{c|ccc}
\hline
Method                        & $n=100$         & $n=200$         & $n=400$         \\\hline
\multicolumn{1}{c|}{Contrast} &2.77 (0.58) &1.787 (0.566) &0.973 (0.377) \\
Over-parameterization         &3.314 (0.755) &2.285 (0.507) &1.352 (0.363) \\\hline
\end{tabular}
\end{table}

First, we compare the contrast-based penalization and the over-parameterization method in terms of the sum of squared $\ell_2$-estimation error of $\bbetaks{1}$ and $\bbetaks{2}$ in Table \ref{table: model 1 sse}. It can be seen that the contrast-based penalization has lower estimation errors than the over-parameterization method.

We also compare the prediction performance of the contrast-based $\ell_1$-penalized multinomial regression with other benchmarks such as Support Vector Machines (SVMs), Linear Discriminant Analysis (LDA), Random Forests, and $k$-Nearest Neighbors ($k$NN). The average misclassification error rates of different methods computed on an independent test set of size $1000$ over $200$ replications are reported in Table \ref{table: model 1 error}. We can see that the $\ell_1$-penalized multinomial regression performs the best among all the methods in all settings.

\begin{table}[!h]
\caption{The average misclassification error rates on test data in Model 1, for $\ell_1$-penalized multinomial regression, SVMs, LDA, Random Forests, and $k$NN.}
\label{table: model 1 error}
\centering
\begin{tabular}{c|ccc}
\hline
Method        & $n=100$                 & $n=200$                 & $n=400$                 \\ \hline
Penalized multinomial regression & 32.37\% (2.23\%)  & 30.08\% (1.80\%)  & 28.52\% (1.51\%)  \\
SVMs             & 52.86\% (3.65\%)  & 45.74\% (2.23\%)  & 40.03\% (1.82\%)  \\
LDA              & 50.36\% (2.49\%)  & 63.26\% (2.32\%)  & 46.87\% (2.08\%)  \\
Random Forests                & 39.67\% (3.08\%)  & 35.52\% (2.41\%)  & 32.42\% (1.62\%)  \\
$k$NN             & 57.32\% (2.52\%)  & 54.70\% (2.20\%)  & 51.66\% (2.03\%)  \\\hline
\end{tabular}
\end{table}

\begin{table}[!h]
\caption{The average coverage probability of $95\%$ CIs on the coefficients corresponding to the signal set $S = \{\{1,2,3\}, \{4,5,6\}\}$ and noise set $S^c = \{\{4,5,6,\ldots, p\}, \{1,2,3, 7, 8,\ldots, p\}\}$ in Model $1$, for the debiased Lasso and the vector bootstrap method.}
\label{table: model 1 coverage prob}
\centering
\begin{tabular}{cc|ccc}
\hline
Method                                              & Measure                               & $n=100$         & $n=200$         & $n=400$         \\ \hline
\multicolumn{1}{c}{\multirow{4}{*}{Debiased-Lasso}} & Average coverage probability on $S$ &0.954 (0.082) &0.928 (0.108) &0.939 (0.107) \\
\multicolumn{1}{c}{}                   & Average coverage probability on $S^c$ &0.991 (0.007) &0.983 (0.008) &0.976 (0.009) \\
\multicolumn{1}{c}{}                 & Average length of CI ($S$)            &2.415 (1.311) &1.52 (0.136) &1.157 (0.066) \\
\multicolumn{1}{c}{}                   & Average length of CI ($S^c$)          &3.116 (1.529) &1.599 (0.102) &1.227 (0.044) \\\hline
\multirow{4}{*}{Bootstrap}                & Average coverage probability on $S$   &0.877 (0.132) &0.914 (0.125) &0.928 (0.116) \\
& Average coverage probability on $S^c$ &1 (0) &1 (0) &1 (0.001) \\
& Average length of CI ($S$)            &1.831 (0.259) &1.658 (0.16) &1.314 (0.098) \\
& Average length of CI ($S^c$)           &0.192 (0.013) &0.23 (0.013) &0.269 (0.019) \\\hline
\end{tabular}
\end{table}

Next, we compare the debiased Lasso with the vector bootstrap method on confidence intervals. We compute the average coverage probability of $95\%$ CIs on the coefficients corresponding to the signal set $S = \{\{1,2,3\}, \{4,5,6\}\}$ and noise set $S^c = \{\{4,5,6,\ldots, p\}, \{1,2,3, 7, 8,\ldots, p\}\}$, where $\{1,2,3\}$ and $\{4,5,6\}$ are signal sets of $\bbetaks{1}$ and $\bbetaks{2}$, respectively. The average CI lengths are calculated separately for the CI of signals and the CI of noises. The average is taken over $200$ simulations. The results are summarized in Table \ref{table: model 1 coverage prob}.  It can be seen that the debiasing method has a larger coverage probability on $S$ than the bootstrap method when $n = 100$. Their coverage probabilities on $S$ in the case of $n = 200$ and $400$ are similar, although the CIs produced by the debiased Lasso are shorter. For the coverage probability on $S^c$, the bootstrap suffers from the super-efficiency phenomenon, where the coverage rate of CIs whose length is close to zero equals one. This is because the Lasso shrinks these coefficients to zero in most replications. In contrast, the debiased Lasso provides reasonable CIs with high coverage probability. This is consistent with the findings of \cite{van2014asymptotically} in the case of linear regression.

\begin{table}[!h]
\caption{The average Type-I error rate and power of individual testing $H_0: \beta^{(k)*}_j = 0$ v.s. $H_1: \beta^{(k)*}_j \neq 0$ at level $5\%$ in Model $1$, for the debiased Lasso and the vector bootstrap method.}
\label{table: model 1 individual testing}
\centering
\begin{tabular}{cc|ccc}
\hline
Method                             & Measure & $n=100$         & $n=200$         & $n=400$         \\ \hline
\multicolumn{1}{c}{\multirow{2}{*}{Debiased-Lasso}} & Type-I error &0.009 (0.007) &0.016 (0.008) &0.024 (0.009) \\
\multicolumn{1}{c}{}                   & Power &0.463 (0.161) &0.712 (0.158) &0.895 (0.116) \\\hline
\multirow{2}{*}{Bootstrap}                & Type-I error &0 (0) &0 (0) &0 (0.001) \\
& Power &0.215 (0.148) &0.504 (0.165) &0.863 (0.131) \\\hline
\end{tabular}
\end{table}

Then, we compare the debiased Lasso with the vector bootstrap method in the individual testing  $H_0: \beta^{(k)*}_j = 0$ v.s. $H_1: \beta^{(k)*}_j \neq 0$ for $j = 1:p$ and $k = 1:(K-1)$, at level $5\%$. The Type-I error rate is calculated by taking the average of Type-I error rates on all noises, and the power is calculated by taking the average of power on all signals. The results can be found in Table \ref{table: model 1 individual testing}. Note that both the debiased Lasso and the vector bootstrap method can successfully control the Type-I error rate under $5\%$ level, but the bootstrap method has much lower power than the debiased Lasso when $n = 100$ and $200$.

\begin{table}[!h]
\caption{The average family-wise error rate (FWER) and power of multiple testing $H_0: \beta^{(k)*}_j = 0$ v.s. $H_1: \beta^{(k)*}_j \neq 0$ for $j=1:p$ and $k=1:(K-1)$ in Model $1$, for the debiased Lasso and the multiple splitting method.}
\label{table: model 1 multiple testing}
\centering
\begin{tabular}{cc|ccc}
\hline
Method                             & Measure & $n=100$         & $n=200$         & $n=400$         \\ \hline
\multicolumn{1}{c}{\multirow{2}{*}{Debiased-Lasso}} & FWER &0.005 (0.071) &0.005 (0.071) &0 (0) \\
\multicolumn{1}{c}{}                   & Power &0.027 (0.061) &0.12 (0.121) &0.318 (0.144) \\\hline
\multirow{2}{*}{Multiple-Spliting}                & FWER &0.015 (0.122) &0.01 (0.1) &0.04 (0.196) \\
& Power &0.233 (0.144) &0.329 (0.111) &0.496 (0.146) \\\hline
\end{tabular}
\end{table}

We also compare the debiased Lasso with the multiple splitting method for multiple testing, where we want to control the family-wise error rate (FWER), i.e., the probability of at least one false discovery, under $5\%$ level. The $p$-values provided by the debiased Lasso are adjusted by a Bonferroni correction. The FWER is the proportion of $200$ simulations where at least one false discovery is reported. The power is calculated by taking the average of the proportion of correctly rejected null hypotheses on the signals over $200$ simulations, which is motivated by \cite{van2014asymptotically}. The results are summarized in Table \ref{table: model 1 multiple testing}. Due to the Bonferroni correction, the debiased Lasso is too conservative in terms of the FWER control, hence does not have immense power. On the other hand, the multiple splitting method successfully controls the FWER with considerably higher statistical power than the debiased Lasso. 

In Section \ref{subsubsec: numerical model 5} of the appendix, we explore a variant of Model 1 where the signal strength is stronger than the original model. We observe a substantial increase in the power of the debiased Lasso, especially in the context of multiple testing.  It is not surprising that the power may be low when the sample size $n$ is small, which implies that some weak signals may be missed. In the literature on Lasso theory for variable selection consistency (e.g., \cite{zhao2006model, wainwright2009sharp}), the signal strength $\min_{j \in S}|\beta^*_j|$ is typically required to be larger than a polynomial function of the penalty parameter $\lambda$, which usually decreases as $n$ grows. As one reviewer noted, missing important variables can be costly in some applications, especially when $\ell_1$-penalized multinomial regression is used for preliminary variable screening. In practice, using smaller $\lambda$ values can help avoid missing weak signals.

\begin{figure}[!h]
    \centering
    \includegraphics[width=\linewidth]{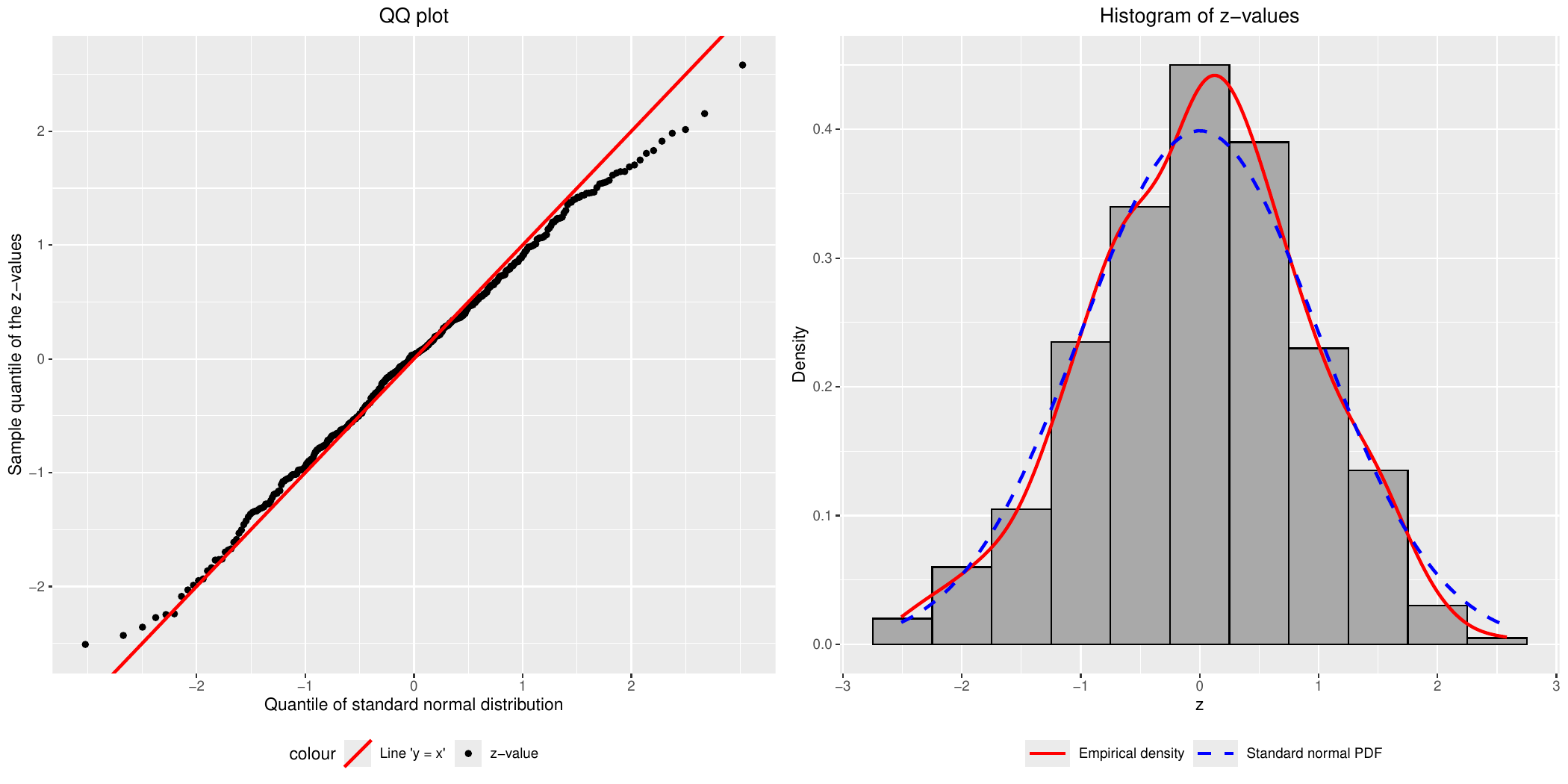}
    \caption{The QQ plot of the sample quantiles of $z$-values $\Big\{\frac{\sqrt{n}(\widehat{b}_j - \beta^*_j)}{\widehat{\bTheta}_j^\top \hSigma \widehat{\bTheta}_j}\Big\}_{j=1}^{(K-1)p}$ in the left panel, and the histogram of $z$-values in the right panel, where $\bbeta^* = ((\bbetaks{1})^\top, (\bbetaks{2})^\top)^\top$. Both are derived from one realization of the setting $(n, K, p) = (400, 3, 200)$ in Model 1.}
    \label{fig: qq-hist}
\end{figure}

\begin{figure}[!h]
    \centering
    \includegraphics[width=0.7\linewidth]{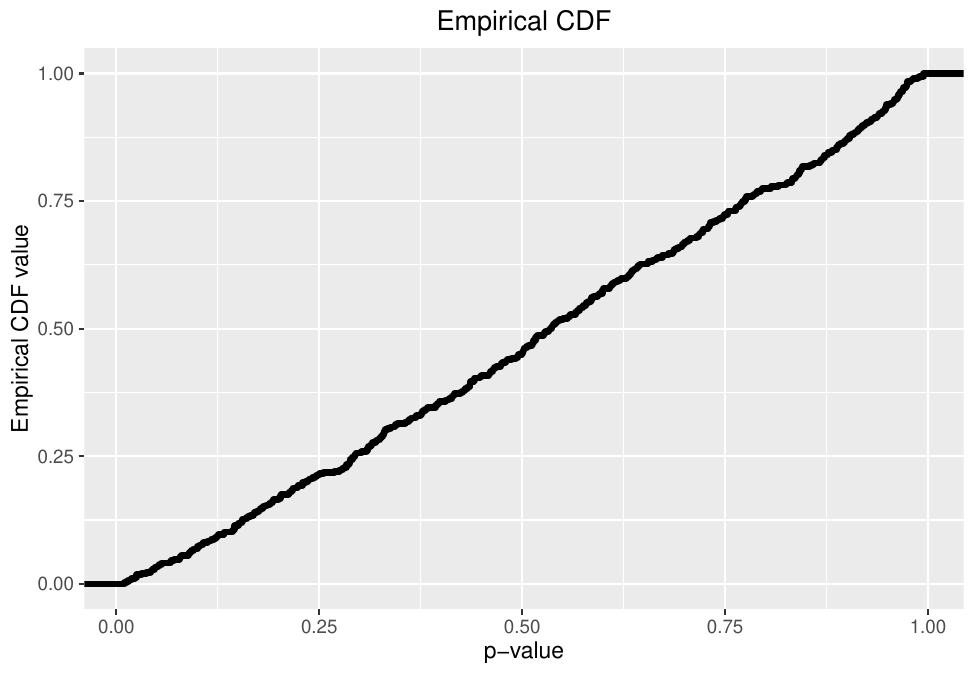}
    \caption{The empirical CDF of computed $p$-values (restricted to zero entries of $\bbeta^*$), where $\bbeta^* = ((\bbetaks{1})^\top, (\bbetaks{2})^\top)^\top$. It is derived from one realization of the setting $(n, K, p) = (400, 3, 200)$ in Model 1.}
    \label{fig: pvalue-diag}
\end{figure}

Finally, we present the QQ plot of sample quantiles of $z$-values $\Big\{\frac{\sqrt{n}(\widehat{b}_j - \beta^*_j)}{\widehat{\bTheta}_j^\top \hSigma \widehat{\bTheta}_j}\Big\}_{j=1}^{(K-1)p}$ and the histogram of $z$-values in Figure \ref{fig: qq-hist}. It is evident that the sample quantiles of $z$ values approximately align with the quantiles of the standard normal distribution, and the histogram fits well with the standard normal density. We also plot the empirical CDF of computed $p$-values (restricted to zero entries of $\bbeta^*$), in Figure \ref{fig: pvalue-diag}. We can see that the computed $p$-values are approximately uniformly distributed on $[0, 1]$. These empirical diagnostic results align with our theoretical findings.

\subsubsection{Model 2}\label{subsubsec: numerical model 2}
In this model, we generate independent observations $\{(\bx_i, y_i)\}_{i=1}^n$ similarly to Model $1$. We first generate the predictor $\bx_i$ from $N(\bm{0}_p, \bSigma_{p \times p})$ with $\bSigma = (0.75^{|i-j|})_{p \times p}$, then generate $y_i$ from \eqref{eq: multinomial model} with $K=4$ classes and contrast coefficients $\bbetaks{1} = (1, 1, 1, \bm{0}_{p-3})$, $\bbetaks{2} = (\bm{0}_{3}, 1, 1, 1, \bm{0}_{p-6})$, $\bbetaks{3} = (\bm{0}_{6}, 1, 1, 1, \bm{0}_{p-9})$, where $p = 200$. We consider different sample size settings $n = 100, 200, 400$. 

\begin{table}[!h]
\caption{The sum of squared $\ell_2$-estimation error of $\bbetaks{1}$, $\bbetaks{2}$, and $\bbetaks{3}$ in Model $2$, for the contrast-based penalization and the over-parameterization method.}
\label{table: model 2 sse}
\centering
\begin{tabular}{c|ccc}
\hline
Method                        & $n=100$         & $n=200$         & $n=400$         \\ \hline
\multicolumn{1}{c|}{Contrast} &4.607 (0.926) &2.93 (0.789) &1.696 (0.496) \\
Over-parameterization         &5.559 (1.292) &3.563 (0.795) &2.134 (0.482) \\\hline
\end{tabular}
\end{table}

The sum of squared $\ell_2$-estimation error of $\bbetaks{1}$, $\bbetaks{2}$, and $\bbetaks{3}$ is summarized in Table \ref{table: model 2 sse}, where it is evident that the contrast-based penalization achieves lower estimation error than the over-parameterized method.

We also compare the prediction performance of the contrast-based $\ell_1$-penalized multinomial regression with other benchmarks, and the average misclassification error rates of different methods are reported in Table \ref{table: model 2 error}. The results indicate that the $\ell_1$-penalized multinomial regression outperforms all the other methods.

\begin{table}[!h]
\caption{The average misclassification error rates on test data in Model 2, for $\ell_1$-penalized multinomial regression, SVMs, LDA, Random Forests, and $k$NN.}
\label{table: model 2 error}
\centering
\begin{tabular}{c|ccc}
\hline
Method        & $n=100$                 & $n=200$                 & $n=400$                 \\ \hline
Penalized multinomial regression & 40.02\% (2.61\%) & 36.33\% (2.09\%) & 34.46\% (1.59\%) \\
SVMs & 63.59\% (4.01\%) & 55.30\% (2.61\%) & 48.74\% (1.92\%) \\
LDA & 59.65\% (2.24\%) & 71.59\% (2.02\%) & 56.46\% (1.94\%) \\
Random Forests & 50.01\% (3.65\%) & 43.79\% (2.58\%) & 40.28\% (1.90\%) \\
$k$NN & 66.11\% (2.24\%) & 63.53\% (1.91\%) & 60.80\% (1.91\%) \\\hline
\end{tabular}
\end{table}

\begin{table}[!h]
\caption{The average coverage probability of $95\%$ CIs on the coefficients corresponding to the signal set $S = \{\{1,2,3\}, \{4,5,6\}, \{7,8,9\}\}$ and noise set $S^c = \{\{4,5,6,\ldots, p\}, \{1,2,3, 7, 8,\ldots, p\}, \{1,2,3,4,5,6, 10, 11,\ldots, p\}\}$ in Model $2$, for the debiased Lasso and the vector bootstrap method.}
\label{table: model 2 coverage prob}
\centering
\begin{tabular}{cc|ccc}
\hline
Method                                              & Measure                               & $n=100$         & $n=200$         & $n=400$         \\ \hline
\multicolumn{1}{c}{\multirow{4}{*}{Debiased-Lasso}} & Average coverage probability on $S$ &0.978 (0.049) &0.934 (0.083) &0.946 (0.078) \\
\multicolumn{1}{c}{}                   & Average coverage probability on $S^c$ &0.993 (0.006) &0.986 (0.006) &0.979 (0.006) \\
\multicolumn{1}{c}{}                 & Average length of CI ($S$)            &3.156 (2.158) &1.697 (0.175) &1.348 (0.075) \\
\multicolumn{1}{c}{}                   & Average length of CI ($S^c$)          &2.761 (0.711) &1.723 (0.104) &1.398 (0.047) \\\hline
\multirow{4}{*}{Bootstrap}                & Average coverage probability on $S$   &0.848 (0.117) &0.895 (0.112) &0.918 (0.112) \\
& Average coverage probability on $S^c$ &1 (0) &1 (0) &1 (0.001) \\
& Average length of CI ($S$)            &1.689 (0.199) &1.598 (0.145) &1.312 (0.078) \\
& Average length of CI ($S^c$)           &0.174 (0.01) &0.21 (0.011) &0.25 (0.014) \\\hline
\end{tabular}
\end{table}

Similar to Model $1$, we compute the average coverage probability of $95\%$ CIs on the coefficients corresponding to the signal set and noise set for Model $2$ in Table \ref{table: model 2 coverage prob}. The debiased Lasso exhibits a higher coverage probability on the signal set $S$ than the bootstrap method.

\begin{table}[!h]
\caption{The average Type-I error rate and power of individual testing $H_0: \beta^{(k)*}_j = 0$ v.s. $H_1: \beta^{(k)*}_j \neq 0$ in Model $2$ at level $5\%$, for the debiased Lasso and the vector bootstrap method.}
\label{table: model 2 individual testing}
\centering
\begin{tabular}{cc|ccc}
\hline
Method                             & Measure & $n=100$         & $n=200$         & $n=400$         \\ \hline
\multicolumn{1}{c}{\multirow{2}{*}{Debiased-Lasso}} & Type-I error &0.007 (0.006) &0.014 (0.006) &0.021 (0.006) \\
\multicolumn{1}{c}{}                   & Power &0.308 (0.122) &0.629 (0.141) &0.778 (0.12) \\\hline
\multirow{2}{*}{Bootstrap}                & Type-I error &0 (0) &0 (0) &0 (0.001) \\
& Power &0.159 (0.1) &0.43 (0.126) &0.791 (0.124) \\\hline
\end{tabular}
\end{table}

In Table \ref{table: model 2 individual testing}, we summarize the average Type-I error rate and the power for the debiased Lasso with the vector bootstrap method in the individual testing  $H_0: \beta^{(k)*}_j = 0$ v.s. $H_1: \beta^{(k)*}_j \neq 0$ for $j = 1:p$ and $k = 1:(K-1)$ at level $5\%$. The debiased Lasso demonstrates significantly larger power than the bootstrap method while successfully controlling the Type-I error rate under $5\%$.

\begin{table}[!h]
\caption{The average family-wise error rate (FWER) and power of multiple testing $H_0: \beta^{(k)*}_j = 0$ v.s. $H_1: \beta^{(k)*}_j \neq 0$ for $j=1:p$ and $k=1:(K-1)$ in Model $2$, for the debiased Lasso and the multiple splitting method.}
\label{table: model 2 multiple testing}
\centering
\begin{tabular}{cc|ccc}
\hline
Method                             & Measure & $n=100$         & $n=200$        & $n=400$         \\ \hline
\multicolumn{1}{c}{\multirow{2}{*}{Debiased-Lasso}} & FWER &0.005 (0.071) &0 (0) &0 (0) \\
\multicolumn{1}{c}{}                   & Power &0.006 (0.024) &0.058 (0.069) &0.146 (0.096) \\\hline
\multirow{2}{*}{Multiple-Spliting}                & FWER &0.44 (0.498) &0.61 (0.489) &0.895 (0.307) \\
& Power &0.089 (0.093) &0.132 (0.084) &0.224 (0.09) \\\hline
\end{tabular}
\end{table}

We summarize the average FWER and power of multiple testing for the debiased Lasso with Bonferroni correction and the multiple splitting method in Table \ref{table: model 2 multiple testing}. The multiple splitting method fails to control the FWER under 5\%, whereas the debiased Lasso successfully controls the FWER, although it might be over-conservative.

\subsubsection{Model 3}\label{subsubsec: numerical model 3}
In this model, we generate each of the observations $\{(\bx_i, y_i)\}_{i=1}^n$ independently using a linear discriminant analysis (LDA) setting. First, the class label $y_i$ is generated from the discrete distribution on $\{1, 2, 3\}$ with probability $0.3, 0.3, 0.4$, respectively. Then given $y_i = k$, the predictor $\bx_i$ is sampled from $N(\bmu^{(k)*}, \bSigma)$ with $\bmu^{(k)*} = \bmu^{(3)*} + \bSigma \bbetaks{k}$ for $k=1, 2$, where each coordinate of $\bmuks{3}$ is randomly sampled from $\textup{Unif}(\{-1, 1\})$, $\bSigma = (0.5^{|i-j|})_{p \times p}$, and $p =200$. The vectors $\bbetaks{k}_{S^{(1)}} = (1, -1, 1)^\top$, $\bbetaks{k}_{S^{(2)}} = (1, 1, -1)^\top$, where $S^{(1)}$ and $S^{(2)}$ are two sets of cardinality $3$ randomly sampled from $1:p$ without replacement. Following a simulation setting in \cite{van2014asymptotically}, we set a random seed for the randomness in the model parameters for all $200$ simulations, which gives $S^{(1)} = \{197,92,152\}$ and $S^{(2)} = \{173,170,191\}$. Thus, the data generation model remains the same across $200$ simulations. In this model, the Gaussian means $\bmuks{1}, \bmuks{2}, \bmuks{3}$ are dense, but the contrast coefficients (or discriminant coefficients, as referred to in LDA models) $\bbetaks{1}, \bbetaks{2}$ are sparse. We consider different sample size settings $n = 100, 200, 400$. 

\begin{table}[!h]
\caption{The sum of squared $\ell_2$-estimation error of $\bbetaks{1}$ and $\bbetaks{2}$, in Model $3$, for the contrast-based penalization and the over-parameterization method.}
\label{table: model 3 sse}
\centering
\begin{tabular}{c|ccc}
\hline
Method                        & $n=100$         & $n=200$         & $n=400$         \\ \hline
\multicolumn{1}{c|}{Contrast} &9.291 (4.103) &5.46 (2.577) &3.379 (1.508) \\
Over-parameterization         &11.691 (4.578) &6.986 (2.967) &4.163 (1.783) \\\hline
\end{tabular}
\end{table}

The results in Table \ref{table: model 3 sse} for the sum of squared $\ell_2$-estimation error of $\bbetaks{1}$ and $\bbetaks{2}$ are similar to those of Models $1$ and $2$, where the contrast-based penalization performs better than the over-parameterization method. 

We also compare the prediction performance of the contrast-based $\ell_1$-penalized multinomial regression with other benchmarks, and the average misclassification error rates of different methods are reported in Table \ref{table: model 3 error}. We can see that the $\ell_1$-penalized multinomial regression outperforms all the other methods.

\begin{table}[!h]
\caption{The average misclassification error rates on test data in Model 3, for $\ell_1$-penalized multinomial regression, SVMs, LDA, Random Forests, and $k$NN.}
\label{table: model 3 error}
\centering
\begin{tabular}{c|ccc}
\hline
Method                     & $n=100$               & $n=200$               & $n=400$               \\ \hline
Penalized multinomial regression & 54.58\% (4.12\%)      & 52.95\% (2.54\%)      & 51.74\% (1.95\%)      \\
SVMs                              & 62.81\% (3.78\%)      & 60.16\% (4.82\%)      & 57.96\% (4.85\%)      \\
LDA                              & 60.66\% (3.73\%)      & 64.86\% (3.09\%)      & 59.49\% (4.75\%)      \\
Random Forests                   & 58.76\% (3.37\%)      & 56.89\% (3.18\%)      & 55.69\% (2.39\%)      \\
$k$NN                            & 60.51\% (2.85\%)      & 59.08\% (2.78\%)      & 58.00\% (2.67\%)      \\
\hline
\end{tabular}
\end{table}

\begin{table}[!h]
\caption{The average coverage probability of $95\%$ CIs on the coefficients corresponding to the signal set $S = \{\{197,92,152\}, \{173,170,191\}\}$ and noise set $S^c = \{(1:p) \backslash \{197,92,152\}, (1:p) \backslash \{173,170,191\}\}$ in Model $3$, for the debiased Lasso and the vector bootstrap method.}
\label{table: model 3 coverage prob}
\centering
\begin{tabular}{cc|ccc}
\hline
Method                                              & Measure                               & $n=100$         & $n=200$         & $n=400$         \\ \hline
\multicolumn{1}{c}{\multirow{4}{*}{Debiased-Lasso}} & Average coverage probability on $S$ &0.963 (0.085) &0.952 (0.089) &0.945 (0.085) \\
\multicolumn{1}{c}{}                   & Average coverage probability on $S^c$ &0.995 (0.005) &0.987 (0.006) &0.977 (0.009) \\
\multicolumn{1}{c}{}                 & Average length of CI ($S$)            &1.838 (0.454) &1.172 (0.124) &0.841 (0.045) \\
\multicolumn{1}{c}{}                   & Average length of CI ($S^c$)          &1.958 (0.363) &1.174 (0.077) &0.831 (0.027) \\\hline
\multirow{4}{*}{Bootstrap}                & Average coverage probability on $S$   &0.738 (0.176) &0.887 (0.121) &0.925 (0.112) \\
& Average coverage probability on $S^c$ &1 (0) &1 (0.001) &0.999 (0.001) \\
& Average length of CI ($S$)            &1.183 (0.135) &1.024 (0.075) &0.795 (0.059) \\
& Average length of CI ($S^c$)           &0.197 (0.01) &0.234 (0.01) &0.259 (0.014) \\\hline
\end{tabular}
\end{table}

Table \ref{table: model 3 coverage prob} summarizes the average coverage probability of $95\%$ CIs for the debiased Lasso and the vector bootstrap method. It shows that the debiased Lasso has a higher coverage probability on $S$ than the multiple splitting method.

\begin{table}[!h]
\caption{The average Type-I error rate and power of individual testing $H_0: \beta^{(k)*}_j = 0$ v.s. $H_1: \beta^{(k)*}_j \neq 0$ at level $5\%$ in Model $3$, for the debiased Lasso and the vector bootstrap method.}
\label{table: model 3 individual testing}
\centering
\begin{tabular}{cc|ccc}
\hline
Method                             & Measure & $n=100$         & $n=200$         & $n=400$         \\ \hline
\multicolumn{1}{c}{\multirow{2}{*}{Debiased-Lasso}} & Type-I error &0.005 (0.005) &0.013 (0.006) &0.023 (0.009) \\
\multicolumn{1}{c}{}                   & Power &0.546 (0.186) &0.893 (0.127) &0.992 (0.036) \\\hline
\multirow{2}{*}{Bootstrap}                & Type-I error &0 (0) &0 (0.001) &0.001 (0.001) \\
& Power &0.368 (0.188) &0.855 (0.134) &0.998 (0.017) \\\hline
\end{tabular}
\end{table}

Table \ref{table: model 3 individual testing} displays the average Type-I error rate and power of individual testing for the debiased Lasso and the vector bootstrap method. Both the debiased Lasso and the vector bootstrap can control the Type-I error rate under the $5\%$ level, but the debiased Lasso exhibits greater power than the bootstrap method when $n = 100$ and $200$.

\begin{table}[!h]
\caption{The average family-wise error rate (FWER) and power of multiple testing $H_0: \beta^{(k)*}_j = 0$ v.s. $H_1: \beta^{(k)*}_j \neq 0$ for $j=1:p$ and $k=1:(K-1)$ in Model $3$, for the debiased Lasso and the multiple splitting method.}
\label{table: model 3 multiple testing}
\centering
\begin{tabular}{cc|ccc}
\hline
Method        & Measure & $n=100$         & $n=200$         & $n=400$         \\ \hline
\multicolumn{1}{c}{\multirow{2}{*}{Debiased-Lasso}} & FWER &0 (0) &0.005 (0.071) &0.005 (0.071) \\
\multicolumn{1}{c}{}                   & Power &0.039 (0.077) &0.267 (0.164) &0.75 (0.165) \\\hline
\multirow{2}{*}{Multiple-Spliting}                & FWER &0.105 (0.307) &0.06 (0.238) &0.03 (0.171) \\
& Power &0.157 (0.141) &0.658 (0.152) &0.995 (0.029) \\\hline
\end{tabular}
\end{table}

In Table \ref{table: model 3 multiple testing}, we show the average FWER and the power of the multiple testing for the debiased Lasso and the multiple splitting method. The multiple splitting method fails to control the FWER when $n = 100$ and $200$ but achieves overwhelming power and successful control for FWER when $n=400$. In contrast, the debiased Lasso with Bonferroni correction consistently maintains good control of FWER but might be too conservative for large $n$, leading to lower statistical power.

\subsubsection{Model 4}\label{subsubsec: numerical model 4}
Similar to Model $3$, we generate each of the observations $\{(\bx_i, y_i)\}_{i=1}^n$ independently using a linear discriminant analysis (LDA) framework. First, we sample the class label $y_i$ from the discrete distribution on $\{1, 2, 3, 4\}$ with probability $0.3, 0.2, 0.3, 0.2$, respectively. Given $y_i = k$, the predictor $\bx_i$ is sampled from $N(\bmu^{(k)*}, \bSigma)$ with $\bmu^{(k)*} = \bmu^{(4)*} + \bSigma \bbetaks{k}$ for $k=1, 2, 3$, where each coordinate of $\bmuks{4}$ is randomly sampled from $\textup{Unif}(\{-1, 1\})$, $\bSigma = (0.5^{|i-j|})_{p \times p}$, and $p =200$. The vectors $\bbetaks{k}_{S^{(1)}} = (1, -1, 1)^\top$, $\bbetaks{k}_{S^{(2)}} = (1, 1, -1)^\top$, $\bbetaks{k}_{S^{(3)}} = (-1, 1, 1)^\top$, where $S^{(1)}$, $S^{(2)}$, and $S^{(3)}$ are three sets of cardinality $3$ randomly sampled from $1:p$ without replacement. We fix a random seed to ensure consistency in model parameters across all $200$ simulations, resulting in $S^{(1)} = \{197,92,152\}$, $S^{(2)} = \{173,170,191\}$, and $S^{(3)} = \{23,73,148\}$. We consider different sample size settings with $n = 100, 200, 400$.

\begin{table}[!h]
\caption{The sum of squared $\ell_2$-estimation error of $\bbetaks{1}$, $\bbetaks{2}$, and $\bbetaks{3}$, in Model $4$, for the contrast-based penalization and the over-parameterization method.}
\label{table: model 4 sse}
\centering
\begin{tabular}{c|ccc}
\hline
Method                        & $n=100$          & $n=200$         & $n=400$         \\ \hline
\multicolumn{1}{c|}{Contrast} &7.331 (2.772) &4.433 (1.706) &2.543 (0.864) \\
Over-parameterization         &9.858 (5.624) &5.902 (2.717) &3.465 (1.435) \\\hline
\end{tabular}
\end{table}

Table \ref{table: model 4 sse} summarizes the sum of squared $\ell_2$-estimation error of the contrast coefficients, demonstrating that contrast-based penalization performs better than over-parameterization.

We also compare the prediction performance of the contrast-based $\ell_1$-penalized multinomial regression with other benchmarks, and the average misclassification error rates of different methods are reported in Table \ref{table: model 4 error}. The results show that the $\ell_1$-penalized multinomial regression outperforms all other methods.

\begin{table}[!h]
\caption{The average misclassification error rates on test data in Model 4, for $\ell_1$-penalized multinomial regression, SVMs, LDA, Random Forests, and $k$NN.}
\label{table: model 4 error}
\centering
\begin{tabular}{c|ccc}
\hline
Method       & $n=100$                 & $n=200$                 & $n=400$                 \\ \hline
Penalized multinomial regression & 56.01\% (5.05\%)        & 50.70\% (4.44\%)        & 47.50\% (3.42\%)        \\
SVMs                              & 70.02\% (3.80\%)        & 66.46\% (4.21\%)        & 62.92\% (5.48\%)        \\
LDA                              & 68.63\% (3.42\%)        & 73.33\% (2.89\%)        & 65.76\% (4.71\%)        \\
Random forests                   & 64.50\% (3.84\%)        & 61.04\% (3.72\%)        & 58.43\% (3.90\%)        \\
$k$-nearest neighbors            & 68.13\% (2.58\%)        & 66.10\% (2.62\%)        & 64.61\% (2.86\%)        \\
\hline
\end{tabular}
\end{table}

\begin{table}[!h]
\caption{The average coverage probability of $95\%$ CIs on the coefficients corresponding to the signal set $S = \{\{197,92,152\}, \{173,170,191\}, \{23,73,148\}\}$ and noise set $S^c = \{(1:p) \backslash \{197,92,152\}, (1:p) \backslash \{173,170,191\}, (1:p) \backslash \{23,73,148\}\}$ in Model $4$, for the debiased Lasso and the vector bootstrap method.}
\label{table: model 4 coverage prob}
\centering
\begin{tabular}{cc|ccc}
\hline
Method   & Measure     & $n=100$         & $n=200$         & $n=400$         \\ \hline
\multicolumn{1}{c}{\multirow{4}{*}{Debiased-Lasso}} & Average coverage probability on $S$ &0.968 (0.064) &0.958 (0.063) &0.962 (0.065) \\
\multicolumn{1}{c}{}                   & Average coverage probability on $S^c$ &0.996 (0.003) &0.99 (0.005) &0.981 (0.006) \\
\multicolumn{1}{c}{}                 & Average length of CI ($S$)            &2.146 (0.673) &1.242 (0.12) &0.936 (0.046) \\
\multicolumn{1}{c}{}                   & Average length of CI ($S^c$)          &2.096 (0.399) &1.277 (0.085) &0.961 (0.034) \\\hline
\multirow{4}{*}{Bootstrap}                & Average coverage probability on $S$   &0.66 (0.148) &0.814 (0.129) &0.921 (0.086) \\
& Average coverage probability on $S^c$ &1 (0) &1 (0.001) &1 (0.001) \\
& Average length of CI ($S$)            &1.115 (0.119) &1.018 (0.067) &0.803 (0.043) \\
& Average length of CI ($S^c$)           &0.174 (0.009) &0.213 (0.008) &0.242 (0.01) \\\hline
\end{tabular}
\end{table}

Table \ref{table: model 4 coverage prob} presents the coverage probability of $95\%$ CIs generated by the debiased Lasso and the vector bootstrap method. The debiased Lasso provides CIs for signals in $S$ with higher coverage probabilities compared to those of the bootstrap method.

\begin{table}[!h]
\caption{The average Type-I error rate and power of individual testing $H_0: \beta^{(k)*}_j = 0$ v.s. $H_1: \beta^{(k)*}_j \neq 0$ at level $5\%$ in Model $4$, for the debiased Lasso and the vector bootstrap method.}
\label{table: model 4 individual testing}
\centering
\begin{tabular}{cc|ccc}
\hline
Method                             & Measure & $n=100$         & $n=200$         & $n=400$         \\ \hline
\multicolumn{1}{c}{\multirow{2}{*}{Debiased-Lasso}} & Type-I error &0.004 (0.003) &0.01 (0.005) &0.019 (0.006) \\
\multicolumn{1}{c}{}                   & Power &0.433 (0.15) &0.854 (0.104) &0.981 (0.052) \\\hline
\multirow{2}{*}{Bootstrap}                & Type-I error &0 (0) &0 (0.001) &0 (0.001) \\
& Power &0.246 (0.126) &0.75 (0.129) &0.991 (0.031) \\\hline
\end{tabular}
\end{table}

We summarize the average Type-I error rate and power of individual testing in Table \ref{table: model 4 individual testing}. It is evident that both the debiased Lasso and the vector bootstrap method control the Type-I error rate under $5\%$. However, the debiased Lasso exhibits significantly greater power than the bootstrap method when $n=100$ and $200$.

\begin{table}[!h]
\caption{The average family-wise error rate (FWER) and power of multiple testing $H_0: \beta^{(k)*}_j = 0$ v.s. $H_1: \beta^{(k)*}_j \neq 0$ for $j=1:p$ and $k=1:(K-1)$ in Model $4$, for the debiased Lasso and the multiple splitting method.}
\label{table: model 4 multiple testing}
\centering
\begin{tabular}{cc|ccc}
\hline
Method                             & Measure & $n=100$         & $n=200$         & $n=400$         \\ \hline
\multicolumn{1}{c}{\multirow{2}{*}{Debiased-Lasso}} & FWER &0 (0) &0.005 (0.071) &0 (0) \\
\multicolumn{1}{c}{}                   & Power &0.012 (0.034) &0.163 (0.104) &0.516 (0.166) \\\hline
\multirow{2}{*}{Multiple-Splitting}                & FWER &0.37 (0.484) &0.78 (0.415) &1 (0) \\
& Power &0.087 (0.119) &0.227 (0.14) &0.616 (0.068) \\\hline
\end{tabular}
\end{table}

Finally, Table \ref{table: model 4 multiple testing} shows the results for the multiple testing. The multiple splitting method fails to control the FWER, whereas the debiased Lasso successfully does so while maintaining relatively high power.

\subsection{A real example: Identifying important variables associated with different subtypes of MCI/dementia in NACC database}\label{subsec: real example numerical}

\begin{table}[!h]
\caption{Contrasts, odds ratios, and $p$-values of top 10 predictors between FTLD and CN obtained by the debiased $\ell_1$-penalized multinomial regression.  Brackets indicate dummy variables for categorical variables.}
\label{table: ftld}
\centering
\begin{tabular}{l|rrrl}
\hline
Feature     & \multicolumn{1}{l}{Coefficient} & \multicolumn{1}{l}{Odds ratio} & \multicolumn{1}{l}{$p$-value} & Description   \\ \hline
TRAVEL3     & 33.9214  & $5.39 \times 10^{14}$          & 0             & \makecell[l]{In the past four weeks, did the subject have any difficulty or need help with: Traveling out \\of the neighborhood, driving, or arranging to take public transportation [3 = Dependent]}       \\\hline  
PSP1        & 48.3045                         & $9.51 \times 10^{20}$          & 0                             & \makecell[l]{Presumptive etiologic diagnosis - primary supranuclear palsy (PSP) [1 = Yes]} \\\hline  
NACCFFTD1   & 4.5809                          & 97.6059                        & 0                             & \makecell[l]{In this family, is there evidence for an FTLD mutation?  \noindent[1 = Yes]}\\\hline 
EVENTS2     & 28.9003                         & $3.56 \times 10^{12}$          & 0                             & \makecell[l]{In the past four weeks, did the subject have any difficulty or need help with: Keeping track \\ of current events \noindent[2 = Requires assistance]} \\\hline  
COGSTAT2    & -1.953                          & 0.1418                         & 0                             & \makecell[l]{Per clinician, based on the neuropsychological examination, the subject's cognitive status is \\deemed [2=Normal for age]}   \\\hline  
PAYATTN1    & 6.3784                          & 589.0052                       & 0                             & \makecell[l]{In the past four weeks, did the subject have any difficulty or need help with: Paying attention  \\to  and understanding a TV program, book, or magazine [1 = Has difficulty, but does by self]} \\\hline  
GAMES2      & 20.4844                         & $7.88 \times 10^8$             & 0                             & \makecell[l]{In the past four weeks, did the subject have any difficulty or need help with: Playing a game \\of skill such as bridge or chess, working on a hobby [2 = Requires assistance]}                \\\hline  
COGSTAT1    & -2.6038                         & 0.074                          & 0                             & \makecell[l]{Per clinician, based on the neuropsychological examination, the subject's cognitive status is \\deemed [1=better than normal for age]}    \\\hline  
COGSTAT3    & -1.9584                         & 0.1411                         & 0.0004                        & \makecell[l]{Per clinician, based on the neuropsychological examination, the subject's cognitive status is \\ deemed [3 = One or two test scores abnormal]}                                                  \\\hline  
HOMEHOBB0.5 & 5.877                           & 356.7285                       & 0.0007                        & \makecell[l]{Home and hobbies [0.5 = Questionable impairment]}  \\ \hline
\end{tabular}
\end{table}

\begin{table}[!h]
\caption{Contrasts, odds ratios, and $p$-values of top 10 predictors between LBD/PD and CN obtained by the debiased $\ell_1$-penalized multinomial regression.  Brackets indicate dummy variables for categorical variables.}
\label{table: lbd}
\centering
\begin{tabular}{l|rrrl}
\hline
Feature   & \multicolumn{1}{l}{Coefficient} & \multicolumn{1}{l}{Odds ratio} & \multicolumn{1}{l}{$p$-value} & Description   \\ \hline
HALLSEV3  & 55.7257                         & $1.59 \times 10^{24}$          & $< 10^{-4}$                & \makecell[l]{Hallucinations severity [3 = Severe (very marked or prominent; a dramatic change)]} \\\hline  
PDOTHR1   & 6.486                           & 655.9019                       & $< 10^{-4}$                & Other parkinsonian disorder [1 = Recent/Active]     \\\hline  
PD1       & 4.4647                          & 86.8966                        & $< 10^{-4}$                & Parkinson's disease (PD) [1 = Recent/Active]  \\\hline  
NACCDAYS  & 0.5998                          & 1.8218                         & $< 10^{-4}$                & Days from initial visit to most recent visit      \\\hline  
HALLSEV2  & 10.8981                         & 54071.5757                     & $< 10^{-4}$                & \makecell[l]{Hallucinations severity [2 = Moderate (significant, but not a dramatic change)]}     \\\hline  
NACCBEHF8 & 4.4825                          & 88.4524                        & $< 10^{-4}$                & \makecell[l]{Indicate the predominant symptom that was first recognized as a decline in the subject's \\behavior [8 = REM sleep behavior disorder]}     \\\hline  
FRSTCHG2  & 2.3303                          & 10.2807                        & 0.0008                         & \makecell[l]{Indicate the predominant domain that was first recognized as changed in the subject [2 = \\behavior]}  \\\hline
GAMES3    & 11.6056                         & $1.10 \times 10^{5}$           & 0.0016                        & \makecell[l]{In the past four weeks, did the subject have any difficulty or need help with: Playing a game of \\ skill such as bridge or chess, working on a hobby [3 = Dependent]} \\\hline  
SLEEPAP1  & -3.8154                         & 0.022                          & 0.0027                        & Sleep apnea present [1 = Yes]  \\\hline  
DEPD1     & 1.1913                          & 3.2915                         & 0.004                         & Depression or dysphoria in the last month [1 = Yes]        \\ \hline
\end{tabular}
\end{table}

\begin{table}[!h]
\caption{Contrasts, odds ratios, and $p$-values of top 10 predictors between VBI and CN obtained by the debiased $\ell_1$-penalized multinomial regression.  Brackets indicate dummy variables for categorical variables.}
\label{table: vbi}
\centering
\begin{tabular}{l|rrrl}
\hline
Feature        & \multicolumn{1}{l}{Coefficient} & \multicolumn{1}{l}{Odds ratio} & \multicolumn{1}{l}{$p$-value} & Description   \\ \hline
NACCDAYS       & 0.6244                          & 1.8672                         & 0                             & Days from initial visit to most recent visit \\\hline
MINTTOTS\_abn1 & 8.471     & 4774.4576                      & 0                   & \makecell[l]{Multilingual Naming Test (MINT) - Total score [abn1 = Abnormal]}  \\\hline  
TRAILB         & 0.3822                          & 1.4655                         & 0                             & \makecell[l]{Trail Making Test Part b - Total number of seconds to complete}    \\\hline  
APNEA9         & 4.6957                          & 109.4747                       & 0.0001                        & \makecell[l]{Sleep apnea history reported at Initial Visit [9 = Unknown]}   \\\hline  
CBSTROKE2      & 1.3345                          & 3.7981                         & 0.0001                        & Stroke [2 = Remote/Inactive]   \\\hline  
DIABET1        & 8.546                           & 5146.2553                      & 0.0001                        & Diabetes present at visit [1 = Yes, Type I]   \\\hline  
BEAGIT1        & 4.5007                          & 90.0779                        & 0.0002                        & \makecell[l]{Subject currently manifests meaningful change in behavior - Agitation [1 = Yes]}    \\\hline  
ENERGY1        & 0.6031                          & 1.8277                         & 0.0004                        & Do you feel full of energy? [1 = No]    \\\hline  
CBTIA1         & 1.8537                          & 6.3837                         & 0.0005                        & Transient ischemic attack (TIA) [1 = Recent/Active]        \\\hline  
DIGFORCT\_abn1 & -7.5341                         & 0.005                          & 0.0006                        & \makecell[l]{Number Span Test: Forward - Number of correct trials [abn1 = Abnormal]}   \\ \hline
\end{tabular}
\end{table}

We have described the dataset background, pre-processing procedure, and the contrast coefficient between AD and CN in Section \ref{subsec: intro example}. In this section, we continue by presenting and discussing the results of contrast coefficients between FTLD and CN, LBD/PD and CN, and VBI and CN, which can be found in Tables \ref{table: ftld}-\ref{table: vbi}, respectively. 

Similar to the effect on AD and CN, the number of days from the initial visit to the most recent visit is also a strong predictor for LBD/PD and CN, and VBI and CN. With other factors fixed, a long time from the initial visit means more likely progressing into AD, LBD/PD, and VBI. For FTLD, the presence of primary supranuclear palsy and family FTLD immutation are significant predictors. This aligns with the findings in \cite{greaves2019update} that around 30\% of FTLD patients have a strong family history, and many FTLD cases are related to genetics. Additionally, the presumptive etiologic diagnosis of primary supranuclear palsy (PSP) is a strong predictor. Such a relationship has been found in the literature. For example, it has been known that PSP shares pathology with FTLD \citep{giagkou2019progressive}. Moreover, difficulties with traveling, playing games, and engaging in hobbies are also predictors of FTLD, suggestive of the subtle, non-memory cognitive symptoms that may arise in preclinical stages. For LBD/PD, hallucinations, REM sleep behavior disorder, and depression are significant predictors against CN, which aligns with findings in the literature \citep{boeve1998rem, onofrj2013visual}. For VBI, a history of vascular events (stroke, transient ischemic attack) and risk factors (sleep apnea, diabetes) appear across several variables as predictors, as does diabetes, a strong indicator of vascular risk. Similar conclusions are noted in \cite{ahtiluoto2010diabetes} and \cite{kalaria2016stroke}.

\begin{sidewaysfigure}
	\hspace{-3cm}
	\includegraphics[width=1.12\textwidth]{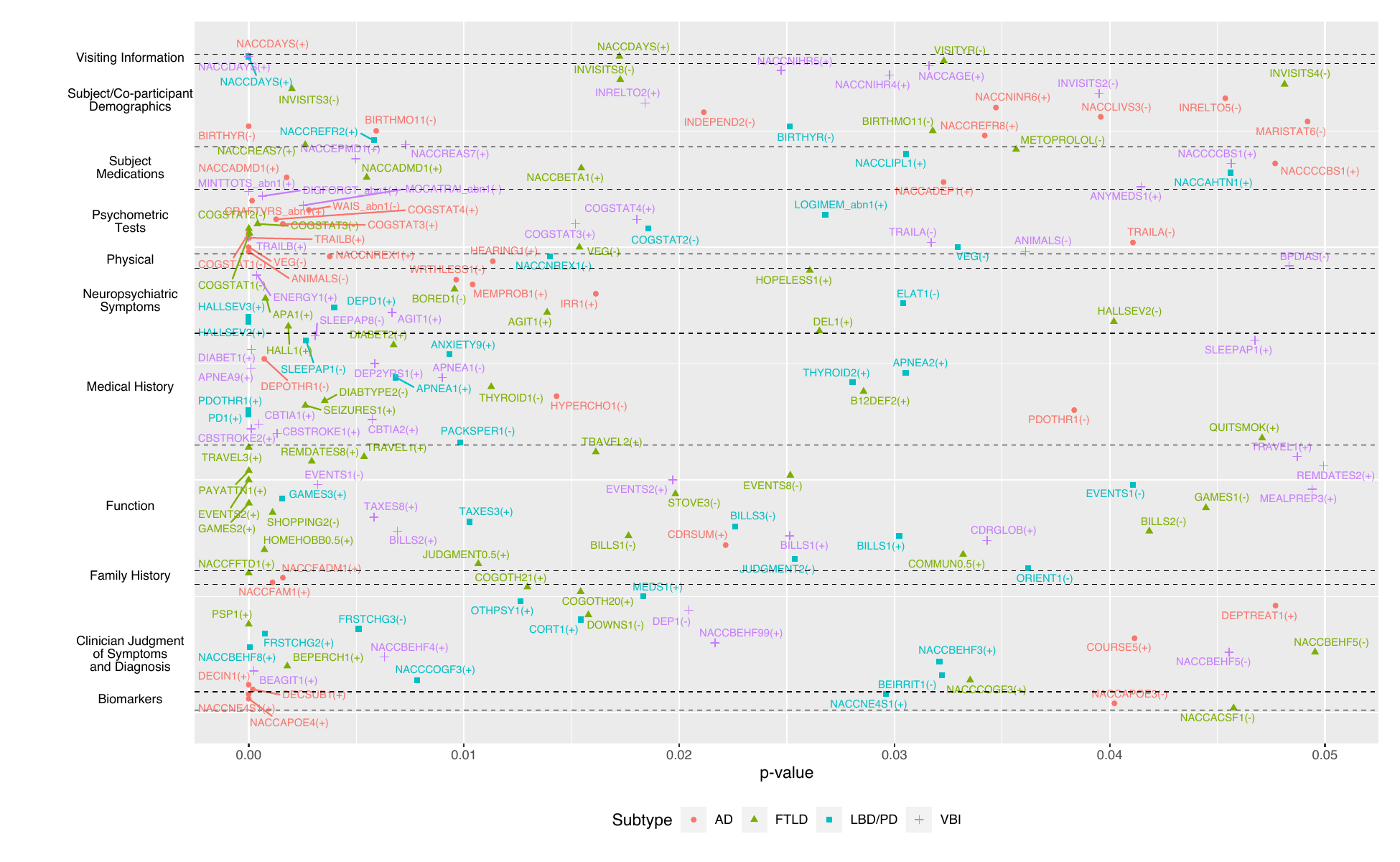}
	\caption{$p$-values of significant predictors (under 5\% significance level) associated with the contrast between 4 dementia subtypes and CN, grouped into 11 categories}
	\label{fig: p-value}
\end{sidewaysfigure}

\begin{figure}[!ht]
	\includegraphics[width=1\textwidth]{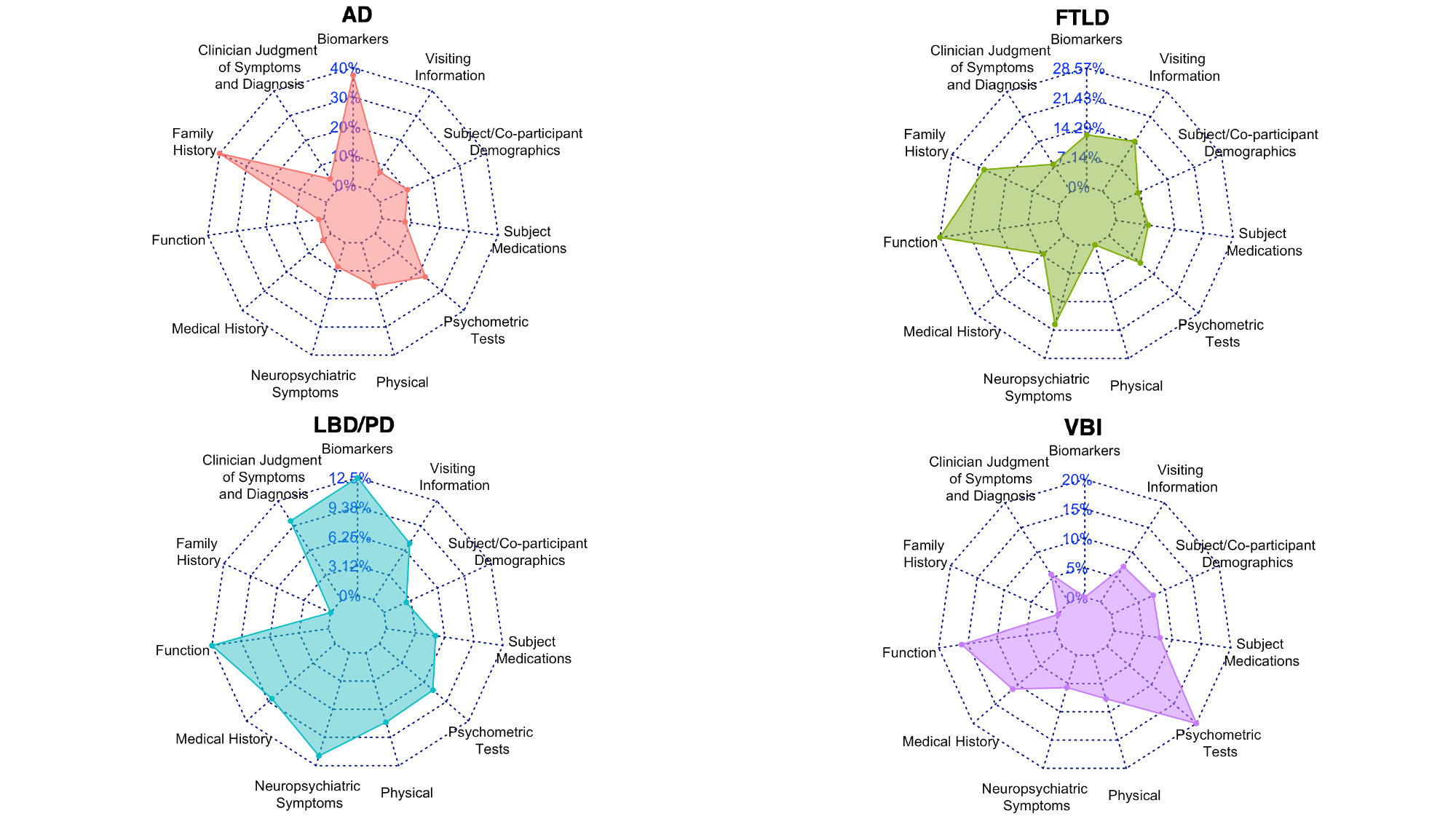}
	\caption{Radar charts of significant predictors (under 5\% significance level) in 11 categories for 4 dementia subtypes. Each percentage represents the number of significant predictors in a category relative to the total number of predictors in that category, reflecting the significance of predictors in each category for each subtype.}
	\label{fig: radar_chart}
\end{figure}

Finally, similar to \cite{tian2022risk}, we group the predictors with individual $p$-values less than $5\%$ into 10 categories, including demographics, subject medications, psychometric tests, etc., and plot them with their $p$-values in Figure \ref{fig: p-value}. We also present radar charts of significant predictors (under $5\%$ level) in different categories between each dementia subtype and CN in Figure \ref{fig: radar_chart}, where each percentage represents the number of significant predictors in a category relative to the total number of predictors in that category. This reflects the significance of predictors in each category for each subtype. Another series of bar charts illustrates significant predictors (under the $5\%$ significance level) across 11 categories for four dementia subtypes, where each percentage represents the number of significant predictors in a category relative to the total number of significant predictors for the subtype, reflecting the composition of significant predictors for each subtype.

\begin{figure}[!ht]
	\includegraphics[width=1\textwidth]{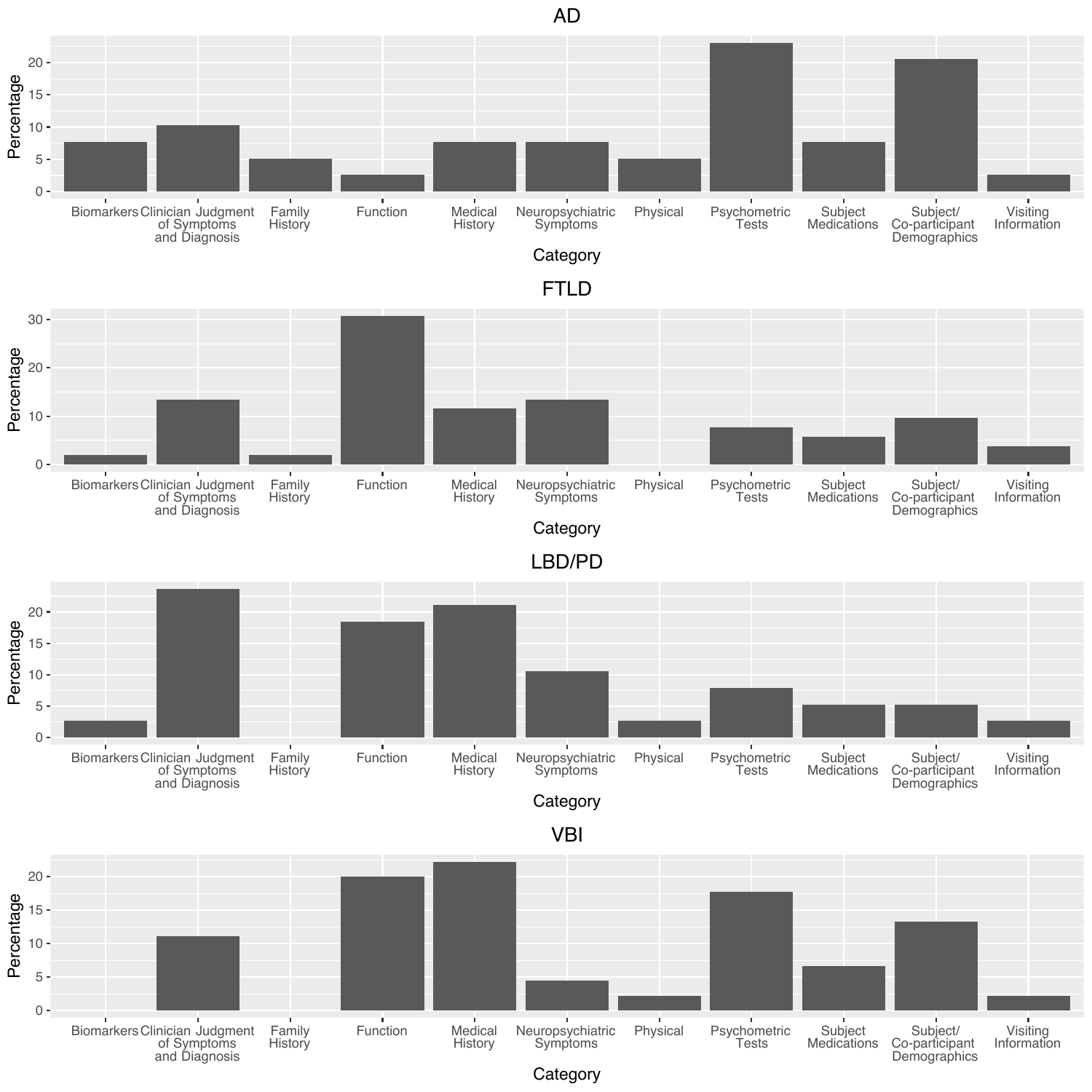}
	\caption{Bar charts of significant predictors (under 5\% significance level) in 11 categories for 4 dementia subtypes. Each percentage represents the number of significant predictors in a category relative to the total number of significant predictors for the subtype, reflecting the composition of significant predictors for each subtype.}
	\label{fig: bar_chart}
\end{figure}

Combining these three figures, we can obtain some interesting findings. From Figure \ref{fig: radar_chart}, we can see that over 20\% of family history variables are important for AD and FTLD, while no family history variables are significant for LBD/PD and VBI. On the other hand, over 10\% of function predictors are significant for FTLD, LBD/PD, and VBI, while none are important for AD. From Figure \ref{fig: bar_chart}, we can conclude that psychometric tests and demographics contain the most useful information to identify AD, as over 40\% significant predictors belong to these two categories. Similarly, function, neuropsychiatric symptoms, clinician diagnosis, and medical history are most useful for identifying FTLD; Clinician diagnosis, medical history, and function predictors are most important for identifying LBD/PD; and Medical history, function, and psychometric tests are most useful for identifying VBI.

\section{Discussion}\label{sec: discussion}
This paper analyzes the estimation and prediction error of the $\ell_1$-penalized multinomial regression model. We also extend the debiased Lasso to the multinomial case to provide valid statistical inference, including confidence intervals and $p$-values for hypothesis testing. Additionally, we address cases of model misspecification and non-identically distributed data. The debiased Lasso is applied to a real dementia dataset, where we identify important variables associated with the progression into different subtypes of dementia, yielding highly interpretable results. The estimation, inference, and prediction methods for the contrast-based $\ell_1$-penalized multinomial regression model have been implemented in a new R package \texttt{pemultinom}, which is available on CRAN. We believe this package will be beneficial to practitioners working with multi-label data.

There are several interesting future research avenues. First, extending the debiasing method and theory to a more general high-dimensional multi-response setting would be valuable. Second, in this work, we do not consider the sparsity structure between the contrasts associated with different classes. As mentioned in Section \ref{subsec: intro hd}, \cite{abramovich2021multiclass} discussed a row-wise sparsity structure, and we believe it is possible to extend the current analysis to that case. Finally, \cite{dezeure2017high} suggested combining the debiasing method with a residual bootstrap procedure, which theoretically requires weaker conditions and empirically performs better than the debiasing method, particularly in simultaneous hypothesis testing problems. Exploring the application of this idea to the multinomial regression version of the debiased Lasso would be an intriguing direction for future research.

\section*{Acknowledgments}
All the computations were conducted through the High-Performance Computing (HPC) service provided by Columbia University and New York University. This research was partially supported by NIH Grant P30AG066512, NIH Grant 1R21AG074205-01, NSF Grant DMS-2324489, a grant from the New York University School of Global Public Health, and the NYU University Research Challenge Fund. The authors would like to thank the anonymous referees, the Associate Editor, and the Editor for their constructive comments that improved the quality of this paper.

\clearpage
\bibliography{wileyNJD-AMA}

\appendix
We stack the coefficients $\{\bbetaks{k}\}_{k=1}^{K-1}$ into a single long vector $\bbeta^*$ of dimension $(K-1)p$. Similarly, we stack the estimates $\{\hbbetak{k}\}_{k=1}^{K-1}$ into a $(K-1)p$-dimensional vector $\hbbeta$. Given any estimate $\bbeta = \{\bbetak{k}\}_{k=1}^{K-1}$, we define the conditional probability $\tp(Y=k|X=\bx) = \frac{\exp(\bx^\top \bbetak{k})}{1+\sum_{k=1}^{K-1}\exp(\bx^\top \bbetak{k})}$ as $p_k(\bx;\bbeta)$, for $ k = 1:(K-1)$. Similarly, we define $\hDeltak{k} = \hbbetak{k}-\bbetaks{k}$ and $\hDelta = \{\hDeltak{k}\}_{k=1}^{K-1}$.

\section{Another Debiased Procedure}\label{sec: relax sparsity appendix}
In Remark \ref{rmk: asmp 3}, we mentioned that the $\ell_0$-sparsity condition of $\bTheta_j$ in Assumption \ref{asmp: sparsity gamma}.(\rom{1}) can be relaxed. In this section, we develop a variant of the inference procedure introduced in the main text and demonstrate its validity without the $\ell_0$-sparsity requirement of $\bTheta_j$.

We begin by splitting the data into two equal halves, $(\bX^{[1]}, \bY^{[1]}) \cup (\bX^{[2]}, \bY^{[2]})$. For simplicity, we assume $n$ is even in this section.

We obtain two point estimators of $\bbetaks{t}$ on two folds of data through \eqref{eq: lasso estimation}, respectively, and denote them as $\hbbeta^{[1]}$ and $\hbbeta^{[2]}$. For $\bSigma$, consider the empirical estimator $\hSigma^{[r]} \coloneqq \frac{1}{n/2}\sum_{i=1}^{n/2} \bm{B}(\bx_i^{[r]};\hbbeta^{[r]})$, with $r = 1, 2$.

Consider another debiased Lasso estimator given by
\begin{equation}
	\widehat{\bm{b}} \coloneqq \hbbeta^{[1]} + \frac{1}{n}\bm{M}\begin{pmatrix}
		(\bX^{[1]})^\top(\bY^{[1](1)}-\bm{p}_1(\bX^{[1]};\hbbeta^{[1]})) \\
		\vdots \\
		(\bX^{[1]})^\top(\bY^{[1](K-1)}-\bm{p}_{K-1}(\bX^{[1]};\hbbeta^{[1]}))
	\end{pmatrix},
\end{equation}
where $\bm{M} \in \mathbb{R}^{[(K-1)p] \times [(K-1)p]}$, and the $j$-th row of $\bm{M}$ is the solution to the following linear program:
\begin{align}
    &\min_{\bm{m} \in \mathbb{R}^{(K-1)p}} \onenorm{\bm{m}} \\
    &\textup{s.t. } \infnorm{\hSigma^{[2]}\bm{m} - \bm{e}_j} \leq \eta. \label{eq: LP}
\end{align}
The choice of $\eta$ will be specified later. The form of \eqref{eq: LP} is similar to the Dantzig selector \citep{candes2007dantzig}, although no $\ell_0$-sparsity of the estimond corresponding to $\bm{m}$ is required.

We impose the following assumption to replace Assumption \ref{asmp: sparsity gamma} in the main text. Note that the new assumption is weaker because the $\ell_0$-sparsity of $\bgamma_j^* = (\gamma^*_{j, 1}, \ldots, \gamma^*_{j, (K-1)p-1})^\top = (\bSigma_{-j,-j})^{-1}\bSigma_{-j,j}$ is not necessary, and the requirement on $\max_{j=1:[(K-1)p]}\onenorm{\bTheta_{j}}$ is less stringent.

\begin{assumption}\label{asmp: Theta supp}
	Denote $\bSigma = \te [\nabla^2 \mL_n(\bbeta^*)]$ and $\bTheta = \bSigma^{-1}$. Then $\max_{j=1:[(K-1)p]}\twonorm{\bTheta_j} \leq C < \infty$.
\end{assumption}

We are now ready to present the following theorem.

\begin{theorem}\label{thm: clt second}
	Let $\lambda = C\sqrt{\frac{\log p}{n}}$ in \eqref{eq: lasso estimation} and $\eta = CK^{5/2}\sqrt{\frac{s\log p}{n}}$ in \eqref{eq: gamma estimation} with a large constant $C>0$. Under Assumptions \ref{asmp: x}, \ref{asmp: sparsity beta}, and \ref{asmp: Theta supp}, when $n \gg \onenorm{\bTheta^*}^2K^{10}s^2(\log p)^2 +  K^{11}s^3(\log p)^2$,
	\begin{equation}
		\frac{\sqrt{n/2}(\widehat{b}_{j}-\beta^*_{j})}{\sqrt{\bm{M}_j^\top\hSigma^{[1]}\bm{M}_j}} \overset{\textup{d}}{\to} N(0, 1),
	\end{equation}
	for $j = 1, \ldots, (K-1)p$.
\end{theorem}

\section{Beyond the Almost Surely Boundedness Assumption}\label{sec: relax a.s. boundedness appendix}
In Remark \ref{rmk: asmp 3}, we mentioned that Assumption \ref{asmp: sparsity gamma}.(\rom{2}) can be removed with a more stringent sample size requirement. The following theorem justifies this claim.

\begin{theorem}\label{thm: clt beyond boundedness}
	Let $\lambda = C\sqrt{\frac{\log (Kp)}{n}}$ in \eqref{eq: lasso estimation} and $\lambda_j = CK^{5/2}\sqrt{\frac{s\log (Kp)}{n}}$ in \eqref{eq: gamma estimation} with a large constant $C>0$. Under Assumptions \ref{asmp: x}-\ref{asmp: sparsity gamma} but without Assumption \ref{asmp: sparsity gamma}.(\rom{3}), when $n \gg K^{20}(ss_0)^2(\log (Kp))^2 + K^{11}ss_0^2\log (Kp)$,
	\begin{equation}
		\frac{\sqrt{n}(\widehat{b}_{j}-\beta^*_{j})}{\sqrt{\hTheta_{j}^\top\hSigma\hTheta_{j}}} \overset{\textup{d}}{\to} N(0, 1),
	\end{equation}
	for $j = 1, \ldots, (K-1)p$.
\end{theorem}

\section{Additional Numerical Studies}\label{sec: additioal numerical appendix}

\subsection{Additional Simulations}\label{subsec: additioal simulation appendix}

\subsubsection{Model 5}\label{subsubsec: numerical model 5}
In this subsection, we explore a simulation example that is identical to Model 1, except that $\bbetaks{1} = (2, 2, 2, \bm{0}_{p-3})$ and $\bbetaks{2} = (0, 0, 0, 2, 2, 2, \bm{0}_{p-6})$. 

Table \ref{table: model 5 sse} summarizes the sum of squared $\ell_2$-estimation error of the contrast-based penalization and the over-parameterization method, where we can see that the contrast-based penalization leads to a smaller estimation error rate. 

\begin{table}[!h]
\caption{The sum of squared $\ell_2$-estimation error of $\bbetaks{1}$, $\bbetaks{2}$, and $\bbetaks{3}$, in Model $4$, for the contrast-based penalization and the over-parameterization method.}
\label{table: model 5 sse}
\centering
\begin{tabular}{c|ccc}
\hline
Method                        & $n=100$          & $n=200$         & $n=400$         \\ \hline
\multicolumn{1}{c|}{Contrast} &9.009 (2.465) &5.682 (1.505) &3.384 (0.98) \\
Over-parameterization         &10.859 (2.144) &7.212 (1.471) &4.469 (0.955) \\\hline
\end{tabular}
\end{table}

The average misclassification error rates for different methods are summarized in Table \ref{table: model 5 error}, where $\ell_1$-penalized multinomial regression performs the best among all approaches. 

\begin{table}[!h]
\caption{The average misclassification error rates on test data in Model 5, for $\ell_1$-penalized multinomial regression, SVMs, LDA, Random Forests, and $k$NN.}
\label{table: model 5 error}
\centering
\begin{tabular}{cccc}
\toprule
Method/$n$                       & $n=100$                 & $n=200$                 & $n=400$                 \\ 
\midrule
Penalized multinomial regression & 20.64\% (2.54\%)        & 18.01\% (1.67\%)        & 16.44\% (1.36\%)        \\
SVMs                              & 47.88\% (3.96\%)        & 38.99\% (2.59\%)        & 31.59\% (2.09\%)        \\
LDA                              & 43.04\% (2.71\%)        & 61.45\% (2.68\%)        & 38.88\% (1.96\%)        \\
Random forests                   & 29.13\% (3.51\%)        & 24.66\% (2.17\%)        & 21.84\% (1.65\%)        \\
$k$-nearest neighbors            & 53.94\% (2.30\%)        & 50.77\% (2.22\%)        & 47.02\% (2.08\%)        \\
\bottomrule
\end{tabular}
\end{table}

Table \ref{table: model 5 coverage prob} demonstrates that the debiased Lasso achieves a higher average coverage probability on $S$ than the bootstrap method when $n = 100$ and $200$, but has a lower average coverage probability on $S$ when $n = 400$.

\begin{table}[!h]
\caption{The average coverage probability of $95\%$ CIs on the coefficients corresponding to the signal set $S = \{\{1,2,3\}, \{4,5,6\}\}$ and noise set $S^c = \{\{4,5,6,\ldots, p\}, \{1,2,3, 7, 8,\ldots, p\}\}$ in Model $5$, for the debiased Lasso and the vector bootstrap method.}
\label{table: model 5 coverage prob}
\centering
\begin{tabular}{cc|ccc}
\hline
Method   & Measure     & $n=100$         & $n=200$         & $n=400$         \\ \hline
\multicolumn{1}{c}{\multirow{4}{*}{Debiased-Lasso}} & Average coverage probability on $S$ &0.917 (0.13) &0.884 (0.132) &0.897 (0.131) \\
\multicolumn{1}{c}{}                   & Average coverage probability on $S^c$ &0.999 (0.002) &0.997 (0.003) &0.992 (0.005) \\
\multicolumn{1}{c}{}                 & Average length of CI ($S$)            &5.491 (3.001) &2.118 (0.586) &1.53 (0.159) \\
\multicolumn{1}{c}{}                   & Average length of CI ($S^c$)          &4.841 (2.861) &2.355 (0.555) &1.645 (0.133) \\\hline
\multirow{4}{*}{Bootstrap}                & Average coverage probability on $S$   &0.776 (0.166) &0.875 (0.141) &0.942 (0.107) \\
& Average coverage probability on $S^c$ &1 (0) &1 (0.001) &1 (0.001) \\
& Average length of CI ($S$)            &2.495 (0.294) &2.185 (0.149) &1.763 (0.111) \\
& Average length of CI ($S^c$)           &0.181 (0.012) &0.217 (0.011) &0.26 (0.013) \\\hline
\end{tabular}
\end{table}

Table \ref{table: model 5 individual testing} shows that the debiased Lasso and the bootstrap method achieve comparable power while successfully controlling the Type-I error under $5\%$, with the debiased Lasso slightly outperforming the bootstrap method when $n = 100$. Compared to the results in Table \ref{table: model 1 individual testing} for Model 1, the power for Model 5 is significantly higher due to the increased signal strength.

The results of multiple testing are summarized in Table \ref{table: model 5 multiple testing}.

\begin{table}[!h]
\caption{The average Type-I error rate and power of individual testing $H_0: \beta^{(k)*}_j = 0$ v.s. $H_1: \beta^{(k)*}_j \neq 0$ at level $5\%$ in Model $5$, for the debiased Lasso and the vector bootstrap method.}
\label{table: model 5 individual testing}
\centering
\begin{tabular}{cc|ccc}
\hline
Method                             & Measure & $n=100$         & $n=200$         & $n=400$         \\ \hline
\multicolumn{1}{c}{\multirow{2}{*}{Debiased-Lasso}} & Type-I error &0.001 (0.002) &0.002 (0.003) &0.009 (0.005) \\
\multicolumn{1}{c}{}                   & Power &0.5 (0.201) &0.892 (0.131) &0.997 (0.029) \\\hline
\multirow{2}{*}{Bootstrap}                & Type-I error &0 (0) &0 (0.001) &0 (0.001) \\
& Power &0.441 (0.149) &0.862 (0.122) &1 (0) \\\hline
\end{tabular}
\end{table}

\begin{table}[!h]
\caption{The average family-wise error rate (FWER) and power of multiple testing $H_0: \beta^{(k)*}_j = 0$ v.s. $H_1: \beta^{(k)*}_j \neq 0$ for $j=1:p$ and $k=1:(K-1)$ in Model $5$, for the debiased Lasso and the multiple splitting method.}
\label{table: model 5 multiple testing}
\centering
\begin{tabular}{cc|ccc}
\hline
Method                             & Measure & $n=100$         & $n=200$         & $n=400$         \\ \hline
\multicolumn{1}{c}{\multirow{2}{*}{Debiased-Lasso}} & FWER &0 (0) &0 (0) &0 (0) \\
\multicolumn{1}{c}{}                   & Power &0.038 (0.076) &0.298 (0.168) &0.778 (0.152) \\\hline
\multirow{2}{*}{Multiple-Spliting}                & FWER &0.045 (0.208) &0.015 (0.122) &0.01 (0.1) \\
& Power &0.057 (0.114) &0.2 (0.186) &0.86 (0.122) \\\hline
\end{tabular}
\end{table}

\subsubsection{Model 6}\label{subsubsec: numerical model 6}
In this subsection, we explore a simulation example that is identical to Model 1, except that $\bbetaks{1} = (\bm{1}_{10}, \bm{0}_{p-10})$ and $\bbetaks{2} = (\bm{0}_{10}, \bm{1}_{10}, \bm{0}_{p-20})$.  

The sum of squared $\ell_2$-estimation errors of the contrast-based penalization and the over-parameterization method are summarized in Table \ref{table: model 6 sse}, which demonstrates the superiority of the contrast-based penalization.

\begin{table}[!h]
\caption{The sum of squared $\ell_2$-estimation error of $\bbetaks{1}$, $\bbetaks{2}$, and $\bbetaks{3}$, in Model $6$, for the contrast-based penalization and the over-parameterization method.}
\label{table: model 6 sse}
\centering
\begin{tabular}{c|ccc}
\hline
Method                        & $n=100$          & $n=200$         & $n=400$         \\ \hline
\multicolumn{1}{c|}{Contrast} &12.041 (1.54) &8.84 (1.195) &6.01 (1.018) \\
Over-parameterization         &13.02 (1.523) &9.813 (1.435) &6.711 (0.87) \\\hline
\end{tabular}
\end{table}

Table \ref{table: model 6 error} reports the average misclassification error rates on test data, showing that $\ell_1$-penalized multinomial regression outperforms the other methods across all sample size settings.

\begin{table}[!h]
\caption{The average misclassification error rates on test data in Model 6, for $\ell_1$-penalized multinomial regression, SVMs, LDA, Random Forests, and $k$NN.}
\label{table: model 6 error}
\centering
\begin{tabular}{c|ccc}
\hline
Method                      & $n=100$                 & $n=200$                 & $n=400$                 \\ \hline
Penalized multinomial regression & 20.05\% (2.37\%)        & 16.03\% (1.60\%)        & 13.79\% (1.19\%)        \\
SVMs                              & 35.07\% (4.43\%)        & 25.94\% (2.29\%)        & 20.36\% (1.49\%)        \\
LDA                              & 37.96\% (2.74\%)        & 60.69\% (2.70\%)        & 36.39\% (1.89\%)        \\
Random forests                   & 23.34\% (3.24\%)        & 19.17\% (1.94\%)        & 16.86\% (1.35\%)        \\
$k$-nearest neighbors            & 45.14\% (2.83\%)        & 39.99\% (2.58\%)        & 34.74\% (2.27\%)        \\
\hline
\end{tabular}
\end{table}

Table \ref{table: model 6 coverage prob} presents the average coverage probability of the debiased Lasso and the bootstrap method. The results indicate that the debiased Lasso provides better coverage than the bootstrap method on $S$.

\begin{table}[!h]
\caption{The average coverage probability of $95\%$ CIs on the coefficients corresponding to the signal set $S = \{\{1, 2, \ldots, 10\}, \{11, 12, \ldots, 20\}\}$ and noise set $S^c = \{\{11, 12, \ldots, p\}, \{1,2, \ldots, 10, 21, 22,\ldots, p\}\}$ in Model $6$, for the debiased Lasso and the vector bootstrap method.}
\label{table: model 6 coverage prob}
\centering
\begin{tabular}{cc|ccc}
\hline
Method   & Measure     & $n=100$         & $n=200$         & $n=400$         \\ \hline
\multicolumn{1}{c}{\multirow{4}{*}{Debiased-Lasso}} & Average coverage probability on $S$ &0.993 (0.022) &0.964 (0.043) &0.923 (0.064) \\
\multicolumn{1}{c}{}                   & Average coverage probability on $S^c$ &1 (0.001) &0.999 (0.001) &0.998 (0.003) \\
\multicolumn{1}{c}{}                 & Average length of CI ($S$)            &4.542 (1.856) &2.801 (0.811) &1.748 (0.197) \\
\multicolumn{1}{c}{}                   & Average length of CI ($S^c$)          &4.083 (1.377) &2.997 (0.882) &1.859 (0.203) \\\hline
\multirow{4}{*}{Bootstrap}                & Average coverage probability on $S$   &0.74 (0.082) &0.849 (0.077) &0.892 (0.071) \\
& Average coverage probability on $S^c$ &1 (0) &1 (0) &1 (0) \\
& Average length of CI ($S$)            &1.512 (0.133) &1.568 (0.105) &1.477 (0.078) \\
& Average length of CI ($S^c$)           &0.162 (0.011) &0.196 (0.012) &0.238 (0.011) \\\hline
\end{tabular}
\end{table}

Tables \ref{table: model 6 individual testing} and \ref{table: model 6 multiple testing} summarize the results of individual testing and multiple testing, respectively.

\begin{table}[!h]
\caption{The average Type-I error rate and power of individual testing $H_0: \beta^{(k)*}_j = 0$ v.s. $H_1: \beta^{(k)*}_j \neq 0$ at level $5\%$ in Model $6$, for the debiased Lasso and the vector bootstrap method.}
\label{table: model 6 individual testing}
\centering
\begin{tabular}{cc|ccc}
\hline
Method                             & Measure & $n=100$         & $n=200$         & $n=400$         \\ \hline
\multicolumn{1}{c}{\multirow{2}{*}{Debiased-Lasso}} & Type-I error &0 (0.001) &0.001 (0.001) &0.003 (0.003) \\
\multicolumn{1}{c}{}                   & Power &0.041 (0.045) &0.17 (0.078) &0.446 (0.085) \\\hline
\multirow{2}{*}{Bootstrap}                & Type-I error &0 (0) &0 (0) &0 (0) \\
& Power &0.059 (0.046) &0.198 (0.073) &0.499 (0.087) \\\hline
\end{tabular}
\end{table}

\begin{table}[!h]
\caption{The average family-wise error rate (FWER) and power of multiple testing $H_0: \beta^{(k)*}_j = 0$ v.s. $H_1: \beta^{(k)*}_j \neq 0$ for $j=1:p$ and $k=1:(K-1)$ in Model $6$, for the debiased Lasso and the multiple splitting method.}
\label{table: model 6 multiple testing}
\centering
\begin{tabular}{cc|ccc}
\hline
Method                             & Measure & $n=100$         & $n=200$         & $n=400$         \\ \hline
\multicolumn{1}{c}{\multirow{2}{*}{Debiased-Lasso}} & FWER &0 (0) &0 (0) &0 (0) \\
\multicolumn{1}{c}{}                   & Power &0 (0) &0.001 (0.009) &0.017 (0.03) \\\hline
\multirow{2}{*}{Multiple-Spliting}                & FWER &0 (0) &0.015 (0.122) &0 (0) \\
& Power &0 (0.005) &0.003 (0.025) &0.053 (0.07) \\\hline
\end{tabular}
\end{table}

\subsubsection{Model 7}\label{subsubsec: numerical model 7}
In this subsection, we explore a setting that is the same as Model 2, except that $\bbetaks{1} = (2, 2, 2, \bm{0}_{p-3})$, $\bbetaks{2} = (\bm{0}_{3}, 2, 2, 2, \bm{0}_{p-6})$, and $\bbetaks{3} = (\bm{0}_{6}, 2, 2, 2, \bm{0}_{p-9})$.

Table \ref{table: model 7 sse} summarizes the sum of squared $\ell_2$-estimation errors of the contrast-based penalization and the over-parameterization method, showing that the contrast-based penalization performs better.

\begin{table}[!h]
\caption{The sum of squared $\ell_2$-estimation error of $\bbetaks{1}$, $\bbetaks{2}$, and $\bbetaks{3}$, in Model $7$, for the contrast-based penalization and the over-parameterization method.}
\label{table: model 7 sse}
\centering
\begin{tabular}{c|ccc}
\hline
Method                        & $n=100$          & $n=200$         & $n=400$         \\ \hline
\multicolumn{1}{c|}{Contrast} &14.669 (2.627) &8.93 (2.007) &5.239 (1.41) \\
Over-parameterization         &17.366 (2.426) &11.319 (1.916) &7.232 (1.328) \\\hline
\end{tabular}
\end{table}

Table \ref{table: model 7 error} reports the average misclassification error rates on test data, demonstrating that $\ell_1$-penalized multinomial regression dominates the other methods across all sample size settings.

\begin{table}[!h]
\caption{The average misclassification error rates on test data in Model 7, for $\ell_1$-penalized multinomial regression, SVMs, LDA, Random Forests, and $k$NN.}
\label{table: model 7 error}
\centering
\begin{tabular}{cccc}
\toprule
Method/$n$                       & $n=100$                 & $n=200$                 & $n=400$                 \\ 
\midrule
Penalized multinomial regression & 25.97\% (2.63\%)        & 22.06\% (1.68\%)        & 19.98\% (1.41\%)        \\
SVMs                              & 59.18\% (4.28\%)        & 48.07\% (2.64\%)        & 39.79\% (2.13\%)        \\
LDA                              & 52.67\% (2.77\%)        & 69.83\% (2.65\%)        & 47.78\% (2.05\%)        \\
Random forests                   & 38.52\% (3.64\%)        & 32.22\% (2.59\%)        & 28.17\% (2.07\%)        \\
$k$-nearest neighbors            & 62.74\% (2.20\%)        & 59.46\% (2.16\%)        & 55.56\% (2.15\%)        \\
\bottomrule
\end{tabular}
\end{table}

The average coverage probability of $95\%$ CIs is summarized in Table \ref{table: model 7 coverage prob}, where the debiased Lasso shows a larger coverage probability than the bootstrap method.

\begin{table}[!h]
\caption{The average coverage probability of $95\%$ CIs on the coefficients corresponding to the signal set $S = \{\{1,2,3\}, \{4,5,6\}, \{7,8,9\}\}$ and noise set $S^c = \{\{4,5,6,\ldots, p\}, \{1,2,3, 7, 8,\ldots, p\}, \{1,2,3,4,5,6, 10, 11,\ldots, p\}\}$ in Model $7$, for the debiased Lasso and the vector bootstrap method.}
\label{table: model 7 coverage prob}
\centering
\begin{tabular}{cc|ccc}
\hline
Method   & Measure     & $n=100$         & $n=200$         & $n=400$         \\ \hline
\multicolumn{1}{c}{\multirow{4}{*}{Debiased-Lasso}} & Average coverage probability on $S$ &0.922 (0.093) &0.901 (0.112) &0.915 (0.1) \\
\multicolumn{1}{c}{}                   & Average coverage probability on $S^c$ &1 (0.001) &0.998 (0.002) &0.994 (0.004) \\
\multicolumn{1}{c}{}                 & Average length of CI ($S$)            &6.145 (4.395) &2.521 (0.991) &1.758 (0.158) \\
\multicolumn{1}{c}{}                   & Average length of CI ($S^c$)          &5.634 (3.871) &2.65 (0.72) &1.889 (0.132) \\\hline
\multirow{4}{*}{Bootstrap}                & Average coverage probability on $S$   &0.684 (0.141) &0.824 (0.128) &0.906 (0.113) \\
& Average coverage probability on $S^c$ &1 (0) &1 (0) &1 (0.001) \\
& Average length of CI ($S$)            &2.298 (0.255) &2.107 (0.149) &1.696 (0.092) \\
& Average length of CI ($S^c$)           &0.163 (0.009) &0.197 (0.01) &0.237 (0.011) \\\hline
\end{tabular}
\end{table}

Tables \ref{table: model 7 individual testing} and \ref{table: model 7 multiple testing} report the results for individual testing and multiple testing, respectively.  The debiased Lasso outperforms the bootstrap and multiple-splitting methods by successfully controlling the Type I error and FWER while achieving higher power.

\begin{table}[!h]
\caption{The average Type-I error rate and power of individual testing $H_0: \beta^{(k)*}_j = 0$ v.s. $H_1: \beta^{(k)*}_j \neq 0$ at level $5\%$ in Model $7$, for the debiased Lasso and the vector bootstrap method.}
\label{table: model 7 individual testing}
\centering
\begin{tabular}{cc|ccc}
\hline
Method                             & Measure & $n=100$         & $n=200$         & $n=400$         \\ \hline
\multicolumn{1}{c}{\multirow{2}{*}{Debiased-Lasso}} & Type-1 error &0 (0.001) &0.002 (0.002) &0.006 (0.004) \\
\multicolumn{1}{c}{}                   & Power &0.303 (0.149) &0.77 (0.136) &0.973 (0.064) \\\hline
\multirow{2}{*}{Bootstrap}                & Type-1 error &0 (0) &0 (0) &0 (0.001) \\
& Power &0.371 (0.118) &0.795 (0.12) &0.994 (0.025) \\\hline
\end{tabular}
\end{table}

\begin{table}[!h]
\caption{The average family-wise error rate (FWER) and power of multiple testing $H_0: \beta^{(k)*}_j = 0$ v.s. $H_1: \beta^{(k)*}_j \neq 0$ for $j=1:p$ and $k=1:(K-1)$ in Model $7$, for the debiased Lasso and the multiple splitting method.}
\label{table: model 7 multiple testing}
\centering
\begin{tabular}{cc|ccc}
\hline
Method                             & Measure & $n=100$         & $n=200$         & $n=400$         \\ \hline
\multicolumn{1}{c}{\multirow{2}{*}{Debiased-Lasso}} & FWER &0 (0) &0 (0) &0 (0) \\
\multicolumn{1}{c}{}                   & Power &0.011 (0.036) &0.134 (0.111) &0.516 (0.155) \\\hline
\multirow{2}{*}{Multiple-Spliting}                & FWER &0.05 (0.218) &0.085 (0.28) &0.985 (0.122) \\
& Power &0.013 (0.057) &0.019 (0.051) &0.398 (0.114) \\\hline
\end{tabular}
\end{table}

\subsubsection{Model 8}\label{subsubsec: numerical model 8}
In this subsection, we explore a setting that is identical to Model 2, except that $\bbetaks{1} = (\bm{1}_{10}, \bm{0}_{p-10})$, $\bbetaks{2} = (\bm{0}_{10}, \bm{1}_{10}, \bm{0}_{p-6})$, and $\bbetaks{3} = (\bm{0}_{20}, \bm{1}_{10}, \bm{0}_{p-30})$.

Table \ref{table: model 8 sse} summarizes the sum of squared $\ell_2$-estimation errors of the contrast-based penalization and the over-parameterization method, showing that contrast-based penalization performs better.

\begin{table}[!h]
\caption{The sum of squared $\ell_2$-estimation error of $\bbetaks{1}$, $\bbetaks{2}$, and $\bbetaks{3}$, in Model $8$, for the contrast-based penalization and the over-parameterization method.}
\label{table: model 8 sse}
\centering
\begin{tabular}{c|ccc}
\hline
Method                        & $n=100$          & $n=200$         & $n=400$         \\ \hline
\multicolumn{1}{c|}{Contrast} &71.432 (1.891) &66.436 (1.547) &63.046 (1.132) \\
Over-parameterization         &57.311 (26.924) &39.248 (17.762) &28.162 (13.787) \\\hline
\end{tabular}
\end{table}

Table \ref{table: model 8 error} reports the average misclassification error rates on test data, indicating that $\ell_1$-penalized multinomial regression dominates the other methods across all sample size settings.

\begin{table}[!h]
\caption{The average misclassification error rates on test data in Model 8, for $\ell_1$-penalized multinomial regression, SVMs, LDA, Random Forests, and $k$NN.}
\label{table: model 8 error}
\centering
\begin{tabular}{c|ccc}
\hline
Method                       & $n=100$                 & $n=200$                 & $n=400$                 \\ 
\hline
Penalized multinomial regression & 44.57\% (2.65\%)        & 38.44\% (2.30\%)        & 33.82\% (2.12\%)        \\
SVMs                              & 63.45\% (6.71\%)        & 56.71\% (5.56\%)        & 53.27\% (5.96\%)        \\
LDA                              & 64.78\% (5.25\%)        & 72.14\% (3.46\%)        & 62.49\% (6.07\%)        \\
Random forests                   & 58.48\% (4.83\%)        & 54.51\% (4.67\%)        & 51.74\% (4.11\%)        \\
$k$-nearest neighbors            & 61.63\% (3.64\%)        & 59.06\% (3.70\%)        & 55.60\% (3.69\%)        \\
\hline
\end{tabular}
\end{table}

Table \ref{table: model 8 coverage prob} summarizes the average coverage probability of $95\%$ CIs for the debiased Lasso and the vector bootstrap method, where the debiased Lasso provides better coverage than the bootstrap method. 

\begin{table}[!h]
\caption{The average coverage probability of $95\%$ CIs on the coefficients corresponding to the signal set $S = \{\{1,2, \ldots, 10\}, \{11, 12, \ldots, 20\}, \{21, 22, \ldots, 30\}\}$ and noise set $S^c = \{\{11, 12,\ldots, p\}, \{1, 2, \ldots, 10, 21, 22,\ldots, p\}, \{1, 2, \ldots, 20, 41, 42, \ldots, p\}\}$ in Model $8$, for the debiased Lasso and the vector bootstrap method.}
\label{table: model 8 coverage prob}
\centering
\begin{tabular}{cc|ccc}
\hline
Method   & Measure     & $n=100$         & $n=200$         & $n=400$         \\ \hline
\multicolumn{1}{c}{\multirow{4}{*}{Debiased-Lasso}} & Average coverage probability on $S$ &0.997 (0.01) &0.994 (0.018) &0.98 (0.03) \\
\multicolumn{1}{c}{}                   & Average coverage probability on $S^c$ &1 (0) &1 (0.001) &0.999 (0.001) \\
\multicolumn{1}{c}{}                 & Average length of CI ($S$)            &6.203 (3.579) &3.331 (1.596) &1.79 (0.331) \\
\multicolumn{1}{c}{}                   & Average length of CI ($S^c$)          &7.467 (9.405) &3.643 (1.853) &1.845 (0.342) \\\hline
\multirow{4}{*}{Bootstrap}                & Average coverage probability on $S$   &0.485 (0.078) &0.603 (0.08) &0.685 (0.072) \\
& Average coverage probability on $S^c$ &1 (0) &1 (0.001) &1 (0.001) \\
& Average length of CI ($S$)            &0.979 (0.067) &1 (0.042) &0.88 (0.025) \\
& Average length of CI ($S^c$)           &0.143 (0.008) &0.165 (0.01) &0.193 (0.012) \\\hline
\end{tabular}
\end{table}

Tables \ref{table: model 8 individual testing} and \ref{table: model 8 multiple testing} summarize the results of individual testing and multiple testing. For multiple testing, both debiased Lasso and multiple-splitting are overly conservative, successfully controlling the FWER but resulting in near-zero power.

\begin{table}[!h]
\caption{The average Type-I error rate and power of individual testing $H_0: \beta^{(k)*}_j = 0$ v.s. $H_1: \beta^{(k)*}_j \neq 0$ at level $5\%$ in Model $8$, for the debiased Lasso and the vector bootstrap method.}
\label{table: model 8 individual testing}
\centering
\begin{tabular}{cc|ccc}
\hline
Method                             & Measure & $n=100$         & $n=200$         & $n=400$         \\ \hline
\multicolumn{1}{c}{\multirow{2}{*}{Debiased-Lasso}} & Type-I error &0 (0) &0 (0.001) &0.001 (0.001) \\
\multicolumn{1}{c}{}                   & Power &0.019 (0.033) &0.124 (0.08) &0.506 (0.099) \\\hline
\multirow{2}{*}{Bootstrap}                & Type-I error &0 (0) &0 (0.001) &0 (0.001) \\
& Power &0.101 (0.052) &0.408 (0.078) &0.793 (0.069) \\\hline
\end{tabular}
\end{table}

\begin{table}[!h]
\caption{The average family-wise error rate (FWER) and power of multiple testing $H_0: \beta^{(k)*}_j = 0$ v.s. $H_1: \beta^{(k)*}_j \neq 0$ for $j=1:p$ and $k=1:(K-1)$ in Model $8$, for the debiased Lasso and the multiple splitting method.}
\label{table: model 8 multiple testing}
\centering
\begin{tabular}{cc|ccc}
\hline
Method                             & Measure & $n=100$         & $n=200$         & $n=400$         \\ \hline
\multicolumn{1}{c}{\multirow{2}{*}{Debiased-Lasso}} & FWER &0 (0) &0 (0) &0 (0) \\
\multicolumn{1}{c}{}                   & Power &0 (0) &0 (0) &0.012 (0.02) \\\hline
\multirow{2}{*}{Multiple-Spliting}                & FWER &0 (0) &0 (0) &0 (0) \\
& Power &0 (0.002) &0 (0) &0 (0) \\\hline
\end{tabular}
\end{table}

\newpage
\section{Proofs}

\subsection{Lemmas and their proofs}
\begin{lemma}\label{lem: rsc}
	Denote 
	\begin{equation}
		\mE_n(\bDelta) = \mL_n(\bbeta^* + \bDelta) - \mL_n(\bbeta^*) - \nabla \mL_n(\bbeta^*)^\top\bDelta.
	\end{equation}
	Under Assumptions \ref{asmp: x} and \ref{asmp: sparsity beta}, there exist constants $c_1$, $c_2 > 0$, such that
	\begin{equation}
		\mE_n(\bDelta) \geq \frac{1}{K^2}\left(c_1\twonorm{\bDelta}^2 - c_2\twonorm{\bDelta}\onenorm{\bDelta}\sqrt{\frac{\log p}{n}}\right), \quad \forall \twonorm{\bDelta} \leq 1,
	\end{equation}
	with probability at least $1-C(Kp)^{-1}$.
\end{lemma}

\begin{lemma}\label{lem: to prove rsc}
	Suppose $K \geq 2$. For any $\{p_k\}_{k=1}^{K}$ with all $p_k \geq 0$, $\sum_{k=1}^K p_k = 1$ and any $\{x_k\}_{k=1}^{K-1} \in \mathbb{R}^{K-1}$,
	\begin{equation}
		(x_1, \ldots, x_{K-1})^\top \begin{pmatrix}
			p_1(1-p_1) &-p_1p_2 &\ldots &-p_1p_{K-1} \\
			-p_1p_2 &p_2(1-p_2) &\ldots &-p_2p_{K-1} \\
			\vdots &\vdots  &\vdots &\vdots \\
			-p_1p_{K-1} &-p_2p_{K-1} &\ldots &p_{K-1}(1-p_{K-1})
		\end{pmatrix}
		\begin{pmatrix}
				x_1 \\
				\vdots \\
				x_{K-1}
		\end{pmatrix} \geq p_K\sum_{k=1}^{K-1}p_kx_k^2.
	\end{equation}
\end{lemma}

\begin{lemma}\label{lem: kl}
	Given $\bbeta = \big((\bbetak{1})^\top, \ldots, (\bbetak{K-1})^\top \big)^\top \in \mathbb{R}^{(K-1)p}$, denote the joint distribution of $(\bx, y)$ as $\tp_{\bx,y;\bbeta}$. Under Assumption \ref{asmp: x}, for any $\bbeta = \big((\bbetak{1})^\top, \ldots, (\bbetak{K-1})^\top \big)^\top \in \mathbb{R}^{(K-1)p}$ and $\widetilde{\bbeta} = \big((\widetilde{\bbeta}^{(1)})^\top, \ldots, (\widetilde{\bbeta}^{(K-1)})^\top \big)^\top \in \mathbb{R}^{(K-1)p}$, it holds that
	\begin{equation}
		\textup{KL}(\tp_{\bx,y;\bbeta}\|\tp_{\bx,y;\widetilde{\bbeta}}) \leq C\sigma^2\twonorm{\bbeta - \widetilde{\bbeta}}^2\cdot \left(\frac{1}{K}e^{2T}+\exp\left\{-\frac{C'T^2}{\min_k \twonorm{\bbetak{k}}^2}\right\}\right),
	\end{equation}
	with any $T > 0$.
\end{lemma}

\begin{lemma}\label{lem: estimation error gamma}
    Under Assumptions \ref{asmp: x}-\ref{asmp: sparsity gamma}, when $n \gg K^{18}s^2(\log (Kp))^2(\log (np))^2 + K^{10}s_0^2(\log (Kp))^2$, for any $j = 1, \ldots, (K-1)p$,
    \begin{align}
        \twonorm{\hgamma_j - \bgamma^*_j} \leq CK^{9/2}\sqrt{\frac{s(\log (Kp))(\log(np))}{n}} &+ CK^{5/2}\sqrt{\frac{s_0\log (Kp)}{n}},\quad  \onenorm{\hgamma_j - \bgamma^*_j} \leq CK^7s\sqrt{\frac{\log (Kp)}{n}}\log(np) + CK^3s_0\sqrt{\frac{\log (Kp)}{n}}, \\
        \norm{\htau^2_j - (\tau^*_j)^2}, \norm{\htau^{-2}_j - (\tau^*_j)^{-2}} &\leq CK^{9/2}\sqrt{\frac{s(\log (Kp))(\log(np))}{n}} + CK^{5/2}\sqrt{\frac{s_0\log (Kp)}{n}},
    \end{align}
    with probability $1-C'\exp\{-C''n\}-C'(np)^{-1}-C'(Kp)^{-1}$.
\end{lemma}
\begin{proof}[Proof of Lemma \ref{lem: rsc}]
By Taylor expansion, we know that
\begin{equation}
	\mE_n(\bDelta) = \bDelta^\top\left(\frac{1}{n}\sum_{i=1}^n \int_0^1 B(\bx_i; \bbeta^* + t\bDelta) dt \right)\bDelta,
\end{equation}
where $$B(\bx; \bbeta) \coloneqq \begin{pmatrix}
	p_1(\bx;\bbeta)[1-p_1(\bx;\bbeta)]\bx\bx^\top & -p_1(\bx;\bbeta)p_2(\bx;\bbeta)\bx\bx^\top &\ldots &-p_1(\bx;\bbeta)p_{K-1}(\bx;\bbeta)\bx\bx^\top \\
	-p_1(\bx;\bbeta)p_2(\bx;\bbeta)\bx\bx^\top &p_2(\bx;\bbeta)[1-p_2(\bx;\bbeta)]\bx\bx^\top &\ldots &-p_2(\bx;\bbeta)p_{K-1}(\bx;\bbeta)\bx\bx^\top \\
	\vdots &\vdots  &\vdots &\vdots \\
	  -p_1(\bx;\bbeta)p_{K-1}(\bx;\bbeta)\bx\bx^\top &-p_2(\bx;\bbeta)p_{K-1}(\bx;\bbeta)\bx\bx^\top &\ldots &p_{K-1}(\bx;\bbeta)[1-p_{K-1}(\bx;\bbeta)]\bx\bx^\top
\end{pmatrix}.$$
By Lemma \ref{lem: to prove rsc},
\begin{equation}
	\mE_n(\bDelta) \geq \frac{1}{n}\sum_{i=1}^n \int_0^1 p_K(\bx_i; \bbeta^*+t\bDelta)\sum_{k=1}^{K-1}p_{k}(\bx_i; \bbeta^*+t\bDelta)((\bDeltak{k})^\top \bx_i)^2 dt.
\end{equation}
Fix $\bDelta$ with $\twonorm{\bDelta} = \delta \in (0, 1]$, and set $\tau = \tilde{k}\delta$, $\phi_{\tau}(u) = u^2\mathds{1}(|u| \leq 2\tau)$, which leads to
\begin{align}
	\mE_n(\bDelta) \geq \frac{1}{n}\sum_{i=1}^n \int_0^1 p_K(\bx_i; \bbeta^*+t\bDelta)\sum_{k=1}^{K-1}p_{k}(\bx_i; \bbeta^*+t\bDelta)\varphi_{\tau}((\bDeltak{k})^\top \bx_i)\mathds{1}\Bigg(\sum_{k=1}^{K-1}e^{(\bbetaks{k})^\top\bx_i} \leq T_1\Bigg)\\
	\cdot \mathds{1}(|(\bbetaks{k})^\top\bx_i| \leq T_1) dt \\
	\geq \gamma \frac{1}{n}\sum_{i=1}^n\sum_{k=1}^{K-1}\varphi_{\tau}((\bDeltak{k})^\top \bx_i)\mathds{1}\Bigg(\sum_{k=1}^{K-1}e^{(\bbetaks{k})^\top\bx_i} \leq T_1\Bigg)\mathds{1}(|(\bbetaks{k})^\top\bx_i| \leq T_2),
\end{align}
where 
\begin{align}
	\gamma &\coloneqq  \min_{\sum_{k=1}^{K-1}e^{(\bbetaks{k})^\top\bx_i} \leq T_1, |(\bbetaks{k})^\top\bx_i| \leq T_2, |(\bDeltak{k})^\top \bx_i| \leq \tilde{k}}\min_k p_K(\bx_i; \bbeta^*+t\bDelta)p_{k}(\bx_i; \bbeta^*+t\bDelta) \\
	&\geq \frac{e^{-(T_2+\tilde{k})}}{(T_1e^{\tilde{k}})^2}.
\end{align}
Define
\begin{align}
	Z_n(r,\delta) &= \sup_{\substack{\twonorm{\bDelta}\leq\delta \\ \onenorm{\bDelta} \leq r}}\Bigg|\frac{1}{n}\sum_{i=1}^n\sum_{k=1}^{K-1}\varphi_{\tau}((\bDeltak{k})^\top \bx_i)\mathds{1}\Bigg(\sum_{k=1}^{K-1}e^{(\bbetaks{k})^\top\bx_i} \leq T_1\Bigg)\mathds{1}(|(\bbetaks{k})^\top\bx_i| \leq T_2) \\
	&- \te \Bigg[\sum_{k=1}^{K-1}\varphi_{\tau}((\bDeltak{k})^\top \bx)\mathds{1}\Bigg(\sum_{k=1}^{K-1}e^{(\bbetaks{k})^\top\bx} \leq T_1\Bigg)\mathds{1}(|(\bbetaks{k})^\top\bx| \leq T_2)\Bigg]\Bigg|.
\end{align}
Note that 
\begin{align}
	&\te \Bigg[\sum_{k=1}^{K-1}\varphi_{\tau}((\bDeltak{k})^\top \bx)\mathds{1}\Bigg(\sum_{k=1}^{K-1}e^{(\bbetaks{k})^\top\bx} \leq T_1\Bigg)\mathds{1}(|(\bbetaks{k})^\top\bx| \leq T_2)\Bigg] \\
	&= \underbrace{\te \Bigg[\sum_{k=1}^{K-1}\varphi_{\tau}((\bDeltak{k})^\top \bx)\Bigg]}_{[1]}- \underbrace{\te\Bigg[\sum_{k=1}^{K-1}\varphi_{\tau}((\bDeltak{k})^\top \bx)\mathds{1}\Bigg(\sum_{k=1}^{K-1}e^{(\bbetaks{k})^\top\bx} > T_1\Bigg)\Bigg]}_{[2]} \\
	&\quad - \underbrace{\te\Bigg[\sum_{k=1}^{K-1}\varphi_{\tau}((\bDeltak{k})^\top \bx)\mathds{1}(|(\bbetaks{k})^\top\bx| > T_2)\Bigg]}_{[3]}. \label{eq: zn 0}
\end{align}
And
\begin{align}
	[1] &\geq \te\Bigg[\sum_{k=1}^{K-1}((\bDeltak{k})^\top\bx)^2\cdot \mathds{1}(|(\bDeltak{k})^\top\bx| \leq \tau)\Bigg] \\
	&\geq \te\Bigg[\sum_{k=1}^{K-1}((\bDeltak{k})^\top\bx)^2\Bigg]-\te\Bigg[\sum_{k=1}^{K-1}((\bDeltak{k})^\top\bx)^2\cdot \mathds{1}(|(\bDeltak{k})^\top\bx| > \tau)\Bigg] \\
	&\geq \twonorm{\bDelta}^2 \lambda_{\min}(\bSigma_{\bx}) - \sum_{k=1}^{K-1}\sqrt{\te ((\bDeltak{k})^\top\bx)^4}\sqrt{\tp(|(\bDeltak{k})^\top\bx| > \tau)} \\
	&\geq \twonorm{\bDelta}^2 \lambda_{\min}(\bSigma_{\bx}) - \sum_{k=1}^{K-1}C\sigma^2\twonorm{\bDeltak{k}}^2\cdot \exp\left\{-\frac{\tau^2}{2\twonorm{\bDeltak{k}}^2\sigma^2}\right\} \\
	&\geq \twonorm{\bDelta}^2\cdot \left[\lambda_{\min}(\bSigma_{\bx}) - C\sigma^2\exp\left\{-\frac{\tau^2}{2\max_k\twonorm{\bDeltak{k}}^2\sigma^2}\right\}\right] \\
	&\geq \twonorm{\bDelta}^2\cdot c\lambda_{\min}(\bSigma_{\bx}), \label{eq: zn 1}
\end{align}
if $\tau = \tilde{k}\delta \geq C'\delta\sigma^2[\log(\lambda_{\min}^{-1}(\bSigma_{\bx})) + \log \sigma^2]$ with a large constant $C' > 0$.
\begin{align}
	[2] &\leq \sum_{k=1}^{K-1}\sqrt{\te ((\bDeltak{k})^\top\bx)^4}\sqrt{\tp\left(\sum_{k=1}^{K-1}e^{(\bbetaks{k})^\top\bx} > T_1\right)}\\
	&\leq \sum_{k=1}^{K-1}C\sigma^2\twonorm{\bDeltak{k}}^2\cdot\sqrt{\frac{\sum_{k=1}^{K-1}\te e^{(\bbetaks{k})^\top\bx}}{T_1}} \\
	&\leq C\sigma^2\twonorm{\bDelta}^2 \cdot \sqrt{\frac{K}{T_1}}\cdot \exp\left\{C\max_k \twonorm{\bbetaks{k}}^2\right\} \\
	&\leq \twonorm{\bDelta}^2\cdot \frac{1}{4}c\lambda_{\min}(\bSigma_{\bx}), \label{eq: zn 2}
\end{align}
if $T_1 = C' K\sigma^4 \exp\left\{C'\max_k \twonorm{\bbetaks{k}}^2\right\}\lambda_{\min}^{-2}(\bSigma_{\bx})$ with a large constant $C' > 0$.
\begin{align}
	[3] &\leq \sum_{k=1}^{K-1}C\sigma^2\twonorm{\bDeltak{k}}^2\cdot\sqrt{\tp(|(\bbetaks{k})^\top\bx| > T_2)} \\
	&\leq C\twonorm{\bDelta}^2 \cdot \exp\left\{-\frac{T_2^2}{2\max_k\twonorm{\bbetaks{k}}^2\sigma^2}\right\} \\
	&\leq \twonorm{\bDelta}^2\cdot \frac{1}{4}c\lambda_{\min}(\bSigma_{\bx}), \label{eq: zn 3}
\end{align}
if $T_2 = C'\sigma^2\max_k \twonorm{\bbetaks{k}}^2[1+\log(\lambda_{\min}^{-1}(\bSigma_{\bx}))]$. 

Putting \eqref{eq: zn 0}, \eqref{eq: zn 1}, \eqref{eq: zn 2} and \eqref{eq: zn 3} together, 
\begin{equation}
	C_e(\bDelta) \coloneqq \te \Bigg[\sum_{k=1}^{K-1}\varphi_{\tau}((\bDeltak{k})^\top \bx)\mathds{1}\Bigg(\sum_{k=1}^{K-1}e^{(\bbetaks{k})^\top\bx} \leq T_1\Bigg)\mathds{1}(|(\bbetaks{k})^\top\bx| \leq T_2)\Bigg] \geq \twonorm{\bDelta}^2\cdot \frac{1}{2}c\lambda_{\min}(\bSigma_{\bx}).
\end{equation}
By symmetrization,
\begin{align}
	&\te Z_n(r,\delta) \\
	&\leq 2\te_{\bx, \bepsilon}\left[\sup_{\substack{\twonorm{\bDelta}\leq\delta \\ \onenorm{\bDelta}\leq r}}\frac{1}{n}\sum_{i=1}^n\sum_{k=1}^{K-1}\epsilon_{i,k}\varphi_{\tau}((\bDeltak{k})^\top \bx_i)\mathds{1}\Bigg(\sum_{k=1}^{K-1}e^{(\bbetaks{k})^\top\bx_i} \leq T_1\Bigg)\mathds{1}(|(\bbetaks{k})^\top\bx_i| \leq T_2)\right] \\
	&\leq 8\tilde{k}\delta\te_{\bx, \bepsilon}\left[\sup_{\substack{\twonorm{\bDelta}\leq\delta \\ \onenorm{\bDelta}\leq r}}\frac{1}{n}\sum_{i=1}^n\sum_{k=1}^{K-1}\epsilon_{i,k}\cdot((\bDeltak{k})^\top \bx_i)\right] \label{eq: contraction lipschitz}\\
	&= 8\tilde{k}\delta\te_{\bx, \bepsilon}\left[\sup_{\substack{\twonorm{\bDelta}\leq\delta \\ \onenorm{\bDelta}\leq r}}\sum_{k=1}^{K-1}\Big\<\bDeltak{k}, \frac{1}{n}\sum_{i=1}^n\epsilon_{i,k}\bx_i\Big\>\right] \\
	&\leq 8\tilde{k}\delta\te_{\bx, \bepsilon}\left[\sup_{\substack{\twonorm{\bDelta}\leq\delta \\ \onenorm{\bDelta}\leq r}}\sum_{k=1}^{K-1}\onenorm{\bDeltak{k}}\infnorma{\frac{1}{n}\sum_{i=1}^n\epsilon_{i,k}\bx_i}\right] \\
	&\leq 8\tilde{k}\delta r \cdot \max_k \infnorma{\frac{1}{n}\sum_{i=1}^n\epsilon_{i,k}\bx_i},
\end{align}
where \eqref{eq: contraction lipschitz} is by the contraction inequality for Rademacher  variables \citep{ledoux1991probability, wainwright2019high}. And by Theorem 3.26 in \cite{wainwright2019high}, for any $\delta' > 0$,
\begin{equation}\label{eq: zn concentration}
	\tp(Z_n(r, \delta) \geq \te Z_n(r, \delta) + \delta') \leq \exp\left\{-\frac{n\delta'^2}{4\cdot (2K\tilde{k}\delta^2)^2}\right\},
\end{equation}
implying that
\begin{align}
	Z_n(r, \delta) &\leq 8\tilde{k}\delta r \cdot \max_k \infnorma{\frac{1}{n}\sum_{i=1}^n\epsilon_{i,k}\bx_i} + \delta' \\
	&\leq C\tilde{k}\delta r \cdot \sqrt{\frac{\log (Kp)}{n}} + CK\tilde{k}\delta^2 \sqrt{\frac{\log (Kp)}{n}}, 
\end{align}
with probability at least $1-\exp\left\{-\frac{n(\delta')^2}{16K^2\tilde{k}^2\delta^2}\right\} = 1-C'(Kp)^{-1}$, by letting $\delta' = C''K\tilde{k}\delta\sqrt{\frac{\log (Kp)}{n}}$ with a large constant $C'' > 0$. This implies that
\begin{align}
	\mE_n(\bDelta) &\geq \gamma \left(\frac{1}{2}c\delta^2\lambda_{\min}(\bSigma_{\bx})- CK\tilde{k}\delta^2 \sqrt{\frac{\log (Kp)}{n}}-C\tilde{k}\delta r \cdot \sqrt{\frac{\log (Kp)}{n}}\right) \\
	&\geq \frac{e^{-(T_2+\tilde{k}\delta)}}{(T_1e^{\tilde{k}\delta})^2}\cdot \left(\frac{1}{2}c\lambda_{\min}(\bSigma_{\bx}) - CK\tilde{k} \sqrt{\frac{\log (Kp)}{n}}-C\tilde{k}r \cdot \sqrt{\frac{\log (Kp)}{n}}\right) \\
	&\geq \underbrace{\frac{c'\lambda_{\min}^4(\bSigma_{\bx})}{\exp\left\{C'\max_k \twonorm{\bbetaks{k}}^2\right\}K^2\sigma^8}\exp\{-C\sigma^2[\log(\lambda^{-1}_{\min}(\bSigma_{\bx})) + \log \sigma^2]\}}_{C_{\star}}\\
	&\quad \cdot \left(\frac{1}{2}c\lambda_{\min}(\bSigma_{\bx})\delta^2-CK\tilde{k}\delta^2 \sqrt{\frac{\log (Kp)}{n}}-C\tilde{k}\delta r \cdot \sqrt{\frac{\log (Kp)}{n}} \right).
\end{align}
Denote a random event $$
\mathcal{A} = \left\{\mE_n(\bDelta) \geq C_{\star}\left[\left(\frac{1}{2}c\lambda_{\min}(\bSigma_{\bx})-CK\tilde{k} \sqrt{\frac{\log (Kp)}{n}}\right)\twonorm{\bDelta}^2-C\tilde{k}r \cdot \sqrt{\frac{\log (Kp)}{n}}\twonorm{\bDelta}\onenorm{\bDelta} \right]\right\}.
$$
Consider the following peeling argument. Let the event $\mathcal{S}_{k,l} = \{\bDelta \in \mathbb{R}^{(K-1)p}: 2^{k-1} \leq \frac{\onenorm{\bDelta}}{\twonorm{\bDelta}} \leq 2^k, 2^{l-1} \leq \twonorm{\bDelta} \leq 2^l\}$, then it is easy to see that $\mathcal{A}^c \subseteq \bigcup_{k=1}^N\bigcup_{l=1}^{\infty}\mathcal{S}_{k,l}$ with $N = \lceil \frac{\log (Kp)}{2\log 2}\rceil$. Suppose $\widehat{\bDelta} \in \mathcal{S}_{k,l}$ for some $k$ and $l$ is not in $\mathcal{A}$. Then since
\begin{align}
	&C_{\star}\left[\left(\frac{1}{2}c\lambda_{\min}(\bSigma_{\bx})-CK\tilde{k} \sqrt{\frac{\log (Kp)}{n}}\right)\twonorm{\widehat{\bDelta}}^2-C\tilde{k} \cdot \sqrt{\frac{\log (Kp)}{n}}\twonorm{\widehat{\bDelta}}\onenorm{\widehat{\bDelta}} \right] \\
	&\geq \mE_n(\widehat{\bDelta}) \\
	&\geq C_{\star}C_{e}(\widehat{\bDelta}) - Z_n(2^{k+l}, 2^l) \\
	&\geq C_{\star}\twonorm{\widehat{\bDelta}}^2\cdot \frac{1}{2}c\lambda_{\min}(\bSigma_{\bx}) - Z_n(2^{k+l}, 2^l),
\end{align}
we must have
\begin{equation}
	Z_n(2^{k+l}, 2^l) \geq C_{\star}CK\tilde{k}\sqrt{\frac{\log (Kp)}{n}} \cdot 2^{2l-2} + C_{\star}C\tilde{k}\sqrt{\frac{\log (Kp)}{n}}\cdot 2^{k-1} \cdot 2^{2l-2}.
\end{equation}
Therefore by \eqref{eq: zn concentration}, for any $k \leq N$ and $l \geq 1$,
\begin{align}
	\tp(\mathcal{A}^c \cap \mathcal{S}_{k,l}) &\leq \tp\left(Z_n(2^{k+l}, 2^l) \geq C_{\star}CK\tilde{k}\sqrt{\frac{\log (Kp)}{n}} \cdot 2^{2l-2} + C_{\star}C\tilde{k}\sqrt{\frac{\log (Kp)}{n}}\cdot 2^{k-1} \cdot 2^{2l-2}\right) \\
	&\leq \exp\left\{-C'K^2\log (Kp) \cdot 2^{4l-4}\right\} \\
	&\leq C''\frac{1}{Kp \log (Kp)}\cdot 16^{-l}.
\end{align}
Hence
\begin{align}
	\tp(\mathcal{A}^c) &\leq \sum_{k=1}^N\sum_{l=1}^{\infty}\tp(\mathcal{A}^c \cap \mathcal{S}_{k,l}) \\
	&\leq N\cdot C''\frac{1}{Kp\log (Kp)}\cdot \sum_{l=1}^{\infty}16^{-l} \\
	&\leq C\cdot \frac{1}{Kp},
\end{align}
which completes our proof.
\end{proof}

\begin{proof}[Proof of Lemma \ref{lem: to prove rsc}]
	By Cauchy-Schwarz inequality,
	\begin{equation}
		\text{LHS} \geq \sum_{k=1}^{K-1}p_kx_k^2 - \left(\sum_{k=1}^{K-1}p_kx_k\right)^2 = \left(\sum_{k=1}^{K-1}p_kx_k^2\right)\left(\sum_{k=1}^{K-1}p_k\right)- \left(\sum_{k=1}^{K-1}p_kx_k\right)^2 + p_K\sum_{k=1}^{K-1}p_kx_k^2 \geq p_K\sum_{k=1}^{K-1}p_kx_k^2.
	\end{equation}
\end{proof}

\begin{proof}[Proof of Lemma \ref{lem: kl}]
	First, notice that
	\begin{equation}
		\textup{KL}(\tp_{X,Y;\bbeta}\|\tp_{X,Y;\widetilde{\bbeta}}) = -\sum_{k=1}^{K-1}\te_X\left[\tp(Y=k|X=\bx;\bbeta)\log\left[\frac{\tp(Y=k|X=\bx;\widetilde{\bbeta})}{\tp(Y=k|X=\bx;\bbeta)}\right]\right],
	\end{equation}
	where 
	\begin{align}
		&\log\left[\frac{\tp(Y=k|X=\bx;\widetilde{\bbeta})}{\tp(Y=k|X=\bx;\bbeta)}\right] \\
		&= \frac{(\nabla_{\bbeta}\tp(Y=k|X=\bx;\bbeta)|_{\bbeta})^\top}{\tp(Y=k|X=\bx;\bbeta)}(\widetilde{\bbeta}-\bbeta) + \frac{1}{2}(\widetilde{\bbeta}-\bbeta)^\top\nabla^2_{\bbeta} \tp(Y=k|X=\bx;\bbeta)|_{\bbeta=\widetilde{\bbeta}_t}(\widetilde{\bbeta}-\bbeta),
	\end{align}
	and $\widetilde{\bbeta}_t = (1-t)\widetilde{\bbeta} + \bbeta$ with some $t \in [0, 1]$. It implies that
	\begin{align}
		&\textup{KL}(\tp_{X,Y;\bbeta}\|\tp_{X,Y;\widetilde{\bbeta}}) \\
		&= -\sum_{k=1}^{K-1}\te_X\left[\tp(Y=k|X=\bx;\bbeta)(\widetilde{\bbeta}-\bbeta)^\top\nabla^2_{\bbeta} \tp(Y=k|X=\bx;\bbeta)|_{\bbeta=\widetilde{\bbeta}_t}(\widetilde{\bbeta}-\bbeta)\right].
	\end{align}
	And by basic algebra,
	\begin{align}
		&-\nabla^2_{\bbeta} \tp(Y=k|X=\bx;\bbeta)|_{\bbeta=\widetilde{\bbeta}_t} \\
		&= \begin{pmatrix}
				p_1(\bx;\widetilde{\bbeta}_t)(1-p_1(\bx;\widetilde{\bbeta}_t)) &\ldots &-p_1(\bx;\widetilde{\bbeta}_t)p_{K-1}(\bx;\widetilde{\bbeta}_t) \\
				\vdots &\vdots &\vdots \\
				-p_1(\bx;\widetilde{\bbeta}_t)p_{K-1}(\bx;\widetilde{\bbeta}_t) &\ldots &p_{K-1}(\bx;\widetilde{\bbeta}_t)(1-p_{K-1}(\bx;\widetilde{\bbeta}_t))		
			\end{pmatrix} \otimes \bx\bx^\top \\
			&\leq \diag(p_1(\bx;\widetilde{\bbeta}_t)(1-p_1(\bx;\widetilde{\bbeta}_t)), \ldots, p_{K-1}(\bx;\widetilde{\bbeta}_t)(1-p_{K-1}(\bx;\widetilde{\bbeta}_t))) \otimes \bx\bx^\top, 
	\end{align}
	which leads to
	\begin{align}
		&\textup{KL}(\tp_{X,Y;\bbeta}\|\tp_{X,Y;\widetilde{\bbeta}}) \\
		&\leq \frac{1}{2}\sum_{k=1}^{K-1}(\widetilde{\bbeta}^{(k)}-\bbetak{k})^\top\te_{\bx}\left[\tp(Y=k|X=\bx; \bbeta)\bx\bx^\top\right](\widetilde{\bbeta}^{(k)}-\bbetak{k}) \\
		&\leq \frac{1}{2}\sum_{k=1}^{K-1}\te_{\bx}\left[\tp(Y=k|X=\bx; \bbeta)(\bx^\top(\widetilde{\bbeta}^{(k)}-\bbetak{k}))^2\mathds{1}\left(\max_k |\bx^\top\bbetak{k}| \leq T\right)\right]\\
		&\quad + \frac{1}{2}\sum_{k=1}^{K-1}\te_{\bx}\left[\tp(Y=k|X=\bx; \bbeta)(\bx^\top(\widetilde{\bbeta}^{(k)}-\bbetak{k}))^2\mathds{1}\left(\max_k |\bx^\top\bbetak{k}| > T\right)\right]\\
		&\leq \frac{e^T}{2Ke^T}\sum_{k=1}^{K-1}\te_{\bx}\left[(\bx^\top(\widetilde{\bbeta}^{(k)}-\bbetak{k}))^2\right] + \frac{1}{2}\sum_{k=1}^{K-1}\sqrt{\te_{\bx}\left[(\bx^\top(\widetilde{\bbeta}^{(k)}-\bbetak{k}))^4\right]\cdot \tp\left(\max_k |\bx^\top\bbetak{k}| > T\right)} \\
		&\leq C\sigma^2\left(\frac{1}{K}e^{2T}\twonorm{\widetilde{\bbeta}-\bbeta}^2 + \sum_{k=1}^{K-1}\twonorm{\widetilde{\bbeta}^{(k)}-\bbetak{k}}^2 \cdot \exp\left\{-\frac{C'T^2}{\min_k \twonorm{\bbetak{k}}^2}\right\}\right) \\
		&\leq C\sigma^2\twonorm{\widetilde{\bbeta}-\bbeta}^2\cdot \left(\frac{1}{K}e^{2T} +  \exp\left\{-\frac{C'T^2}{\min_k \twonorm{\bbetak{k}}^2}\right\}\right).
	\end{align}
\end{proof}

\begin{proof}[Proof of Lemma \ref{lem: estimation error gamma}]
	Denote $\hSigma_{\bbeta} = \frac{1}{n}\sum_{i=1}^n \bm{B}(\bx_i;\bbeta)$. By the optimality of $\hgamma_j$, we have
    \begin{equation}
        \frac{1}{2}\hgamma_j^\top \hSigma_{-j,-j}\hgamma_j - \hSigma_{-j,j} + \lambda_j\onenorm{\hgamma_j} \leq \frac{1}{2}(\bgamma^*_j)^\top \hSigma_{-j,-j}\bgamma^*_j - \hSigma_{-j,j} + \lambda_j\onenorm{\bgamma^*_j},
    \end{equation}
    which implies that
    \begin{align}
        &\frac{1}{2}(\hgamma_j-\bgamma^*_j)^\top \hSigma_{-j,-j}(\hgamma_j-\bgamma^*_j) \\
        &\leq (\hSigma_{-j,j} - \hSigma_{-j,-j}\bgamma^*_j)^\top(\hgamma_j-\bgamma^*_j) + \lambda_j(\onenorm{\bgamma^*_j}-\onenorm{\hgamma_j}) \\
        &\leq (\hSigma_{-j,j} - \hSigma_{\bbeta^*,-j, j} - \hSigma_{-j,-j}\bgamma^*_j + \hSigma_{\bbeta^*, -j,-j}\bgamma^*_j)^\top(\hgamma_j-\bgamma^*_j) + (\hSigma_{\bbeta^*,-j, j}-\hSigma_{\bbeta^*, -j,-j}\bgamma^*_j)^\top(\hgamma_j-\bgamma^*_j) + \lambda_j(\onenorm{\bgamma^*_j}-\onenorm{\hgamma_j}).  \label{eq: basic ineq gamma}
    \end{align}
    Let us bound $(\hSigma_{-j,j} - \hSigma_{\bbeta^*,-j, j})^\top(\hgamma_j-\bgamma^*_j)$ on the RHS. WLOG, consider the case $j = 1$. Denote $\hgamma_j = ((\hgamma_j^{(1)})^\top, \ldots, (\hgamma_j^{(K-1)})^\top)^\top$, $\bgamma_j^* = ((\bgamma_j^{(1)*})^\top, \ldots, (\bgamma_j^{(K-1)*})^\top)^\top$, $\bbeta^* = ((\bbetaks{1})^\top, \ldots, (\bbetaks{K-1})^\top)^\top$, and $\hbbeta = ((\hbbetak{1})^\top, \ldots, (\hbbetak{K-1})^\top)^\top$. Then
    \begin{align}
        \big|(\hSigma_{-j,j} - \hSigma_{\bbeta^*,-j, j})^\top(\hgamma_j-\bgamma^*_j)\big| &= \bigg|\frac{1}{n}\sum_{k=1}^{K-1}\sum_{i=1}^n (\hgamma_j^{(k)} - \bgamma_j^{(k)*})^\top \bx_i \cdot \big[g_k(\{\bx_i^\top \hbbetak{k}\}_{k=1}^{K-1})-g_k(\{\bx_i^\top \bbetaks{k})\}_{k=1}^{K-1}\big]\cdot x_{i,1}\bigg| \\
        &\leq \frac{c}{2nK^2}\sum_{k=1}^{K-1}\sum_{i=1}^n \big[(\hgamma_j^{(k)} - \bgamma_j^{(k)*})^\top \bx_i\big]^2 + \frac{K^2}{2nc}\sum_{k=1}^{K-1}\sum_{i=1}^n [\bx_i^\top(\hbbetak{k}-\bbetaks{k})]^2\cdot [\nabla g_k(\bx_i^\top \tilde{\bbeta}_{k, i})]^2\cdot |x_{i,1}|^2, \\ \label{eq: interm}
    \end{align}
    where $g_1(\{\bx_i^\top \hbbetak{k}\}_{k=1}^{K-1}) = p_1(\bx_i; \hbbeta)[1-p_1(\bx_i; \hbbeta)]$, $g_k(\{\bx_i^\top \hbbetak{k}\}_{k=1}^{K-1}) = -p_1(\bx_i; \hbbeta)p_k(\bx_i; \hbbeta)$ for $k \geq 2$. Define five events
    \begin{align}
        \mathcal{A}_1 = &\bigg\{\frac{1}{n}\sum_{i=1}^n (\bx_i^\top \bu)^2 \leq C\twonorm{\bu}^2, \, \forall \bu \in \Pi_1(Ks)\bigg\}, \quad \mathcal{A}_2 = \bigg\{\max_{i=1:n}\infnorm{\bx_i} \leq C'\sqrt{\log (np)}\bigg\}, \\
        &\mathcal{A}_3 = \bigg\{\twonorm{\hbbeta - \bbeta^*} \leq CK^{5/2}\sqrt{\frac{s\log (Kp)}{n}}, \onenorm{\hbbeta - \bbeta^*} \leq CK^3\sqrt{\frac{s\log (Kp)}{n}}\bigg\} \\
        \mathcal{A}_4 = &\bigg\{\frac{1}{n}\sum_{i=1}^n (\bx_i^\top \bu)^2 \leq C\twonorm{\bu}^2, \, \forall \bu \in \Pi_1(K^7s)\bigg\}, \quad \mathcal{A}_5 = \bigg\{\lambda_j = C\sqrt{\frac{\log (Kp)}{n}} \geq 2\infnorm{\hSigma_{\bbeta^*, -j,j} - \hSigma_{\bbeta^*, -j,-j}\bgamma^*_j}\bigg\}.
    \end{align}
    where $\Pi_1(Ks) = \{\bu\in \mathbb{R}^{p}: \twonorm{\bu} \leq 1, \onenorm{\bu} \leq \sqrt{Ks}\}$, $C$ and $C'$ are some large positive constants. We want to argue that $\tp(\mathcal{A}_1 \cap \mathcal{A}_2\cap \mathcal{A}_3 \cap \mathcal{A}_4 \cap \mathcal{A}_5) \geq 1-2C''\exp\{-C'''n\}-C''(np)^{-1}-C''(Kp)^{-1}$. Note that by the sub-Gaussianality of $\bx_{i1}$, we have $\tp(\mathcal{A}_2) \geq 1-C''(np)^{-1}$. By Lemma 3.1 of \cite{plan2013one}, $\Pi_1(Ks) \subseteq 2\overline{\text{conv}}(\Pi_0(Ks))$ where $\Pi_0(Ks) = \{\bu\in \mathbb{R}^{p}:\twonorm{\bu} \leq 1, \zeronorm{\bu} \leq Ks\}$. Then by Lemma 2.3 in \cite{mendelson2008uniform}, $\Pi_0(Ks) \subseteq 2\overline{\text{conv}}(\Lambda)$, where $\Lambda$ is a $1/2$-cover of $\Pi_0(Ks)$ with $\norm{\Lambda} \leq C\binom{p}{Ks}5^{Ks} \leq \exp\{C'n\}$ when $n\geq C'Ks\log p$. By Theorem 2.4 in \cite{mendelson2008uniform}, $\tp(\mathcal{A}_1) \geq 1-C''\exp\{-C'''n\}$ when $n \geq C'Ks\log p$. By a similar argument,  $\tp(\mathcal{A}_4) \geq 1-C''\exp\{-C'''n\}$ when $n \geq C''s\log (pK)$. Additionally, by Theorem \ref{thm: estimation error beta}, we have $\tp(\mathcal{A}_3) \geq 1-C''(Kp)^{-1}$. Regarding $\tp(\mathcal{A}_5)$, note that
    \begin{equation}
        \hSigma_{\bbeta^*, -j, j} - \hSigma_{\bbeta^*, -j,-j}\bgamma^*_j = \hSigma_{\bbeta^*, -j, j} - \bSigma_{-j, j} + (\bSigma_{-j, -j} - \hSigma_{\bbeta^*, -j, -j})\bSigma_{-j, -j}^{-1}\bSigma_{-j, j},
    \end{equation}
    leading to
    \begin{equation}
        \infnorm{\hSigma_{\bbeta^*, -j, j} - \hSigma_{\bbeta^*, -j,-j}\bgamma^*_j} \lesssim  \infnorm{\hSigma_{\bbeta^*, -j, j} - \bSigma_{-j, j}} + \maxnorm{\bSigma_{-j, -j} - \hSigma_{\bbeta^*, -j, -j}}.
    \end{equation}
    Denote $B_{ij} = (\hSigma_{\bbeta^*,-j,-j}-\bSigma_{-j,-j})_{ij}$. It holds that
	\begin{equation}
		\norma{(\hSigma_{\bbeta^*,-j,-j}-\bSigma_{-j,-j})_{11}} 
		\leq \norma{\frac{1}{n}\sum_{i=1}^n x_{i1}^2p_1(\bx_i;\bbeta^*)(1-p_1(\bx_i;\bbeta^*))-\te [x_{i1}^2p_1(\bx_i;\bbeta^*)(1-p_1(\bx_i;\bbeta^*))]}.
	\end{equation}
    Since $p_1(\bx_i;\bbeta^*)(1-p_1(\bx_i;\bbeta^*))$ is always bounded, by the tail bound of sub-exponential variables, we have
	\begin{equation}
		\tp(B_{ij} \geq \delta) \leq C\exp\{-C'n\delta^2\}, \,\, \forall i, j.
	\end{equation}
	Therefore
	\begin{equation}
		\tp\left(\max_{i,j}B_{ij} \geq C\sqrt{\frac{\log (Kp)}{n}}\right) \leq C'[p(K-1)]^2\exp\{-C''\log (Kp)\} \leq C'''(pK)^{-1}.
	\end{equation}
    Hence $\tp\Big(\maxnorm{\bSigma_{-j, -j} - \hSigma_{\bbeta^*, -j, -j}} \leq C\sqrt{\frac{\log (Kp)}{n}}\Big)\geq 1-C'''(pK)^{-1}$. Similarly, we can show $\tp\Big(\infnorm{\hSigma_{\bbeta^*, -j, j} - \bSigma_{-j, j}}\leq C\sqrt{\frac{\log (Kp)}{n}}\Big) \geq 1-C'''(pK)^{-1}$. Therefore $\tp(\mathcal{A}_5) \geq 1-C'''(pK)^{-1}$.
    
    Then $\tp(\mathcal{A}_1 \cap \mathcal{A}_2\cap \mathcal{A}_3 \cap \mathcal{A}_4 \cap \mathcal{A}_5) \geq 1-C''\exp\{-C'''n\}-C''(np)^{-1}-C''(Kp)^{-1}$ follows by the basic inequality $\tp(\mathcal{A}_1 \cap \mathcal{A}_2 \cap \mathcal{A}_3\cap \mathcal{A}_4 \cap \mathcal{A}_5) \geq 1-\tp(\mathcal{A}_1^c) - \tp(\mathcal{A}_2^c)-\tp(\mathcal{A}_3^c)- \tp(\mathcal{A}_4^c)-\tp(\mathcal{A}_5^c)$.

    Under $\mathcal{A}_1 \cap \mathcal{A}_2 \cap \mathcal{A}_3 \cap \mathcal{A}_4 \cap \mathcal{A}_5$, we have
    \begin{align}
        |\bx_i^\top \hbbetak{k}| &\leq |\bx_i^\top (\hbbetak{k} - \bbetaks{k})| + |\bx_i^\top \bbetaks{k}| \\
        &\leq \infnorm{\bx_i} \onenorm{\hbbetak{k} - \bbetaks{k}} + C'' \\
        &\lesssim C'\sqrt{\log(np)}\cdot K^{5/2}\sqrt{\frac{s\log (Kp)}{n}}  +1 \\
        &\lesssim 1,
    \end{align}
    for all $i \in [n]$ and $k \in [K-1]$, when $n\gtrsim K^5s(\log (Kp))(\log (np))^2$. Therefore, conditioned on $\mathcal{A}_1 \cap \mathcal{A}_2 \cap \mathcal{A}_3$, we have $p_k(\bx_i; \hbbeta) \gtrsim 1/K$. Further by Lemma \ref{lem: to prove rsc}, conditioned on $\mathcal{A}_1 \cap \mathcal{A}_2 \cap \mathcal{A}_3 \cap \mathcal{A}_4 \cap \mathcal{A}_5$, we have
    \begin{align}
        \textup{LHS of \eqref{eq: basic ineq gamma}} &= \frac{1}{2}(\hgamma_j-\bgamma^*_j)^\top \hSigma_{-j,-j}(\hgamma_j-\bgamma^*_j) \\
        &\geq \frac{1}{n}\sum_{i=1}^n p_K(\bx_i; \hbbeta) \sum_{k=1}^{K-1}p_k(\bx_i; \hbbeta) [(\hgamma^{(k)}_j-\bgamma^{(k)*}_j)^\top \bx_i]^2 \\
        &\geq \frac{c}{nK^2}\sum_{i=1}^n \sum_{k=1}^{K-1}[(\hgamma^{(k)}_j-\bgamma^{(k)*}_j)^\top \bx_i]^2. \label{eq: LHS C10}
    \end{align}
    Denote $S_j = \textup{supp}(\bgamma^*_j)$. Similar to the arguments in the proof of Theorem \ref{thm: estimation error beta}, when $\lambda_j \geq 2\infnorm{\hSigma_{\bbeta^*, -j,j} - \hSigma_{\bbeta^*, -j,-j}\bgamma^*_j}$, it follows from \eqref{eq: basic ineq gamma} and \eqref{eq: interm} that
    \begin{equation}
        \frac{c}{nK^2}\sum_{i=1}^n \sum_{k=1}^{K-1}[(\hgamma^{(k)}_j-\bgamma^{(k)*}_j)^\top \bx_i]^2 \leq CK^7\frac{s\log (Kp)}{n}\log(np) + \lambda_j\left(\frac{3}{2}\onenorm{(\hgamma_j-\bgamma_j^*)_{S_j}}-\frac{1}{2}\onenorm{(\hgamma_j-\bgamma_j^*)_{S_j^c}}\right).
    \end{equation}
    Plugging $\lambda_j \asymp \sqrt{\frac{\log p}{n}}$ into above, we have
    \begin{equation}
        \onenorm{(\hgamma_j-\bgamma_j^*)_{S_j^c}} \leq 3\onenorm{(\hgamma_j-\bgamma_j^*)_{S_j}} + CK^7s\sqrt{\frac{\log (Kp)}{n}}\log(np).
    \end{equation}
    \underline{\textbf{Case 1:}} $\onenorm{(\hgamma_j-\bgamma_j^*)_{S_j}} > CK^7s\sqrt{\frac{\log (Kp)}{n}}\log(np)$.

    We have $ \onenorm{(\hgamma_j-\bgamma_j^*)_{S_j^c}} \leq 4\onenorm{(\hgamma_j-\bgamma_j^*)_{S_j}}$ and
    \begin{equation}
        \frac{c'}{K^2}\twonorm{\hgamma_j - \bgamma^*_j}^2 \leq \frac{c}{nK^2}\sum_{i=1}^n \sum_{k=1}^{K-1}[(\hgamma^{(k)}_j-\bgamma^{(k)*}_j)^\top \bx_i]^2 \leq C\sqrt{\frac{\log (Kp)}{n}}\onenorm{(\hgamma_j-\bgamma_j^*)_{S_j}} \leq C\sqrt{\frac{Ks_0\log (Kp)}{n}}\twonorm{\hgamma_j-\bgamma_j^*},
    \end{equation}
    where the first inequality comes from Theorem 1.6 of \cite{zhou2009restricted}. It implies that
    \begin{equation}
        \twonorm{\hgamma_j - \bgamma^*_j}\leq C'K^{5/2}\sqrt{\frac{s_0\log (Kp)}{n}},
    \end{equation}
    which entails that
    \begin{equation}
         \onenorm{\hgamma_j - \bgamma^*_j}\leq C'K^3s_0\sqrt{\frac{\log (Kp)}{n}}.
    \end{equation}
    
    \noindent\underline{\textbf{Case 2:}} $\onenorm{(\hgamma_j-\bgamma_j^*)_{S_j^c}} \leq CK^7s\sqrt{\frac{\log (Kp)}{n}}\log(np)$.

    We must have $\onenorm{(\hgamma_j-\bgamma_j^*)_{S_j}} \leq 4CK^7s\sqrt{\frac{\log (Kp)}{n}}\log(np) \leq K^{7/2}\sqrt{s}$, when $n \gtrsim K^7s(\log (Kp))(\log (np))^2$. By Lemma 3.1 of \cite{plan2013one}, $\Pi_1(K^7s) \subseteq 2\overline{\text{conv}}(\Pi_0(K^7s))$. Then by Lemma 2.3 in \cite{mendelson2008uniform}, $\Pi_0(K^7s) \subseteq 2\overline{\text{conv}}(\Lambda)$, where $\Lambda$ is a $1/2$-cover of $\Pi_0(K^7s)$ with $\norm{\Lambda} \leq C\binom{p}{K^7s}5^{K^7s} \leq \exp\{C'n\}$ when $n\geq C'K^7s\log p$. By Theorem 2.4 in \cite{mendelson2008uniform}, with probability at least $1-C''\exp\{-C'''n\}$, we have $\frac{c'}{K^2}\twonorm{\hgamma_j - \bgamma^*_j}^2 \leq \frac{c}{nK^2}\sum_{i=1}^n \sum_{k=1}^{K-1}[(\hgamma^{(k)}_j-\bgamma^{(k)*}_j)^\top \bx_i]^2$. Therefore,
    \begin{equation}
        \frac{c'}{K^2}\twonorm{\hgamma_j - \bgamma^*_j}^2 \leq CK^7\frac{s\log (Kp)}{n}\log(np) + C\sqrt{\frac{Ks_0\log (Kp)}{n}}\twonorm{\hgamma_j - \bgamma^*_j},
    \end{equation}
    which implies that
    \begin{equation}
        \twonorm{\hgamma_j - \bgamma^*_j} \lesssim K^{9/2}\sqrt{\frac{s(\log (Kp))(\log(np))}{n}} + K^{5/2}\sqrt{\frac{s_0\log (Kp)}{n}},\quad  \onenorm{\hgamma_j - \bgamma^*_j} \lesssim K^7s\sqrt{\frac{\log (Kp)}{n}}\log(np).
    \end{equation}
    
    Combining two cases, we have
    \begin{equation}
        \twonorm{\hgamma_j - \bgamma^*_j} \lesssim K^{9/2}\sqrt{\frac{s(\log (Kp))(\log(np))}{n}} + K^{5/2}\sqrt{\frac{s_0\log (Kp)}{n}},\quad  \onenorm{\hgamma_j - \bgamma^*_j} \lesssim K^7s\sqrt{\frac{\log (Kp)}{n}}\log(np) + K^3s_0\sqrt{\frac{\log (Kp)}{n}}, 
    \end{equation}
    with probability $1-C'\exp\{-C''n\}-C'(np)^{-1}-C'(Kp)^{-1}$.
    
    \noindent Next we will bound $\norm{\htau_j^2-(\tau_j^*)^2}$ and $\norm{\htau_j^{-2}-(\tau_j^*)^{-2}}$. Note that $\htau_j^2 = \huSigma_{j,j} - \hSigma_{-j,j}^\top\hgamma_j$, and $(\tau_j^*)^2 = \Sigma_{j,j} - \bSigma_{-j,j}^\top\bgamma_j^*$. Then
	\begin{align}
		\norm{\htau_j^2-(\tau_j^*)^2} &\leq \norm{\huSigma_{j,j}-\Sigma_{j,j}} + \norm{\hSigma_{-j,j}^\top\hgamma_j-\bSigma_{-j,j}^\top\bgamma_j^*} \\
		&\leq \norm{\huSigma_{j,j}-\Sigma_{j,j}} + \norm{(\hSigma_{-j,j}-\bSigma_{-j,j})^\top\hgamma_j} + \norm{\bSigma_{-j,j}^\top(\hgamma_j-\bgamma_j^*)} \\
		&\leq \norm{\huSigma_{j,j}-\Sigma_{j,j}} + \norm{(\hSigma_{-j,j}-\bSigma_{-j,j})^\top\bgamma_j^*} + \infnorm{\hSigma_{-j,j}-\bSigma_{-j,j}}\onenorm{\hgamma_j-\bgamma_j^*} + \twonorm{\bSigma_{-j,j}} \cdot \twonorm{\hgamma_j-\bgamma_j^*}, 
	\end{align}
	where 
	\begin{align}
		\norm{\huSigma_{j,j}-\Sigma_{j,j}} &\lesssim \twonorm{\hbbeta - \bbeta^*} \lesssim  K^{5/2}\sqrt{\frac{s\log (Kp)}{n}},\\
        \norm{(\hSigma_{-j,j}-\bSigma_{-j,j})^\top\bgamma_j^*} &\lesssim \twonorm{\hbbeta - \bbeta^*}\twonorm{\bgamma_j^*} \lesssim K^{5/2}\sqrt{\frac{s\log (Kp)}{n}}, \\
        \infnorm{\hSigma_{-j,j}-\bSigma_{-j,j}} &\lesssim \twonorm{\hbbeta - \bbeta^*}  \lesssim K^{5/2}\sqrt{\frac{s\log (Kp)}{n}}, \\
        \onenorm{\hgamma_j - \bgamma^*_j} &\lesssim K^7s\sqrt{\frac{\log (Kp)}{n}}\log(np) + K^3s_0\sqrt{\frac{\log (Kp)}{n}},\\
        \twonorm{\bSigma_{-j,j}} &\lesssim 1,
	\end{align}
	with probability $1-C'(Kp)^{-1}$, which can be similarly derived as before. This implies
	\begin{align}
		\norm{\htau_j^2-(\tau_j^*)^2} &\lesssim K^{9/2}\sqrt{\frac{s(\log (Kp))(\log(np))}{n}} + K^{5/2}\sqrt{\frac{s\log (Kp)}{n}}\cdot \Bigg(K^7s\sqrt{\frac{\log (Kp)}{n}}\log(np) + K^3s_0\sqrt{\frac{\log (Kp)}{n}}\Bigg)+ K^{5/2}\sqrt{\frac{s_0\log (Kp)}{n}}, \\
        &\lesssim K^{9/2}\sqrt{\frac{s(\log (Kp))(\log(np))}{n}} + K^{5/2}\sqrt{\frac{s_0\log (Kp)}{n}}, \\
		\norm{\htau_j^{-2}-(\tau_j^*)^{-2}} &= \frac{\norm{\htau_j^2-(\tau_j^*)^2}}{\htau_j^2(\tau_j^*)^2} \lesssim \norm{\htau_j^2-(\tau_j^*)^2} \lesssim K^{9/2}\sqrt{\frac{s(\log (Kp))(\log(np))}{n}} + K^{5/2}\sqrt{\frac{s_0\log (Kp)}{n}},
	\end{align}
	with probability $1-C'\exp\{-C''n\}-C'(np)^{-1}-C'(Kp)^{-1}$, where the second inequality of $\norm{\htau_j^2-(\tau_j^*)^2}$ is due to the condition $n \gtrsim K^{10}s^2\log(Kp) + K^6s_0^2\log(Kp)$. Since $\hTheta_j$ consists of $\htau_j^{-2}$ and $\hgamma_j$, we have
    \begin{align}
        \twonorm{\hTheta_j - \bTheta_j^*} &\lesssim \twonorm{\hgamma_j - \bgamma_j^*} + \norm{\htau_j^{-2} - (\tau_j^*)^{-2}} \lesssim K^{9/2}\sqrt{\frac{s(\log (Kp))(\log(np))}{n}} + K^{5/2}\sqrt{\frac{s_0\log (Kp)}{n}}, \\
        \onenorm{\hTheta_j - \bTheta_j^*} &\lesssim \onenorm{\hgamma_j - \bgamma_j^*} + \norm{\htau_j^{-2} - (\tau_j^*)^{-2}} \lesssim K^7s\sqrt{\frac{\log (Kp)}{n}}\log(np) + K^3s_0\sqrt{\frac{\log (Kp)}{n}},
    \end{align}
    with probability $1-C'\exp\{-C''n\}-C'(np)^{-1}-C'(Kp)^{-1}$.
\end{proof}

\subsection{Proofs of theorems}
\begin{proof}[Proof of Theorem \ref{thm: estimation error beta}]
	Denote $\hDelta = \hbbeta - \bbeta^*$.  The negative log-likelihood is
	\begin{equation}
		\mL_n(\bbeta) = -\frac{1}{n}\sum_{i=1}^n\sum_{k=1}^Ky_i^{(k)}\cdot \bx_i^\top\bbetak{k} + \frac{1}{n}\sum_{i=1}^n \log\left(1+\sum_{k=1}^{K-1}e^{\bx_i^\top\bbetak{k}}\right).
	\end{equation}
	The gradient is
	\begin{equation}
		\nabla \mL_n(\bbeta) = -\begin{pmatrix}
			\frac{1}{n}\sum_{i=1}^n\left(\yk{1}_i - \frac{e^{\bx_i^\top\bbetak{1}}}{1+\sum_{k=1}^{K-1}e^{\bx_i^\top\bbetak{k}}}\right)\bx_i \\
			\vdots \\
			\frac{1}{n}\sum_{i=1}^n\left(\yk{k}_i - \frac{1}{1+\sum_{k=1}^{K-1}e^{\bx_i^\top\bbetak{k}}}\right)\bx_i
		\end{pmatrix}_{[(K-1)p]\times 1}.
	\end{equation}
	By the convexity of $\mL_n$,
	\begin{align}
		0 &\leq \mL_n(\bbeta^* + \hDelta) - \mL_n(\bbeta^*) - \nabla \mL_n(\bbeta^*)^\top\hDelta \\
		&\leq - \nabla \mL_n(\bbeta^*)^\top\hDelta + \lambda(\onenorm{\bbeta^*}-\onenorm{\hbbeta})\\
		&\leq \frac{3}{2}\lambda\onenorm{\hDelta_S}-\frac{1}{2}\lambda\onenorm{\hDelta_{S^c}},
	\end{align}
	implying that
	\begin{equation}\label{eq: cone}
		\onenorm{\hDelta_{S^c}} \leq 3 \onenorm{\hDelta_S}.
	\end{equation}
	Define $F_n(\bDelta) = \mL_n(\bbeta^* + \bDelta) - \mL_n(\bbeta^*) + \lambda\onenorm{\bbeta^*+\bDelta}-\lambda\onenorm{\bbeta^*}$, which is a convex function of $\bDelta$. By the optimality of $\hbbeta$, $F_n(\hDelta) \leq 0$. If $\twonorm{\hDelta} > 1$, then $\exists t\in (0, 1)$ such that $\tDelta = t\hDelta$. By convexity $F_n(\tDelta) \leq tF_n(\hDelta) + (1-t)F_n(\bm{0}) \leq 0$. And by \eqref{eq: cone}, $\onenorm{\tDelta_S} \leq 3\onenorm{\tDelta_{S^c}}$. Then
	\begin{align}
		F_n(\tDelta) &\geq \mathcal{E}_n(\tDelta) + \nabla \mL_n(\bbeta^*)^\top\tDelta + \lambda\onenorm{\bbeta^* + \tDelta} - \lambda \onenorm{\bbeta^*} \\
		&\geq \mathcal{E}_n(\tDelta) - \infnorm{\nabla \mL_n(\bbeta^*)}\onenorm{\tDelta} + \lambda(\onenorm{\bbeta^*_S + \tDelta_S}-\onenorm{\bbeta^*_S}+\onenorm{\tDelta_{S^c}}) \\
		&\geq \mathcal{E}_n(\tDelta) - \frac{3}{2}\lambda \onenorm{\tDelta_S} + \frac{1}{2}\lambda\onenorm{\tDelta_{S^c}},
	\end{align}
	because $\infnorm{\nabla \mL_n(\bbeta^*)} \leq C\sqrt{\frac{\log (Kp)}{n}}$ with probability at least $1-C'(Kp)^{-1}$.
	Since $\twonorm{\tDelta} = 1$, by Lemma \ref{lem: rsc},
	\begin{align}
		F_n(\tDelta) &\geq \frac{1}{K^2}\left(c_1-c_2\sqrt{\frac{\log (Kp)}{n}}\onenorm{\tDelta}\right) - \frac{3}{2}\lambda\onenorm{\tDelta_S} + \frac{1}{2}\lambda\onenorm{\tDelta_{S^c}} \\
		&\geq \frac{c_1}{K^2}  - \frac{4c_2}{K^2}\cdot \sqrt{\frac{\log (Kp)}{n}}\cdot \sqrt{Ks}- \frac{3}{2}\lambda\sqrt{Ks} \\
		&\geq \frac{c_1}{K^2} - \frac{3}{2K^2}C\sqrt{\frac{Ks\log (Kp)}{n}} - c_2\sqrt{\frac{Ks\log (Kp)}{n}} \\
		&> 0,
	\end{align}
	when $n \geq C'K^5 s\log p$ with a large constant $C' > 0$. But this is contradicted by the fact that $F_n(\tDelta) \leq 0$. Therefore, we must have $\twonorm{\hDelta} \leq 1$. Then we can apply a similar argument above with Lemma \ref{lem: rsc} to show that
	\begin{align}
		\frac{1}{K^2}\left(c_1\twonorm{\hDelta}^2 - c_2\sqrt{\frac{\log (Kp)}{n}}\onenorm{\hDelta}\twonorm{\hDelta}\right) &\leq \mathcal{E}_n(\hDelta) \\
		&\leq \frac{3}{2}\lambda \onenorm{\hDelta_S} - \frac{1}{2}\lambda\onenorm{\hDelta_{S^c}}\\
		&\leq \frac{3}{2}C\sqrt{\frac{Ks\log (Kp)}{n}}\twonorm{\hDelta},
	\end{align}
	with probability at least $1-C'(Kp)^{-1}$. Recalling \eqref{eq: cone}, we have
	\begin{equation}
		\frac{1}{2K^2}c_1\twonorm{\hDelta}^2 \leq \frac{1}{K^2}\left(c_1-c_2\sqrt{\frac{\log (Kp)}{n}}4\sqrt{Ks}\right)\twonorm{\bDelta}^2 \leq \frac{3}{2}C\sqrt{\frac{Ks\log (Kp)}{n}}\twonorm{\hDelta},
	\end{equation}
	when $n \geq 64\frac{c_1^2}{c_2^2}Ks\log (Kp)$. Therefore,
	\begin{equation}
		\twonorm{\hDelta} \leq \frac{3C}{c_1}K^{5/2}\sqrt{\frac{s\log (Kp)}{n}}, \quad \onenorm{\hDelta} \leq 4\sqrt{Ks}\twonorm{\hDelta} \leq \frac{12C}{c_1}K^3s\sqrt{\frac{\log (Kp)}{n}},
	\end{equation}
	with probability at least $1-C'(Kp)^{-1}$.
\end{proof}

\begin{proof}[Proof of Theorem \ref{thm: lower bdd estimation error beta}]
	Consider $\mathcal{H} = \delta\cdot \{\bm{z} \in \{-1,0,1\}^p: \zeronorm{\bm{z}} = s\}$, then by Lemma 5 in \cite{raskutti2011minimax}, there exists $\widetilde{\mathcal{H}} \subseteq \mathcal{H}$ such that for any $\bz'$, $\bz \in \widetilde{\mathcal{H}}$, we have $\zeronorm{\bz-\bz'} \geq s/2$ and $\log |\widetilde{\mathcal{H}}| \geq C's\log(ep/s)$. Take $\bbeta$, $\widetilde{\bbeta} \in \widetilde{\mathcal{H}}^{\otimes (K-1)}$, by Lemma \ref{lem: kl},
	\begin{align}
		\textup{KL}(\tp_{\bx,y;\bbeta}^{\otimes n}\|\tp_{\bx,y;\widetilde{\bbeta}}^{\otimes n}) &\leq Cn\sigma^2\twonorm{\widetilde{\bbeta}-\bbeta}^2\cdot \left(\frac{1}{K}e^{2T} +  \exp\left\{-\frac{C'T^2}{\min_k \twonorm{\bbetak{k}}^2}\right\}\right) \\
		&\leq CnK\delta^2s \left(\frac{1}{K}e^{2T} +  \exp\left\{-\frac{C'T^2}{\delta^2s}\right\}\right).
	\end{align}
	Then by Fano's lemma \citep{tsybakov2009introduction},
	\begin{align}
		\inf_{\hbbeta}\sup_{\bbeta}\tp\left(\twonorm{\hbbeta-\bbeta}^2 \geq \frac{Ks}{4}\delta^2\right) &\geq 1-\frac{\log 2 + CnK\delta^2s \left(\frac{1}{K}e^{2T} +  \exp\left\{-\frac{C'T^2}{\delta^2s}\right\}\right)}{K\log |\widetilde{\mathcal{H}}|} \\
		&\geq 1-\frac{\log 2 + CnK\frac{K\log (ep/s)}{n}s \left(\frac{1}{K}C'' +  \exp\left\{-\frac{C'''n}{Ks\log (ep/s)}\right\}\right)}{C'Ks\log(ep/s)}\\
		&\geq \frac{1}{4},
	\end{align}
	if $T$ is a small constant and $\delta = c\sqrt{\frac{K\log (ep/s)}{n}}$ with a small constant $c > 0$, when $n \geq CKs(\log (ep/s))(\log K)$, $p \gg s$. The lower bound for $\onenorm{\hbbeta-\bbeta}$ can be similarly shown.
\end{proof}

\begin{proof}[Proof of Theorem \ref{thm: error}]
	It is easy to see that for any classifier $\mC$,
	\begin{equation}
		R(\mathcal{C}) = 1 - \te_{X}[\tp_{Y|X}(\mC(X) = Y)].
	\end{equation}
	This implies that
	\begin{equation}
		R(\mathcal{C}_{\bbeta^*}) = 1 - \te_{\bx}\left[\max_k\tp(Y=k|X=\bx)\right].
	\end{equation}
	Since 
	\begin{align}
		\tp_{Y|X}(\mathcal{C}_{\hbbeta}(X) = Y) &\geq \widehat{\tp}_{Y|X}(\mathcal{C}_{\hbbeta}(X) = Y) - \max_k\norma{\widehat{\tp}(Y=k|X=\bx)-\tp(Y=k|X=\bx)} \\
		&= \max_k \widehat{\tp}(Y=k|X=\bx) - \max_k\norma{\widehat{\tp}(Y=k|X=\bx)-\tp(Y=k|X=\bx)},
	\end{align}
	we have
	\begin{equation}
		R(\mathcal{C}_{\hbbeta}) \leq 1-\te_{\bx}\left[\max_k \widehat{\tp}(Y=k|X=\bx)\right] + \te_{\bx}\left[\max_k\norma{\widehat{\tp}(Y=k|X=\bx)-\tp(Y=k|X=\bx)}\right].
	\end{equation}
	Hence
	\begin{align}
		&R(\mathcal{C}_{\hbbeta}) - R(\mathcal{C}_{\bbeta^*}) \\
		&\leq 2\te_{\bx}\left[\max_k\norma{\widehat{\tp}(Y=k|X=\bx)-\tp(Y=k|X=\bx)}\right] \\
		&= 2\te_{\bx}\left[\max_k\norma{p_k(\bx;\hbbeta)-p_k(\bx;\bbeta^*)}\right] \\
		&= 2\te_{\bx}\left[\max_k\norma{\nabla g_k(t\bx^\top\hbbetak{1} + (1-t)\bx^\top\bbetaks{1}, \ldots, t\bx^\top\bbetaks{K-1} + (1-t)\bx^\top\bbetaks{K-1})\begin{pmatrix}
			\bx^\top(\hbbetak{1}-\bbeta_1^*) \\
			\vdots \\
			\bx^\top(\hbbeta_{K-1}-\bbeta_{K-1}^*)
		\end{pmatrix}}\right]  \\
		&\leq 2\te_{\bx}\Bigg[\max_k\twonorm{\nabla g_k(t\bx^\top\hbbetak{1} + (1-t)\bx^\top\bbetaks{1}, \ldots, t\bx^\top\hbbetak{K-1} + (1-t)\bx^\top\bbetaks{K-1})}\\
		&\quad \cdot \sqrt{\sum_{k=1}^{K-1}(\bx^\top(\hbbetak{k}-\bbetaks{k}))^2}\Bigg] \\
		&\leq C\te_{\bx}\sqrt{\sum_{k=1}^{K-1}(\bx^\top(\hbbetak{k}-\bbetaks{k}))^2} \\
		&\leq C\sqrt{\sum_{k=1}^{K-1}\te_{\bx}(\bx^\top(\hbbetak{k}-\bbetaks{k}))^2} \\
		&\leq C\twonorm{\hbbeta-\bbeta^*} \\
		&\leq CK^{5/2}\sqrt{\frac{s\log (Kp)}{n}},
	\end{align}
	with probability at least $1-C'(Kp)^{-1}$, which completes the proof.
\end{proof}

\begin{proof}[Proof of Theorem \ref{thm: clt}]
	Note that
	\begin{equation}
		\widehat{\bm{b}} - \bbeta^* = (\bm{I} - \hTheta \bm{B}_n)(\hbbeta-\bbeta^*) + \frac{1}{n}(\hTheta-\bTheta)\begin{pmatrix}
			\bX^\top\bepsilon_1 \\
			\bX^\top\bepsilon_2 \\
			\vdots \\
			\bX^\top\bepsilon_{K-1}
		\end{pmatrix} + \frac{1}{n}\bTheta \begin{pmatrix}
			\bX^\top\bepsilon_1 \\
			\bX^\top\bepsilon_2 \\
			\vdots \\
			\bX^\top\bepsilon_{K-1}
		\end{pmatrix},
	\end{equation}
	where 
	\begin{equation}
	\bm{B}_n = \frac{1}{n}\sum_{i=1}^n \int_0^1 B(\bx_i; \bbeta^* + t(\hbbeta-\bbeta^*)) dt,
\end{equation}
$$B(\bx; \bbeta) \coloneqq \begin{pmatrix}
	p_1(\bx;\bbeta)[1-p_1(\bx;\bbeta)]\bx\bx^\top & -p_1(\bx;\bbeta)p_2(\bx;\bbeta)\bx\bx^\top &\ldots &-p_1(\bx;\bbeta)p_{K-1}(\bx;\bbeta)\bx\bx^\top \\
	-p_1(\bx;\bbeta)p_2(\bx;\bbeta)\bx\bx^\top &p_2(\bx;\bbeta)[1-p_2(\bx;\bbeta)]\bx\bx^\top &\ldots &-p_2(\bx;\bbeta)p_{K-1}(\bx;\bbeta)\bx\bx^\top \\
	\vdots &\vdots  &\vdots &\vdots \\
	  -p_1(\bx;\bbeta)p_{K-1}(\bx;\bbeta)\bx\bx^\top &-p_2(\bx;\bbeta)p_{K-1}(\bx;\bbeta)\bx\bx^\top &\ldots &p_{K-1}(\bx;\bbeta)[1-p_{K-1}(\bx;\bbeta)]\bx\bx^\top
\end{pmatrix}.$$
Then
\begin{equation}
	\widehat{b}_j -\beta^*_j = \underbrace{(e_j-\hTheta_{j}^\top\bm{B}_n)(\hbbeta-\bbeta^*)}_{[1]} + \underbrace{\frac{1}{n}(\hTheta_j-\bTheta_j)^\top\begin{pmatrix}
			\bX^\top\bepsilon_1 \\
			\bX^\top\bepsilon_2 \\
			\vdots \\
			\bX^\top\bepsilon_{K-1}
		\end{pmatrix}}_{[2]} + \frac{1}{n}\underbrace{\bTheta_{j}^\top \begin{pmatrix}
			\bX^\top\bepsilon_1 \\
			\bX^\top\bepsilon_2 \\
			\vdots \\
			\bX^\top\bepsilon_{K-1}
		\end{pmatrix}}_{[3]}.
\end{equation}
And
\begin{equation}
	[1] \leq \underbrace{\norma{(\hTheta_j-\bTheta_j)^\top\bm{B}_n(\hbbeta-\bbeta^*)}}_{[1].1} + \underbrace{\norma{\bTheta_j^\top(\bSigma-\bm{B}_n)(\hbbeta-\bbeta^*)}}_{[1].2},
\end{equation}
where 
\begin{align}
	[1].1 &\leq \frac{1}{2}(\hTheta_j-\bTheta_j)^\top\bm{B}_n(\hTheta_j-\bTheta_j) + \frac{1}{2}(\hbbeta-\bbeta^*)^\top\bm{B}_n(\hbbeta-\bbeta^*) \\
	&\leq \frac{1}{2}(\hTheta_j-\bTheta_j)^\top \diag(\bX\bX^\top, \ldots, \bX\bX^\top)(\hTheta_j-\bTheta_j) \\
	&\quad + \frac{1}{2}(\hbbeta-\bbeta^*)^\top \diag(\bX\bX^\top, \ldots, \bX\bX^\top)(\hbbeta-\bbeta^*) \\
	&\lesssim_p C\twonorm{\hTheta_j-\bTheta_j}^2 + C\twonorm{\hbbeta-\bbeta^*}^2\\
	&\lesssim_p K^9\frac{s(\log (Kp))(\log(np))}{n} + K^5\frac{s_0\log (Kp)}{n},
\end{align}
where the second last inequality comes from a similar argument in the proof of Lemma \ref{lem: estimation error gamma} to analyze event $\mathcal{A}_1$, and the last inequality comes of Theorem \ref{thm: estimation error beta} and Lemma \ref{lem: estimation error gamma}. In addition, by a similar argument to prove \eqref{eq: interm}, we have
\begin{align}
    [1].2 &\lesssim \frac{1}{n}\sum_{k=1}^{K-1}\sum_{i=1}^n \big[(\hgamma_j^{(k)} - \bgamma_j^{(k)*})^\top \bx_i\big]^2 + \frac{1}{n}\sum_{k=1}^{K-1}\sum_{i=1}^n [\bx_i^\top(\hbbetak{k}-\bbetaks{k})]^2\cdot |x_i^\top \bTheta_j^{(k)*}|^2 \\
    &\lesssim \twonorm{\hgamma_j - \bgamma^*_j}^2 + \twonorm{\hbbeta - \bbeta^*}^2 \log(nK) \\
    &\lesssim_{\tp} K^9\frac{s(\log (Kp))(\log(np))}{n} + K^5\frac{s_0\log (Kp)}{n}.
\end{align}
Therefore,
\begin{equation}
	[1] \lesssim_{\tp} K^9\frac{s(\log (Kp))(\log(np))}{n} + K^5\frac{s_0\log (Kp)}{n} \ll n^{-1/2},
\end{equation}
when $n \gg K^{18}s^2(\log (Kp))^2(\log(np))^2 + K^{10}s_0^2(\log (Kp))^2$.

Similarly,
\begin{equation}
	[2] \leq \frac{1}{n}\onenorm{\hTheta_j-\bTheta_j}\cdot \infnorma{\begin{pmatrix}
			\bX^\top\bepsilon_1 \\
			\bX^\top\bepsilon_2 \\
			\vdots \\
			\bX^\top\bepsilon_{K-1}
		\end{pmatrix}} \\
		\lesssim_p \bigg[K^7s\sqrt{\frac{\log (Kp)}{n}}\log(np) + K^3s_0\sqrt{\frac{\log (Kp)}{n}}\bigg]\cdot \sqrt{\frac{\log (Kp)}{n}}\\
		\ll n^{-1/2}.
\end{equation}
And
\begin{equation}
	[3] = \frac{1}{n}\sum_{i=1}^n \bx_i^\top\left(\sum_{k=1}^{K-1}\epsilon_{k,i}\btheta_{k}\right) = \bTheta_j^\top\frac{1}{n}\sum_{i=1}^n\begin{pmatrix}
		\epsilon_{1,i}\bx_i \\
		\vdots \\
		\epsilon_{K-1,i}\bx_i
	\end{pmatrix}.
\end{equation}
By the Lindeberg-Feller central limit theorem, we have the following asymptotic normality:
\begin{equation}
	\frac{\sqrt{n}[3]}{\sqrt{\bTheta_j^\top\bm{D}\bTheta_j}} \overset{\textup{d}}{\to} N\left(0, 1\right),
\end{equation}
where $\bm{D} = \cov((\epsilon_1 X^\top, \ldots, \epsilon_{K-1} X^\top)^\top)$. Note that
\begin{equation}
	\bm{D} = \te \begin{pmatrix}
		\epsilon_1^2 XX^\top &\ldots &\epsilon_1\epsilon_{K-1} XX^\top  \\
		\vdots &\vdots &\vdots \\
		\epsilon_1\epsilon_{K-1} XX^\top &\ldots &\epsilon_{K-1}^2 XX^\top
	\end{pmatrix},
\end{equation}
where
\begin{align}
	\te(\epsilon_k^2 XX^\top) &= \te_{X}\left\{\te[(y_k-p_k(X;\bbeta^*))^2|X]XX^\top\right\} = \te_{X}\left[XX^\top p_k(X;\bbeta^*)(1-p_k(X;\bbeta^*))\right],\\
	\te(\epsilon_{k_1}\epsilon_{k_2} XX^\top) &= \te_{X}\left\{\te[(y_{k_1}-p_{k_1}(X;\bbeta^*))(y_{k_2}-p_{k_2}(X;\bbeta^*))|X]XX^\top\right\} \\
	&= -\te_{\bx}\left[XX^\top p_{k_1}(X;\bbeta^*)p_{k_2}(X;\bbeta^*)\right],
\end{align}
for any $k$, $k_1 \neq k_2 \in \{1, \ldots, K-1\}$. Therefore $\bm{D} = \te [\nabla^2 \mL_n(\bbeta^*)] = \bSigma$. It implies that
\begin{align}
	&\norma{\hTheta_j^\top\hSigma\hTheta_j - \bTheta_j^\top\bSigma\bTheta_j} \\
	&\leq \norma{(\hTheta_j-\bTheta_j)^\top\hSigma\hTheta_j} + \norma{\bTheta_j^\top(\hSigma-\bSigma)^\top\hTheta_j} + \norma{\bTheta_j^\top\bSigma(\hTheta_j-\bTheta_j)} \\
    &\leq \norma{(\hTheta_j-\bTheta_j)^\top(\hSigma-\bSigma)(\hTheta_j-\bTheta_j)} + \norma{(\hTheta_j-\bTheta_j)^\top(\hSigma-\bSigma)\bTheta_j} + \norma{(\hTheta_j-\bTheta_j)^\top\bSigma(\hTheta_j-\bTheta_j)}  + \norma{(\hTheta_j-\bTheta_j)^\top\bSigma\bTheta_j}\\
    &\quad + \norma{\bTheta_j^\top(\hSigma-\bSigma)^\top(\hTheta_j-\bTheta_j)} +  \norma{\bTheta_j^\top(\hSigma-\bSigma)^\top\bTheta_j} + \norma{\bTheta_j^\top\bSigma(\hTheta_j-\bTheta_j)}\\
	&\lesssim \onenorm{\hTheta_j-\bTheta_j}^2\maxnorm{\hSigma-\bSigma} + \onenorm{\hTheta_j-\bTheta_j}\maxnorm{\hSigma-\bSigma}\onenorm{\bTheta_j} + \twonorm{\hTheta_j-\bTheta_j}^2 + \twonorm{\hTheta_j-\bTheta_j}\twonorm{\bSigma\bTheta_j}\\
    &\lesssim_{\tp} \maxnorm{\hSigma-\bSigma}\sqrt{Ks_0} + \twonorm{\hTheta_j-\bTheta_j} \\
    &\lesssim_{\tp} K^3\sqrt{\frac{ss_0\log(Kp)}{n}} + K^{9/2}\sqrt{\frac{s(\log (Kp))(\log(np))}{n}} + K^{5/2}\sqrt{\frac{s_0\log (Kp)}{n}} \\
	&= \smallo(1),
\end{align}
because $\onenorm{\hTheta_j-\bTheta_j}  = \smallo_{\tp}(1)$ by Lemma \ref{lem: estimation error gamma}, when $n \gg K^{18}s^2(\log (Kp))^2(\log (np))^2 + K^{10}s_0^2(\log (Kp))^2$. Hence, the desired conclusion in Theorem \ref{thm: clt} holds due to Slutsky's Theorem.
\end{proof}

\begin{proof}[Proof of Theorem \ref{thm: misspecification}]
    The results in Theorems \ref{thm: estimation error beta}, \ref{thm: error}, and \ref{thm: clt} can be derived in a similar way in the model misspecification scenario, so we only point out the difference here. 
    
    When deriving the estimation error bounds in Theorem \ref{thm: estimation error beta}, we need to show $\infnorm{\nabla \mL_n(\bbeta^*)} \lesssim \sqrt{\frac{log(Kp)}{n}}$ with probability at least $1-C(Kp)^{-1}$. Note that
    \begin{equation}\label{eq: nabla L miss}
        \nabla \mL_n(\bbeta^*) = \begin{pmatrix}
            \frac{1}{n}\sum_{i=1}^n [\yk{1}_i - p_1(\bx_i;\bbeta^*)]\bx_i \\
            \vdots \\
            \frac{1}{n}\sum_{i=1}^n [\yk{K-1}_i - p_{K-1}(\bx_i;\bbeta^*)]\bx_i
        \end{pmatrix} = 
        \frac{1}{n}\begin{pmatrix}
			\bX^\top\bepsilon_1 \\
			\bX^\top\bepsilon_2 \\
			\vdots \\
			\bX^\top\bepsilon_{K-1}
		\end{pmatrix},
    \end{equation}
    where $\bepsilon_k \coloneqq \byk{k} - p_k(\bX; \bbeta^*)$ for $k = 1:(K-1)$. By the definition of the best approximator \eqref{eq: beta approx} and the first-order condition of a minimizer of a convex function, we have $\te [\nabla \mL_n(\bbeta^*)] = \bm{0}$, where the expectation is w.r.t. the joint distribution of $(X, Y)$. Moreover, since $Y - p_k(X;\bbeta^*)$ is bounded, it is a sub-Gaussian variable with a bounded variance proxy. Therefore, each coordinate of $\nabla \mL_n(\bbeta^*)$ is a zero-mean sub-exponential variable, which entails that $\infnorm{\nabla \mL_n(\bbeta^*)} \lesssim \sqrt{\frac{log(Kp)}{n}}$ with probability at least $1-C(Kp)^{-1}$. Then, we can follow the same argument in the proof of Theorem \ref{thm: estimation error beta} to obtain the same high-probability bounds. 

    The conclusion in Theorem \ref{thm: error} can be proved in the same way as Theorem \ref{thm: error}, hence we omit the proof.

    Finally, we comment on the proof of asymptotic normality in Theorem \ref{thm: clt}. Note that we can obtain the same results in Lemma \ref{lem: estimation error gamma} based on the same derivation. With the estimation error bounds for $\bTheta$ in Lemma \ref{lem: estimation error gamma} at hand, we can walk through the arguments in the proof of Theorem \ref{thm: clt}. First, note that we can decompose the $\hat{b}_j - \beta^*_j$ as
    \begin{align}
        \widehat{b}_j -\beta^*_j &= \underbrace{(e_j-\hTheta_{j}^\top\bm{B}_n)(\hbbeta-\bbeta^*)}_{[1]} + \underbrace{\frac{1}{n}(\hTheta_j-\bTheta_j)^\top\begin{pmatrix}
			\bX^\top\bepsilon_1 \\
			\bX^\top\bepsilon_2 \\
			\vdots \\
			\bX^\top\bepsilon_{K-1}
		\end{pmatrix}}_{[2]} + \frac{1}{n}\underbrace{\bTheta_{j}^\top \begin{pmatrix}
			\bX^\top\bepsilon_1 \\
			\bX^\top\bepsilon_2 \\
			\vdots \\
			\bX^\top\bepsilon_{K-1}
		\end{pmatrix}}_{[3]},
    \end{align}
    where $\bepsilon_k \coloneqq \byk{k} - p_k(\bX; \bbeta^*)$ for $k = 1:(K-1)$. Our previous analysis of \eqref{eq: nabla L miss} and the estimator error of $\hbbeta$ guarantees that [1] and [2] can be bounded similarly as in the proof of Theorem \ref{thm: clt}, which finally leads to the conclusion that $[1], [2] =\smallo_{\tp}(n^{-1/2})$. 

    Denote $\bSigma = \textup{Cov}((X^\top\epsilon_1, X^\top\epsilon_2, \ldots, X^\top\epsilon_{K-1})^\top)$. Then by CLT, we have
    \begin{equation}
        \frac{1}{\sqrt{\bTheta_j^\top \bSigma \bTheta_j}}\left[\sqrt{n}[3]-\frac{1}{\sqrt{n}}\bTheta_j^\top \te\begin{pmatrix}
            \bX^\top \epsilon_1 \\
            \bX^\top \epsilon_2 \\
            \vdots \\
            \bX^\top \epsilon_{K-1}
        \end{pmatrix}\right] \overset{\textup{d}}{\to} N(0, 1),
    \end{equation}
    where 
    \begin{equation}
        \frac{1}{n}\te\begin{pmatrix}
            \bX^\top \epsilon_1 \\
            \bX^\top \epsilon_2 \\
            \vdots \\
            \bX^\top \epsilon_{K-1}
        \end{pmatrix} = \te [\nabla \mL_n(\bbeta^*)] = \bm{0}.
    \end{equation}
    Then by Slutsky's theorem, we have
    \begin{equation}
        \frac{\sqrt{n}[3]}{\sqrt{\bTheta_j^\top \bSigma \bTheta_j}} \overset{\textup{d}}{\to} N(0, 1).
    \end{equation}
    Combining the analysis on all three terms $[1], [2], [3]$, it holds that
    \begin{equation}\label{eq: clt approx first}
        \frac{\sqrt{n}(\widehat{b}_j-\beta^*_j)}{\sqrt{\bTheta_j^\top \bSigma \bTheta_j}} \overset{\textup{d}}{\to} N(0, 1).
    \end{equation}
    Finally, it suffices to derive $\norm{\hTheta_j^\top \hSigma \hTheta - \bTheta_j^\top \bSigma \bTheta_j} = \smallo_{\tp}(1)$, and the rest would follow from \eqref{eq: clt approx first} and Slutsky's theorem. Note that similar to the arguments in the proof of Theorem \ref{thm: clt}, it holds that
    \begin{align}
	&\norma{\hTheta_j^\top\hSigma\hTheta_j - \bTheta_j^\top\bSigma\bTheta_j} \\
	&\leq \norma{(\hTheta_j-\bTheta_j)^\top\hSigma\hTheta_j} + \norma{\bTheta_j^\top(\hSigma-\bSigma)^\top\hTheta_j} + \norma{\bTheta_j^\top\bSigma(\hTheta_j-\bTheta_j)} \\
    &\leq \norma{(\hTheta_j-\bTheta_j)^\top(\hSigma-\bSigma)(\hTheta_j-\bTheta_j)} + \norma{(\hTheta_j-\bTheta_j)^\top(\hSigma-\bSigma)\bTheta_j} + \norma{(\hTheta_j-\bTheta_j)^\top\bSigma(\hTheta_j-\bTheta_j)}  + \norma{(\hTheta_j-\bTheta_j)^\top\bSigma\bTheta_j}\\
    &\quad + \norma{\bTheta_j^\top(\hSigma-\bSigma)^\top(\hTheta_j-\bTheta_j)} +  \norma{\bTheta_j^\top(\hSigma-\bSigma)^\top\bTheta_j} + \norma{\bTheta_j^\top\bSigma(\hTheta_j-\bTheta_j)}\\
	&\lesssim \onenorm{\hTheta_j-\bTheta_j}^2\maxnorm{\hSigma-\bSigma} + \onenorm{\hTheta_j-\bTheta_j}\maxnorm{\hSigma-\bSigma}\onenorm{\bTheta_j} + \twonorm{\hTheta_j-\bTheta_j}^2 + \twonorm{\hTheta_j-\bTheta_j}\twonorm{\bSigma\bTheta_j}\\
    &\lesssim_{\tp} \maxnorm{\hSigma-\bSigma}\sqrt{Ks_0} + \twonorm{\hTheta_j-\bTheta_j} \\
    &\lesssim_{\tp} K^3\sqrt{\frac{ss_0\log(Kp)}{n}} + K^{9/2}\sqrt{\frac{s(\log (Kp))(\log(np))}{n}} + K^{5/2}\sqrt{\frac{s_0\log (Kp)}{n}} \\
	&= \smallo(1),
\end{align}
which completes the proof.
\end{proof}

\begin{proof}[Proof of Theorem \ref{thm: non id}]
    The results in Theorems \ref{thm: estimation error beta}, \ref{thm: error}, and \ref{thm: clt} can be derived in a similar way for the non-identically distributed data, as we argued in Section \ref{subsubsec: non id}, thus we omit the proof.
\end{proof}

\begin{proof}[Proof of Theorem \ref{thm: clt second}]
    By the definition of $\widehat{\bm{b}}$:
    \begin{align}
        \widehat{\bm{b}} - \bbeta^* &= (\bm{I}_{(K-1)p} - \bm{M}\bm{B}_n^{[1]})(\hbbeta^{[1]} - \bbeta^*) + \frac{1}{n}\bm{M}\begin{pmatrix}
            (\bX^{[1]})^\top \bepsilon^{[1](1)} \\
            \vdots \\
            (\bX^{[1]})^\top \bepsilon^{[1](K-1)}
        \end{pmatrix} \\
        &= (\bm{I}_{(K-1)p} - \bm{M}\hSigma_{\hbbeta^{[2]}}^{[2]})(\hbbeta^{[1]} - \bbeta^*) + \bm{M}(\hSigma_{\hbbeta^{[2]}}^{[2]} - \bm{B}_n^{[1]})(\hbbeta^{[1]} - \bbeta^*) + \frac{1}{n/2}\bm{M}\begin{pmatrix}
            (\bX^{[1]})^\top \bepsilon^{[1](1)} \\
            \vdots \\
            (\bX^{[1]})^\top \bepsilon^{[1](K-1)}
        \end{pmatrix}, 
    \end{align}
    where 
    \begin{equation}
	   \bm{B}_n^{[1]} = \frac{1}{n/2}\sum_{i=1}^{n/2} \int_0^1 \bm{B}(\bx_i^{[1]}; \bbeta^* + t(\hbbeta^{[1]}-\bbeta^*)) dt,
    \end{equation}
    $$\bm{B}(\bx; \bbeta) \coloneqq \begin{pmatrix}
    	p_1(\bx;\bbeta)[1-p_1(\bx;\bbeta)]\bx\bx^\top & -p_1(\bx;\bbeta)p_2(\bx;\bbeta)\bx\bx^\top &\ldots &-p_1(\bx;\bbeta)p_{K-1}(\bx;\bbeta)\bx\bx^\top \\
    	-p_1(\bx;\bbeta)p_2(\bx;\bbeta)\bx\bx^\top &p_2(\bx;\bbeta)[1-p_2(\bx;\bbeta)]\bx\bx^\top &\ldots &-p_2(\bx;\bbeta)p_{K-1}(\bx;\bbeta)\bx\bx^\top \\
    	\vdots &\vdots  &\vdots &\vdots \\
    	  -p_1(\bx;\bbeta)p_{K-1}(\bx;\bbeta)\bx\bx^\top &-p_2(\bx;\bbeta)p_{K-1}(\bx;\bbeta)\bx\bx^\top &\ldots &p_{K-1}(\bx;\bbeta)[1-p_{K-1}(\bx;\bbeta)]\bx\bx^\top
    \end{pmatrix}.$$    
    \underline{\textbf{Step 1:}} Show that when $\eta \asymp K^{5/2}\sqrt{\frac{s\log p}{n}}$, the LP \eqref{eq: LP} is feasible with probability $1-\smallo(1)$.

    Consider $\bm{M} = \bTheta$ with $\bm{m}_j = \bTheta_j$:
    \begin{equation}
        \bTheta\hSigma_{\hbbeta^{[2](k)}}^{[2]} - \bm{I} = \bTheta(\hSigma_{\hbbeta^{[2](k)}}^{[2]} - \hSigma_{\bbeta^*}^{[2]}) + \bTheta \hSigma_{\bbeta^*}^{[2]} - \bm{I}.
    \end{equation}
    There exist functions $\{g_{j_1,j_2}\}$ such that
    \begin{equation}
        (\hSigma_{\hbbeta^{[2](k)}}^{[2]})_{j_1,j_2} = \frac{1}{n/2}\sum_{i=1}^{n/2} g_{j_1,j_2}(\{\bx_i^\top \hbbeta^{[2](k)}\}) (\bx_i^{[2]})_{j_1 \textup{ mod } K}(\bx_i^{[2]})_{j_2 \textup{ mod } K}, \quad (\hSigma_{\bbetaks{k}}^{[2]})_{j_1,j_2} = \frac{1}{n/2}\sum_{i=1}^{n/2} g_{j_1,j_2}(\{(\bx_i^{[2]})^\top \bbetaks{k}\}) (\bx_i^{[2]})_{j_1 \textup{ mod } K}(\bx_i^{[2]})_{j_2 \textup{ mod } K},
    \end{equation}
    where $\twonorm{\nabla g_{j_1,j_2}} \lesssim 1$. Hence, by Cauchy-Schwarz inequality, we have
    \begin{align}
        \big|\bm{e}_1^\top \bTheta(\hSigma_{\hbbeta^{[2]}}^{[2]} - \hSigma_{\bbeta^*}^{[2]}) \bm{e}_1\big| &= \Bigg|\frac{1}{n}\sum_{i=1}^n\sum_{k=1}^{K-1} [\nabla g_{1,1}(\{(1-\nu_i)\bx_i^\top \hbbeta^{[2]} + \nu_i\bx_i^\top \bbetaks{k}\}_{k=1}^{K-1})]^\top \{\bx_i^\top (\hbbeta^{[2](k)} - \bbetaks{k})\}_{k=1}^{K-1} \cdot (\bTheta^{(k)}_1)^\top \bx_i\cdot x_{i,1}\Bigg| \\
        &\lesssim \sqrt{\frac{1}{n}\sum_{i=1}^n\sum_{k=1}^{K-1} [(\bTheta^{(k)}_1)^\top \bx_i]^2 x_{i,1}^2} \cdot \sqrt{\frac{1}{n}\sum_{i=1}^n \sum_{k=1}^{K-1}[\bx_i^\top (\hbbeta^{[2](k)} - \bbetaks{k})]^2} \\
        &\lesssim \twonorm{\hbbeta^{[2]} - \bbeta^*} \\
        &\lesssim K^{5/2}\sqrt{\frac{s\log(Kp)}{n}},
    \end{align}
    with probability at least $1-C'(Kp)^{-1}$. Similarly, we can show that
    \begin{equation}
        \maxnorm{\bTheta(\hSigma_{\hbbeta^{[2]}}^{[2]} - \hSigma_{\bbeta^*}^{[2]})} \lesssim K^{5/2}\sqrt{\frac{s\log(Kp)}{n}}
    \end{equation}
    with probability at least $1-C'(Kp)^{-1}$. Additionally, similar to the argument in the proof of Lemma \ref{lem: estimation error gamma}, we can show that
    \begin{equation}
        \maxnorm{\hSigma_{\bbeta^*}^{[2]} - \bm{I}} \lesssim \sqrt{\frac{\log(Kp)}{n}},
    \end{equation}
    with probability at least $1-C(Kp)^{-1}$. Therefore, with probability at least $1-C(Kp)^{-1}$, we have
    \begin{equation}
        \maxnorm{\bTheta^*\hSigma_{\hbbeta^{[2]}}^{[2]} - \bm{I}} \lesssim K^{5/2}\sqrt{\frac{s\log (Kp)}{n}},
    \end{equation}
    which implies that the LP \eqref{eq: LP} is feasible when $\eta \gtrsim K^{5/2}\sqrt{\frac{s\log (Kp)}{n}}$.

    \noindent \underline{\textbf{Step 2:}} We show that $\big|\bm{m}_j^\top(\hSigma_{\hbbeta}^{[2]} - \bm{B}_n^{[1]})(\hbbeta^{[2]} - \bbeta^*)\big| \lesssim \onenorm{\bTheta_j}\cdot \frac{K^5s\log(Kp)}{n}$ with probability $1-\smallo(1)$.

    Since $\bTheta_j$ is feasible for the LP \eqref{eq: LP}, we must have $\onenorm{\bm{m}_j} \leq \onenorm{\bTheta_j}$. Therefore,
    \begin{equation}
        \big|\bm{m}_j^\top(\hSigma_{\hbbeta^{[2]}}^{[2]} - \bm{B}_n^{[1]})(\hbbeta^{[2]} - \bbeta^*)\big| \leq \big|\bm{m}_j^\top(\hSigma_{\hbbeta^{[2]}}^{[2]} - \hSigma_{\bbeta^*}^{[2]})(\hbbeta^{[2]} - \bbeta^*)\big| + \big|\bm{m}_j^\top(\hSigma_{\bbeta^*}^{[2]} - \bSigma)(\hbbeta^{[2]} - \bbeta^*)\big| + \big|\bm{m}_j^\top(\bSigma - \bm{B}_n^{[1]})(\hbbeta^{[2]} - \bbeta^*)\big|, 
    \end{equation}
    where 
    \begin{align}
        \big|\bm{m}_j^\top(\hSigma_{\hbbeta^{[2]}}^{[2]} - \hSigma_{\bbeta^*}^{[2]})(\hbbeta^{[2]} - \bbeta^*)\big| &\leq \onenorm{\bm{m}_j}\twonorm{\hbbeta^{[2]} - \bbeta^*}^2 \lesssim \onenorm{\bTheta_j}\frac{K^5s\log(Kp)}{n}, \\
        \big|\bm{m}_j^\top(\hSigma_{\bbeta^*}^{[2]} - \bSigma)(\hbbeta^{[2]} - \bbeta^*)\big| &\leq \onenorm{\bm{m}_j}\maxnorm{\hSigma_{\bbeta^*}^{[2]} - \bSigma}\onenorm{\hbbeta^{[2]} - \bbeta^*}  \lesssim \onenorm{\bTheta_j}\frac{K^3s\log(Kp)}{n}, \\
        \big|\bm{m}_j^\top(\bSigma - \bm{B}_n^{[1]})(\hbbeta^{[2]} - \bbeta^*)\big| &\leq \onenorm{\bm{m}_j}\maxnorm{\bSigma - \bm{B}_n^{[1]}}\onenorm{\hbbeta^{[2]} - \bbeta^*}  \lesssim \onenorm{\bTheta_j} \frac{K^5s\log(Kp)}{n},
    \end{align}
    with probability at least $1-C'(Kp)^{-1}$. Therefore,
    \begin{equation}
         \big|\bm{m}_j^\top(\hSigma_{\hbbeta}^{[2]} - \bm{B}_n^{[1]})(\hbbeta^{[2]} - \bbeta^*)\big| \lesssim \onenorm{\bTheta_j}\cdot \frac{K^5s\log(Kp)}{n},
    \end{equation}
    with probability at least $1-C'(Kp)^{-1}$. 

    \noindent \underline{\textbf{Step 3:}} We complete the proof by applying CLT. 

    Note that 
    \begin{equation}
        \|(\bm{e}_j - \bm{m}_j^\top\hSigma_{\hbbeta^{[2]}}^{[2]})(\hbbeta^{[1]} - \bbeta^*)\| \leq \infnorm{\hSigma_{\hbbeta^{[2]}}^{[2]} \bm{m}_j - \bm{e}_j}\onenorm{\hbbeta^{[1]} - \bbeta^*} \lesssim \eta K^3\sqrt{\frac{s\log(Kp)}{n}} \lesssim K^{11/2}s^{3/2}\frac{\log(Kp)}{n},
    \end{equation}
    with probability at least $1-C'(Kp)^{-1}$. Because $\bm{m}_j$ is independent of $\bX^{[1]}$ and $\bepsilon^{[1]}$, conditioned on $(\bX^{[2]}, \bY^{[2]})$, by the CLT,
    \begin{equation}
        \frac{1}{\sqrt{n/2}\sqrt{\bm{m}_j^\top \bSigma \bm{m}_j}}\bm{m}_j^\top\begin{pmatrix}
            (\bX^{[1]})^\top \bepsilon^{[1](1)} \\
            \vdots \\
            (\bX^{[1]})^\top \bepsilon^{[1](K-1)}
        \end{pmatrix} \overset{\textup{d}}{\to} N(0, 1).
    \end{equation}
    By the definition of weak convergence through the expectation of continuous bounded functions, the weak convergence also holds with the $\sigma$-field generated by all the data $(\bX^{[1]}, \bY^{[1]}) \cup (\bX^{[2]}, \bY^{[2]})$.

    Finally, we have
    \begin{align}
        \frac{\sqrt{n/2}(\hat{b}_j - \beta^*_j)}{\sqrt{\bm{m}_j^\top \bSigma \bm{m}_j}} &= \mathcal{O}_{\tp}\bigg(\onenorm{\bTheta_j}\cdot \frac{K^5s\log(Kp)}{\sqrt{n}} + K^{11/2}s^{3/2}\frac{\log(Kp)}{\sqrt{n}}\bigg) +  \frac{1}{\sqrt{n/2}}\bm{m}_j^\top\begin{pmatrix}
            (\bX^{[1]})^\top \bepsilon^{[1](1)} \\
            \vdots \\
            (\bX^{[1]})^\top \bepsilon^{[1](K-1)}
        \end{pmatrix} \\
        &\overset{\textup{d}}{\to} N(0, 1).
    \end{align}
    It suffices to prove $\big|\bm{m}_j^\top \hSigma^{[1]} \bm{m}_j - \bm{m}_j^\top \bSigma \bm{m}_j\big| = \smallo_{\tp}(1)$, which follows from a similar argument used in the proof of Theorem \ref{thm: clt}, hence we omit it here.
\end{proof}

\begin{proof}[Proof of Theorem \ref{thm: clt beyond boundedness}]
Most of the proof follows the same argument in the proof of Theorem \ref{thm: clt}, therefore we only point out the different places. 

First, we need to show the high-probability bound for $\twonorm{\hgamma_j - \bgamma^*_j}$ and $\onenorm{\hgamma_j - \bgamma^*_j}$, respectively. Instead of the way we took to bound $\twonorm{\hgamma_j - \bgamma^*_j}$ in the proof of Theorem \ref{thm: clt}, we use the fact that
\begin{align}
    \infnorm{\hSigma_{-j,j} - \hSigma_{\bbeta^*,-j, j} - \hSigma_{-j,-j}\bgamma^*_j + \hSigma_{\bbeta^*, -j,-j}\bgamma^*_j} &\lesssim \twonorm{\hbbeta - \bbeta^*} \lesssim K^{5/2}\sqrt{\frac{s\log(Kp)}{n}}, \\
    \big|(\hSigma_{-j,j} - \hSigma_{\bbeta^*,-j, j} - \hSigma_{-j,-j}\bgamma^*_j + \hSigma_{\bbeta^*, -j,-j}\bgamma^*_j)^\top(\hgamma_j-\bgamma^*_j)\big| &\lesssim \infnorm{\hSigma_{-j,j} - \hSigma_{\bbeta^*,-j, j} - \hSigma_{-j,-j}\bgamma^*_j + \hSigma_{\bbeta^*, -j,-j}\bgamma^*_j}\onenorm{\hgamma_j - \bgamma^*_j},
\end{align}
with probability at least $1-(Kp)^{-1}$, which implies that
\begin{equation}\label{eq: cone supp}
    \onenorm{(\hgamma_j - \bgamma^*_j)_{S_j^c}} \lesssim \onenorm{(\hgamma_j - \bgamma^*_j)_{S_j}}.
\end{equation}
with probability at least $1-(Kp)^{-1}$.

Second, using \eqref{eq: cone supp}, with a similar analysis in the proof of Lemma \ref{lem: rsc}, we can prove that
\begin{equation}
    \frac{1}{K^2}\twonorm{\hgamma_j - \bgamma^*_j} \lesssim \frac{1}{2n}(\hgamma_j - \bgamma^*_j)^\top\hSigma_{-j,-j}(\hgamma_j - \bgamma^*_j) \lesssim \lambda_j\onenorm{(\hgamma_j - \bgamma^*_j)_{S_j}} \lesssim \lambda_j\sqrt{Ks_0}\twonorm{(\hgamma_j - \bgamma^*_j)_{S_j}},
\end{equation}
with probability at least $1-(Kp)^{-1}$, which implies that
\begin{equation}
    \twonorm{\hgamma_j - \bgamma^*_j} \lesssim K^5\sqrt{ss_0}\sqrt{\frac{\log (Kp)}{n}}, \quad \onenorm{\hgamma_j - \bgamma^*_j} \lesssim K^{11/2}\sqrt{s}s_0\sqrt{\frac{\log (Kp)}{n}},
\end{equation}
with probability at least $1-(Kp)^{-1}$. Then, by following the same argument in the proof of Lemma \ref{lem: estimation error gamma}, we can show that
\begin{equation}
    |\htau^2_j - (\tau^*_j)^2|, |\htau^{-2}_j - (\tau^*_j)^{-2}| \lesssim K^5\sqrt{ss_0}\sqrt{\frac{\log (Kp)}{n}},
\end{equation}
with probability at least $1-(Kp)^{-1}$, when $n \gtrsim K^{11}ss_0^2\log(pK)$. 

Finally, we follow the steps in the proof of Theorem \ref{thm: clt} and argue that $\twonorm{\hgamma_j - \bgamma^*_j}^2\vee |\htau^{-2}_j - (\tau^*_j)^{-2}|^2 \vee |\htau^2_j - (\tau^*_j)^2|^2 \vee \twonorm{\hbbeta - \bbeta^*}^2 \ll n^{-1/2}$ with probability at least $1-(Kp)^{-1}$, when $n \gg K^{20}(ss_0)^2(\log(Kp))^2$, which completes the proof.
\end{proof}

\end{document}